\documentclass[12pt]{article}
\usepackage{authblk}
\usepackage[english]{babel}
\usepackage{amssymb}
\usepackage{amsmath}
\allowdisplaybreaks
\usepackage{amsbsy}
\usepackage{epsfig}
\usepackage{amsthm} 
\usepackage{bbold}         
\usepackage{tikz,tikz-feynman}
\usepackage{textcomp, gensymb}
\usepackage{stackrel}
\usepackage[font=small,labelfont=bf]{caption}
\usepackage{cancel}
\usepackage{xcolor}
\usepackage{cite}
\usepackage{hyperref}
\usetikzlibrary{positioning}
\usetikzlibrary{arrows}
\usetikzlibrary{calc}
\tikzset{shifted path/.style args={from #1 to #2 by #3}{insert path={
let \p1=($(#1.east)-(#1.center)$),
\p2=($(#2.east)-(#2.center)$),\p3=($(#1.center)-(#2.center)$),
\n1={veclen(\x1,\y1)},\n2={veclen(\x2,\y2)},\n3={atan2(\y3,\x3)} in
(#1.{\n3+180+asin(#3/\n1)}) to (#2.{\n3-asin(#3/\n2)})
}}}

\usepackage[capitalise]{cleveref}

\crefname{section}{sec.}{secs.}
\crefname{table}{tab.}{tabs.}
\crefname{figure}{fig.}{figs.}
\crefname{equation}{eq.}{eqs.}
\crefname{appendix}{Appendix\ }{Appendix\ }

\definecolor{darkgreen}{rgb}{0,0.5,0}
\definecolor{darkred}{rgb}{0.7,0,0}

\newcommand{\myslash} [1] {#1 \kern-.5em/}

\newcommand{\tr}{\text{Tr}}

\newcommand{\vergiss}[1]{}

\def\gsim{\raise0.3ex\hbox{$\;>$\kern-0.75em\raise-1.1ex\hbox{$\sim\;$}}}
\def\lsim{\raise0.3ex\hbox{$\;<$\kern-0.75em\raise-1.1ex\hbox{$\sim\;$}}}

\newcommand{\ii}{\mathrm{i}}

\newcommand{\oD}{\overline{D}}
\newcommand{\oG}{\overline{G}}
\newcommand{\oL}{\overline{\Lambda}}
\newcommand{\op}{\overline{\phi}}
\newcommand{\oPi}{\overline{\Pi}}
\newcommand{\oV}{\overline{V}}

\tikzstyle{Gamma}=[circle,draw=black,fill=black,thick,inner sep=0pt,minimum size=6mm]
\tikzstyle{VEV}=[circle,draw=black,thick,inner sep=0pt,minimum size=2mm]
\tikzstyle{Pi}=[rectangle,draw=black,fill=gray,thick,inner sep=0pt,minimum size=6mm]
\tikzstyle{Sigma}=[rectangle,draw=black,fill=cyan,thick,inner sep=0pt,minimum size=6mm]
\tikzstyle{VR}=[rectangle,draw=black,thick,inner sep=0pt,minimum size=6mm]

\title{\textbf{Renormalized equations of motions 
for scalars and fermions in the 2PI formalism}}

\author{A.~Banik\,\thanks{E-mail: \href{mailto:amitayus.banik@uni-wuerzburg.de}{amitayus.banik@uni-wuerzburg.de}}}
\author{H.~Hinrichsen\thanks{E-mail: \href{mailto:hinrichsen@physik.uni-wuerzburg.de}{hinrichsen@physik.uni-wuerzburg.de}}}
\author{W.~Porod\thanks{E-mail: \href{mailto:porod@physik.uni-wuerzburg.de}{porod@physik.uni-wuerzburg.de}}}
\affil{\textit{Institut f\"{u}r Theoretische Physik und Astrophysik, Universit\"{a}t W\"{u}rzburg, D-97074 W\"{u}rzburg, Germany}}

\date{} 

\begin{document}

\maketitle
\begin{abstract}
We present on shell-scheme for the 2PI formalism
with a particular focus on the renormalized
equations of motion. We first revisit the
so-called Hartree approximation where
we give the counterterms for both the broken
and unbroken phase. Moreover, we give
explicit formulas for the renormalized
three- and four-point functions in the broken phase.
We then turn to the sunset approximation, with only scalars and then including fermions. We give explicit formulas for the 
wavefunction and mass counterterms. Moreover,
we show that, in particular, the two-point functions
can be obtained numerically in a fast converging
scheme even for large couplings of order one.
\end{abstract}

\section{Introduction}
The vast majority of physical phenomena, which are vital to our understanding of nature, take place out of equilibrium. Non-equilibrium processes can be found in a wide range of physical domains ranging from particle physics and cosmology to astrophysics and condensed matter systems. A few examples are heavy-ion collisions performed, for instance, at the Large Hadron Collider (LHC) with the aim to produce quark-gluon plasma, the generation of density fluctuations during inflation and the explosive particle production at the end of inflation, as well as phase transitions in the early universe or in condensed matter systems. These phenomena also go beyond the reach of standard perturbation theory, meaning one needs to resort to non-peturbative methods for consistent results.

The challenge of addressing models beyond perturbation theory and tackling out-of-equilibrium aspects can be fulfilled using functional integral techniques based on $n$-particle irreducible ($n$PI) effective actions. In this work, we focus on the two-particle irreducible (2PI) formalism \cite{Jackiw:1974cv,Cornwall:1974vz}, where the expectation value of the field (one-point function) and the propagator (two-point function) constitute the dynamical degrees of freedom. 

Resummation schemes based on the 2PI effective action have been known to show better convergence properties in comparison to other methods, such as in bosonic finite temperature field theory \cite{Blaizot:2000fc,Andersen:2004re,Berges:2004hn} and inflationary preheating \cite{Arrizabalaga:2004iw}. Out-of-equilibrium properties can be formally studied, whereby the resummation feature of the 2PI formalism allows one to obtain approximations uniform in time \cite{PhysRevD.37.2878, Ivanov:1998nv, Berges:2004yj, Berges:2015kfa}. Furthermore, approximations within the 2PI formalism have been shown to be consistent with (global) conservation laws stemming from Noether's theorem, thus guaranteeing charge and energy conservation \cite{Arrizabalaga:2005tf,Berges:2010nk}. 
Finally, using the equations of motion of the 2PI formalism
one can obtain, in principle, the transport equations
\cite{Konstandin:2013caa,Jukkala:2019slc}
which are an important ingredient in the study 
of cosmological phase transitions, see for e.g.~\cite{Prokopec:2003pj,Prokopec:2004ic,Konstandin:2013caa}.

By definition, approximation schemes based on the 2PI formalism involve  resummation of the two-point function to all orders in perturbation theory, when one chooses a particular truncation for the expansion, such as restricting to a given loop order. Despite this selective resummation, 2PI methods show renormalizability through local counterterms, as demonstrated in \cite{Blaizot:2003br, Blaizot:2003an, Carrington:2014lba, Carrington:2017lry} for a symmetric scalar theory with a quartic coupling. The procedure to obtain the counterterms for $\text{O}(N)$ symmetric theories has been explored in \cite{Patkos:2008ik, Patkos:2008sg, Pilaftsis:2013xna, Pilaftsis:2015cka, Pilaftsis:2017enx}. Progress has also been with regard to renormalizability with the inclusion of 
abelian gauge bosons and consistency with Ward identities \cite{Berges:2004hn, Reinosa:2006cm, Reinosa:2007vi, Reinosa:2009tc, Oliveira:2022bar}, as well as the inclusion of fermions \cite{Reinosa:2005pj}. A systematic approach
for the renormalization of the 2PI action 
been presented in \cite{Berges:2005hc} focusing
on scalars.

An aspect which has to our knowledge not been
discussed so far is how to obtain an on-shell
renormalization scheme for the 2PI formalism.
Our aim
is to obtain this for a coupled system of fermions
and scalars.
We will do this with a particular focus on
the renormalized equations of motions which 
in turn, can be used as starting point for transport
equations in cosmological phase transitions.
In the present work, we demonstrate the applicability of the techniques in \cite{Berges:2005hc} to various truncations of the 2PI effective action at two-loop order, and extend the techniques to incorporate fermions in the 2PI formalism. Our main aim is to present the renormalized equations of motion, after which we focus on the necessary steps to obtain the various counterterms.

The paper is organized as follows: in Section \ref{sec:eqm}, 
we first recall relevant features of the 2PI formalism, focusing in particular on obtaining the renormalized equations of motion.
It is well-known that these equations contain 
several counterterms which need to be fixed by appropriate renormalization conditions. To this end, 
we present a suitable on-shell scheme in  
 Section~\ref{sec:gen_renorm}, 
which allows for connecting to physical observables in a straightforward fashion. 
We then apply this scheme to various two-loop truncations of the 2PI effective action.
We start in Section \ref{sec:scalars} with the well-known Hartree approximation. We find, in particular, 
that all counterterms are finite
and give explicit 
relations between the broken and unbroken phases for 
these counterterms. 
In the Hartree approximation, one can give 
analytic expressions for all interesting quantities, though
this does not hold in other cases.
Therefore, we proceed to the so-called scalar sunset 
approximation to demonstrate the intricacies and how 
to resolve them. This model also provides a simple 
example containing a so-called memory integral in the 
equation of motion which are important for the 
thermalisation of a system, see e.g.~\cite{Berges:2015kfa}.
Within this framework, we obtain the gap equation to 
obtain the renormalized two-point function and solve it 
using an iterative procedure. We demonstrate that
this procedure converges very fast even for large 
couplings.

In Section \ref{sec:fermions}, we take the fermionic sunset approximation, which serves as the simplest example to include fermions in the 2PI formalism. Here, we obtain coupled system of gap equations for the fermionic and scalar propagators which we again solve using the aforementioned iterative procedure, finding sufficiently fast convergence for an $\mathcal{O}(1)$ Yukawa coupling. We find that the inclusion of fermions renders none of the counterterms finite. Finally, in Section \ref{sec:outlook}, we present our conclusions and give an outlook. 

\section{The 2PI Formalism and Renormalized Equations of Motion}
\label{sec:eqm}

In order to fix notation, let us first summarize some key aspects of the 2PI formalism in the 
example of a single scalar field~\cite{Cornwall:1974vz}, with the generating functional as starting point
\begin{align}
Z[J_1,J_2] = \int {\cal D}\varphi \exp\left(\ii S[\varphi] + \ii \int_x J_1(x) \varphi(x) + \frac{\ii}{2}\int_{xy}  \varphi(x) J_2(x,y) \varphi(y)  \right)\,,
\end{align}
where $x$ and $y$ denote vectors in a $d$-dimensional Minkowski space with the mostly minus signature and $J_1$ and $J_2$ are two external currents suitably shifted in order to account for a Gaussian initial state. Since one is interested in the non-equilibrium evolution, where a final state is not known, the temporal integration in
\begin{align}
\int_x \equiv \int_{\mathcal{C}} dx_0\int d^{d-1} x
\end{align}
is carried out along a time-ordered Keldysh contour $\mathcal{C}$. In terms of the 
cumulant-generating functional $W[J_1,J_2] = - \ii \ln Z[J_1,J_2]$, the macroscopic field and the connected two-point correlator, defined as $\phi(x)=\langle\varphi(x)\rangle_{\mathcal C}$ and  $G(x,y)=\langle\varphi(x)\varphi(y)\rangle_{\mathcal C}$, are given by
\begin{equation}
\phi(x) = \frac{\delta  W[J_1,J_2]}{\delta J_1(x)} \,, \qquad
G(x,y) = 2 \frac{\delta  W[J_1,J_2]}{\delta J_2(x,y)} - \phi(x) \phi(y) \,.
\end{equation}
It is more practical to describe the system by the Legendre transform of $W[J_1,J_2]$, the so-called 2PI effective action
\begin{align}
\Gamma^{\text{2PI}}[\phi,G] &= W[J_1,J_2] - \int_x  \frac{\delta W[J_1,J_2]}{\delta J_1(x)}J_1(x) - \int_{xy}  \frac{\delta W[J_1,J_2]}{\delta J_2(x,y)}J_2(x,y)\\
&= W[J_1,J_2] - \int_x  \phi(x)J_1(x) - \frac{1}{2} \int_{xy} \Bigl( \phi(x)\phi(y)+ G(x,y) \Bigr)J_2(y,x)
\end{align}
which allows the currents to be expressed as
\begin{equation}
J_1(x)=-\frac{\delta \Gamma^{\text{2PI}}[\phi,G] }{\delta\phi}-\int_y J_2(x,y) \phi(y)\,,\qquad J_2(x,y) = -2 \frac{\delta \Gamma^{\text{2PI}}[\phi,G]}{\delta G(x,y)}\,.
\end{equation}
The  effective action can be split into $ \Gamma^{\text{2PI}}[\phi,G] = \Gamma_1^{\text{2PI}}[\phi,G] + \Gamma_2^{\text{2PI}}[\phi,G]$, where
\begin{equation}
\Gamma_1^{\text{2PI}}[\phi,G]=S[\phi] + \frac{\ii}{2}    \tr\left[\ln G^{-1} \right] 
  + \frac{\ii}{2} \tr\left[\tilde G^{-1}_\phi G \right] -  \underbrace{\frac{\ii}{2} \tr\left[G^{-1} G \right] }_{=\textrm{const}}
\end{equation}
comprises the classical action and all 1-loop contributions while $\Gamma_2^{\text{2PI}}[\phi,G]$ accounts for higher contributions from 2-loop onward. ``$\tr$'' refers to integration over the space-time variables. Here, 
\begin{align}
\tilde G^{-1}_\phi = -\ii \frac{\delta^2S[\phi]}{\delta\phi(x)\delta\phi(y)}
\end{align} 
denotes the classical inverse propagator and the trace stands for integration along the Keldysh contour. In practical calculations, $\Gamma_2^{\text{2PI}}[\phi,G]$ is only evaluated up to given loop order.

In view of the renormalization procedure, it is useful to split the action into a free and an interacting part $S[\phi] = S_0[\phi] + S_{\text{int}}[\phi]$ and to decompose $\tilde G^{-1}_{\phi}=\tilde G^{-1}_{0}+\tilde G^{-1}_{\phi,\textrm{int}}$ \cite{Berges:2005hc}. Correspondingly one can split the effective action into $\Gamma^{\text{2PI}}=\Gamma_0^{\text{2PI}}+\Gamma_{\textrm{int}}^{\text{2PI}}$, where
\begin{align}
\Gamma_0^{\text{2PI}}[\phi,G]&=
S_0[\phi] + \frac{\ii}{2}    \tr\left[\ln G^{-1} \right] + \frac{\ii}{2}\tr[\tilde G^{-1}_{0} G] \,, \\
\Gamma_\textrm{int}^{\text{2PI}}[\phi,G]&=
S_{\text{int}}[\phi] +  \frac{\ii}{2}\tr[\tilde G^{-1}_{\phi,\text{int}} G] + \Gamma_2[\phi,G] + \text{const.}
\end{align}
The stationarity conditions determine the physical one- and two-point functions $\overline{\phi}$ and $\overline{G}$ in the absence of external sources. i.e.~$J_1=0$ and $J_2=0$:
\begin{align}
\label{eq:extr_Gamma_phi}
\frac{\delta \Gamma^{\text{2PI}}[\phi,G]}{\delta \phi(x)} \bigg|_{\overline{\phi},\overline{G}} &= 
\frac{\delta S_{0}[\phi,G]}{\delta \phi(x)} \bigg|_{\overline{\phi}}+\frac{\delta \Gamma^{\text{2PI}}_{\text{int}}[\phi,G]}{\delta \phi(x)} \bigg|_{\overline{\phi},\overline{G}} = 0 \,, \\ 
\frac{\delta \Gamma^{\text{2PI}}[\phi,G]}{\delta G(x,y)} \bigg|_{\overline{\phi},\overline{G}} &=
- \frac{\ii}{2} \overline{G}^{-1}(x,y) + \frac{\ii}{2} \tilde G^{-1}_0(x,y) +
\frac{\delta \Gamma^{\text{2PI}}_{\text{int}}[\phi,G]}{\delta G(x,y)} \bigg|_{\overline{\phi},\overline{G}} =
0 \,.
\label{eq:extr_Gamma_G}
\end{align}  
We note for completeness that these conditions imply that the sources need to be redefined order-by-order
once a perturbative evaluation is performed, as has already been pointed out in \cite{Jackiw:1974cv}
in the context of usual 1PI effective action.

At this stage, it is useful to introduce the self-energy
\begin{align}
\label{eq:def_Pi}
\overline{\Pi}(x,y) &=  2 \ii   \frac{\delta \Gamma^{\text{2PI}}_{\text{int}}[\phi,G]}{\delta G(x,y)}\bigg|_{\overline{\phi},\overline{G}} 
\end{align}
which can be decomposed into a local part, proportional to $\delta_{\cal C}(x-y)$, and a non-local one, see below. The index ${\cal C}$ indicates that  the Delta-distribution takes values along the Keldysh contour.

The one- and two-point functions are obtained from the equations of motions (EOMs) once the corresponding boundary conditions are specified. More concretely, the EOM for $\phi$ is given by \cref{eq:extr_Gamma_phi} whereas the one for the two point function can be obtained from \cref{eq:extr_Gamma_G}  by convoluting it with  $G(y,z)$ yielding
\begin{align}
\label{eq:eom_G_generic}
(\square_x + m^2) G(x,z) + \int_y  \overline{\Pi}(x,y) G(y,z) &= \delta_{\cal C}(x-z) \,.
\end{align}

Fermions can be treated in an analogous way \cite{Cornwall:1974vz,Berges:2004yj}. 
We will denote the Dirac field by $\psi(x)$ and the 
corresponding two-point function by $D(x,y)$. With 
the assumption that the fermionic field does not 
acquire a vacuum expectation value (VEV), 
we obtain the generic equation of motion for $D(x)$ as
\begin{align}
\label{eq:eom_D_generic}
(\ii \myslash{\partial} - M) D(x,z) + \int_y  \overline{\Sigma}(x,y) D(y,z) &= \delta_{\cal C}(x-z) \,,
\end{align}
where $\overline{\Sigma}(x,y)$ is the fermionic self-energy given by
\begin{align}
\label{eq:def_Sigma}
\overline{\Sigma}(x,y) &=  \ii   \frac{\delta \Gamma^{\text{2PI}}_{\text{int}}[\phi,G,D]}{\delta  D(x,y)}\bigg|_{\overline{\phi},\overline{G},\overline{D}} \,. 
\end{align}
Later, we will renormalize the fields based on a procedure that we will outline, retaining the assumption that $\psi$ does not get a VEV.

We take the following classical action as a starting point
\begin{align}
S[\phi,\psi] &=  S_0[\phi,\psi] + S_{\text{int}}[\phi,\psi] \\
&= \int_x \Bigg\{ \frac{1}{2} \partial_\mu \phi(x) \partial^\mu \phi(x) - \frac{m^2}{2} \phi^2(x) + \bar{\psi}(x)(\ii \myslash{\partial} - M) \psi(x)  \nonumber \\ 
&  \qquad \qquad 
- \frac{\alpha}{3!} \phi^3(x) -\frac{\lambda}{4!} \phi^4(x)  - g \bar{\psi}(x)\psi(x) \phi(x)\Bigg\} 
\label{eq:classical_action}
\end{align}
where in the last step we have split the action into a free and an interaction part corresponding to the first and second line of \eqref{eq:classical_action}. We will include all contributions of the effective action up to two-loop order which is sufficient to detail all the intricacies of the renormalization, in particular in the fermionic sector. The 2PI action up to this order is given by
\begin{align}
 \Gamma_{\text{2PI}}[\phi,G,D] =&
\int_x \Bigg\{ \frac{1}{2} \partial_\mu \phi(x) \partial^\mu \phi(x) - \frac{m^2}{2} \phi^2(x) 
- \frac{\alpha}{3!} \phi^3(x) -\frac{\lambda}{4!} \phi^4(x) \nonumber \\
  - &\frac{1}{2} \left( \square_x + m^2 \right) G(x,y) \big|_{x=y} - \frac{1}{2} \alpha \, \phi(x) \, G(x,x)
  - \frac{1}{8} \lambda G^2(x,x)
  - \frac{1}{4} \lambda \,\phi^2(x) \,  G(x,x) 
  \nonumber \\
+ &  \text{tr}\left[\left(\ii \myslash{\partial}_x - M\right) 
 D(x,y)\big|_{x=y}\right] - g \, \phi(x) \, \text{tr}[D (x,x)]
 \Bigg\} \nonumber \\
+ &   \int_x \int_y \left[\frac{\ii}{12} (\alpha + \lambda \phi(x)) (\alpha + \lambda \phi(y)) \,G^3(x,y)
- \frac{\ii}{2} g^2 G(x,y) \, \text{tr}\left[D(x,y)D(y,x)\right] 
\right] \,.
\label{eq:Gamma_2PI_unrenormalized}
\end{align}
where we have already used the fact that terms linear 
in $\psi$ vanish as it does not acquire a VEV. We 
note for completeness, that the last line will give 
rise to the so-called ``memory integrals''. We stress 
that \eqref{eq:Gamma_2PI_unrenormalized} represents 
the unrenormalized action and that 
\eqref{eq:extr_Gamma_phi}, \eqref{eq:eom_G_generic} 
and  \eqref{eq:eom_D_generic} hold for both the 
unrenormalized and renormalized action.

Eventually, we are interested in obtaining the renormalized EOMs. It has been shown in \cite{Berges:2005hc} that one needs not only to renormalize the couplings appearing in the classical action $S$ but also the ones which couple one-point functions to two-point functions, as well as the couplings between two-point functions only. The reason is, that the two-point functions are resummed propagators within the 2PI formalism implying orders in perturbation theory get mixed. However, it has been shown on general grounds in ref.~\cite{Berges:2005hc}, that  this is done in a particular way. This allows one to carry out the renormalization such that only a finite number of counterterms are needed.

As is customary, we first define the renormalized fields from the bare ones, using their respective wave-function renormalizations, as
\begin{align}
&\phi(x) = Z^{\frac{1}{2}}_{\phi,2} \phi_R(x) \,, \quad \quad
G(x,y) = Z_{\phi,0} \, G_R(x,y) \,, \nonumber \\  
&\psi(x) = Z^{\frac{1}{2}}_{\psi,2} \psi_R(x)\,,\quad \quad  D(x,y) = Z_{\psi,0} \, D_R(x,y)\,,
\end{align}
where we have indicated the number of fields $\phi\, (\psi)$ associated with a term by the index $i$ in $Z_{\phi,i} \,(Z_{\psi,i})$. 
We adopt this same notation for the mass and coupling 
counterterms, and obtain
\begin{align}
\label{eq:counterterms}
\begin{array}{ll}
Z_{\phi,2} m^2 = m^2_R + \delta m^2_2 \,,
&
Z_{\phi,2}^{\frac{i}{2}} Z_{\phi,0}^{\frac{3-i}{2}} \alpha = \alpha_R + \delta \alpha_{i} \quad (i=0,1,3) 
\\[4mm] 
Z_{\phi,0} m^2 = m^2_R + \delta m^2_0 \,,
& Z_{\phi,2}^{j/2} Z_{\phi,0}^{(4-j)/2} \lambda = \lambda_{R} + \delta \lambda_j\quad (j=0,2,4)\,, \\[4mm]
Z_{\psi,0} M = M_R + \delta M_0 \,, 
&Z_{\psi,0}  Z_{\phi,2}^{\frac{k}{2}} Z_{\phi,0}^{\frac{1-k}{2}} g = g_{R} + \delta g_k \quad (k=0,1) \,.
\end{array}
\end{align} 
We note for completeness, that one needs an additional counterterm in the action to cancel 
loop-induced contributions to the effective action which are linear in $\phi_R$, i.e.
\begin{align}
- \int_x \delta t_1 \, \phi_R(x) \,.
\end{align}
 
At this stage, we can already give the renormalized equations of motions
\begin{align}
&\left[ (1+\delta Z_{\phi,2})\square_x  +
\widehat{m}^2_2(x) \right]\phi_R(x) = - \delta t_1 - \frac{(\alpha_R + \delta \alpha_1)}{2} G_R(x,x) - (g_R + \delta g_1) \text{tr}[D_R(x,x)] \,,
\label{eq:eom_phi} \\[2mm]
&\left[ (1+\delta Z_{\phi,0})\square_x  +
\widehat{m}^2_0(x) 
\right] G_R(x,y)  = \delta_{\cal C}(x-y)  - \int_z \Pi(x,z)  G_R(z,y)   \,,
\label{eq:eom_G} \\[2mm]
& \left[\ii\left(1+\delta Z_{\psi,0}\right)\myslash{\partial}_x
 -\widehat{M}_0(x) \right]
 D_R(x,y)  = \delta_{\cal C}(x-y)  - \int_z \Sigma(x,z)   D_R(z,y)  \,,
\label{eq:eom_D}
\end{align}
with
\begin{align}
& \widehat{m}^2_2(x) = m^2_R + \delta m^2_2 
+ \frac{1}{2} (\alpha_R + \delta \alpha_3) \phi_R(x)
+ \frac{1}{6} (\lambda_R + \delta \lambda_4) \phi^2_R(x) + \frac{1}{2} (\lambda_R + \delta \lambda_2) G_R(x,x)\,, \\[3mm]
& \widehat{m}^2_0(x)=
 m^2_R + \delta m^2_0 + \frac{1}{4}(\lambda_R + \delta \lambda_0) G_R(x,x)+ \frac{1}{2} (\alpha_R + \delta \alpha_3) \phi_R(x)   + \frac{1}{4} (\lambda_R + \delta \lambda_2)  \phi^2_R(x)\,, \\[3mm]
& \widehat{M}_0(x) = \left(M_R+\delta M_0\right)
+ (g_R+\delta g_1)\phi_R(x) \,,\\[3mm]
& \Pi(x,z) = -\frac{\ii \left[(\alpha_R + \delta \alpha_0)+(\lambda_R +\delta \lambda_1)\phi_R(x)\right]\, \left[(\alpha_R + \delta \alpha_0)+(\lambda_R +\delta \lambda_1)\phi_R(z)\right]}{4} \,G^2_R(x,z) \nonumber \\[2mm]
& \qquad \qquad  +\frac{\ii2(g_0 + \delta g_0)^2}{2} \,\text{tr}[D_R(x,z)D_R(z,x)] \,,\\[3mm]
&\Sigma(x,z) = \frac{\ii (g_R + \delta g_0)^2}{2} \, G_R(x,z) \left[ D_R(x,z) + D_R(z,x) \right]\,.
\end{align}
Here, we have split the self-energies $\overline{\Pi}(x,z)$ and $\overline{\Sigma}(x,z)$ into non-local contributions $\Pi(x,z)$ and $\Sigma(x,z)$ and local ones which are absorbed in $\widehat{m}^2_0(x)$ and $\widehat{M}_0(x)$, respectively. Note that the integral on the right sides of \eqref{eq:eom_G} and \eqref{eq:eom_D} are the previously mentioned memory integrals. 

One can solve \eqref{eq:eom_phi}--\eqref{eq:eom_D} numerically for a given set of initial conditions. However, before one can start such a task, one must determine the yet unspecified counterterms. To this end, we use an on-shell scheme, detailed in the next section.

\section{Renormalization: Generic Aspects}
\label{sec:gen_renorm}
It is known that the Bogoliubov-Parasiuk-Hepp-Zimmermann (BPHZ) procedure \cite{10.1007/BF02392399, Hepp:1966eg, Zimmermann:1969jj}, which 
is used in standard QFT to determine the structure of 
the divergences, does not suffice in the case of the 2PI 
formalism (see, for e.g., \cite{Berges:2005hc} and 
references therein). This is caused by the resummed 
nature of the 2-point functions $G$ and $D$. The 
solution to this challenge is the auxiliary vertex 
functions which can be resummed such that a 
consistent renormalization with only a finite number 
of counterterms is possible. 

The corresponding 2PI kernels, which enter the corresponding Bethe-Salpeter equations (BSE) are \cite{Berges:2005hc,Reinosa:2009tc}
\begin{align}
\oL^{(4)}(x_1,x_2,x_3,x_4) &\equiv 4 \frac{\delta^2 \Gamma^{\text{2PI}}_{\text{int}}}{\delta G(x_1,x_2) \delta G(x_3,x_4)} \bigg|_{\op,\oG,\oD}\,, \\
\Lambda_{\psi\psi}(x_1,x_2,x_3,x_4)_{(ab),(cd)} &\equiv -\frac{\delta^2 \Gamma^{\text{2PI}}_{\text{int}}}{\delta D^{ba}(x_1,x_2)\,\delta D^{cd}(x_3,x_4)} \bigg|_{\op,\oG,\oD}\,, \\
\Lambda^{(4)}_{\psi\phi}(x_1,x_2,x_3,x_4)_{ab} &\equiv -2\frac{\delta^2 \Gamma^{\text{2PI}}_{\text{int}}}{\delta D^{ba}(x_1,x_2)\,\delta G(x_3,x_4)}\bigg|_{\op,\oG,\oD} \\
& = \Lambda^{(4)}_{\phi\psi}(x_3,x_4,x_1,x_2)_{ab} \equiv -2\frac{\delta^2 \Gamma^{\text{2PI}}_{\text{int}}}{\delta G(x_3,x_4)\, \delta D^{ab}(x_1,x_2)}\bigg|_{\op,\oG,\oD}\,,
\end{align}
where we have denoted spinor indices through lowercase Latin alphabets. At this stage, we still need to take into account that a loop with four internal fermions can generate a divergent contribution to the quartic scalar coupling. This can be taken care of by introducing the following modified kernel for scalars
\begin{align}
\tilde{\Lambda}_{\phi\phi}(x_1,x_2,x_3,x_4) & = \oL^{(4)}(x_1,x_2,x_3,x_4) \nonumber \\
& - \ii \int_{y_1\dots y_4} \text{tr}\left[\oD(y_1,y_3)\, \overline{\Lambda}^{(4)}_{\phi\psi}(x_1,x_2,y_1,y_2) \,\oD(y_2,y_4)\,\overline{\Lambda}_{\psi \phi}(y_3,y_4,x_3,x_4) \right] \nonumber\\
    &+ \ii\int_{y_1\dots y_8}\text{tr}\bigg[\oD(y_1,y_3)\,\Lambda^{(4)}_{\phi\psi}(x_1,x_2,y_1,y_2)\,\oD(y_2,y_4)V_{\psi\psi}(y_3,y_4,y_5,y_6)\nonumber \\ 
    & \qquad \qquad \oD(y_5,y_7)\,\Lambda^{(4)}_{\psi \phi}(y_7,y_8,x_3,x_4)\,\oD(y_6,y_8)\bigg] \,.
\end{align}
where the trace in the second line (``tr") is over the spinor indices. Now, we are in the position
to define the vertex functions which are 
given by the following Bethe-Salpeter equations
\begin{align}
\label{eq:BSEVbar}
\oV^{(4)}(x_1,x_2,x_3,x_4) & 
= \tilde{\Lambda}_{\phi\phi}(x_1,x_2,x_3,x_4) \nonumber \\
&+ \frac{\ii}{2} \int_{y_1\dots y_4}
\tilde{\Lambda}_{\phi\phi}(x_1,x_2,y_1,y_2) \oG(y_1,y_3) \oG(y_2,y_4) 
\overline{V}^{(4)}(y_3,y_4,x_3,x_4)\,, \\
\label{eq:BSEVpsibar}
V_{\psi\psi} (x_1,x_2,x_3,x_4)_{ab,cd} &= \Lambda_{\psi \psi}(x_1,x_2,x_3,x_4)_{ab,cd} \nonumber \\
    &+\ii \int_{y_1\dots y_4} \Lambda_{\psi \psi}(x_1,x_2,y_1,y_2)_{ab,ef}\oD(y_1,y_3)_{eg} \nonumber\\
  &\hspace{10mm}V_{\psi\psi} (y_3,y_4,x_3,x_4)_{ef,cd}\oD(y_2,y_4)_{fb}\, \,, \\
V_{\psi\phi}^{(4)} (x_1,x_2,x_3,x_4)_{ab} &= \Lambda^{(4)}_{\psi \phi}(x_1,x_2,x_3,x_4)_{ab} \nonumber \\
    &+\ii \int_{y_1\dots y_4} \Lambda^{(4)}_{\psi \phi}(x_1,x_2,y_1,y_2)_{ae}\oD(y_1,y_3)_{ef} \nonumber\\
  &\hspace{10mm}V^{(4)}_{\psi\phi} (y_3,y_4,x_3,x_4)_{fb}\oG(y_2,y_4)\, \,.  
\end{align}

In the study of the vacuum structure of a theory, it is useful to consider any $\phi$ and evaluate
the corresponding two-point function (Green's function) $G(\phi)$. This leads to a 2PI effective action $\Gamma[\phi] = \Gamma_{\text{2PI}}[\phi, G(\phi)]$ which gives rise to a 2PI improved effective potential. When studying
phase transitions in the 2PI formalism, this is the object to consider when looking for the vacuum states of the theory.
The two-point functions in this case can be evaluated from the stationarity condition \eqref{eq:extr_Gamma_G}, leading to the gap equation for the scalar propagator
\begin{align}
 \overline{G}^{-1}(x,y; \phi) &= G^{-1}_0(x,y; \phi) - \overline{\Pi}(x,y; \phi) 
 \label{eq:gap_eqn_G}
\end{align}
and similarly one for the fermionic propagator
\begin{align}
 \overline{D}^{-1}(x,y; \phi) &= D^{-1}_0(x,y; \phi) - \overline{\Sigma}(x,y; \phi) \,.
 \label{eq:gap_eqn_D}
\end{align}
From this perspective, the stationarity condition for the scalar field becomes
\begin{align}
\Gamma^{(1)}(x) &\equiv \frac{\delta \Gamma^{\text{2PI}}}{\delta \phi(x)}\bigg|_{\overline{\phi},\oG,\oD} \stackrel{!}{=} 0  \\
&=  \frac{\delta \Gamma^{\text{2PI}}}{\delta \phi(x)}\bigg|_{\overline{\phi},\oG,\oD} + \int_{y_1,y_2}\frac{\delta\Gamma^{\text{2PI}}}{\delta G(y_1,y_2)}\bigg|_{\overline{\phi},\oG,\oD}\frac{\delta G(y_1,y_2)}{\delta\phi(x)} + \int_{y_1,y_2}\text{tr}\bigg\{\frac{\delta\Gamma^{\text{2PI}}}{\delta D(y_1,y_2)}\bigg|_{\overline{\phi},\oG,\oD}\frac{\delta D(y_1,y_2)}{\delta\phi(x)}\bigg\}
\end{align}
where $\Gamma^{(1)}$ denotes the physical one-point function. The second and third terms of the last line vanish due to the stationarity conditions. 

Moreover, we will require the $n$-point functions
\begin{align}
\Gamma^{(n)}(x_1,\dots,x_n) = \frac{\delta^n \Gamma^{\text{2PI}}}{\delta \phi(x_1) \dots \delta \phi(x_n)}\bigg|_{\overline{\phi},\oG,\oD} \,.
\end{align}
These are related to the derivatives of the 2PI generating functional $\Gamma^{\text{2PI}}$. However, one has to take care since the two-point function obtained from \eqref{eq:gap_eqn_G} depends on $\phi$ and thus the chain rule must be used. As a consequence, a system of coupled integral equations emerges. In general, these equations have the following form
\begin{align}
\label{eq:npoint2}
\frac{\delta^n \ii \Gamma}{\delta \phi(x_1) \dots \delta \phi(x_n)} = &
{\cal A}^{(n)}(x_1,\dots,x_n) \nonumber \\
+& \int_{z_1\dots z_4} \frac{\delta^2 \ii \Gamma^{\text{2PI}}_{\text{int}}}{\delta \phi(x_1) G(z_1,z_2)}\bigg|_{\op,\oG,\overline{D}} 
\oG(z_1,z_3) \frac{\delta^{n-1} \oPi(z_3,z_4)}{\delta \phi(x_2) \dots \delta \phi(x_n)}
\oG(z_4,z_2) \nonumber \\
+&\int_{z_1 \dots z_4}\text{tr}\bigg\{\frac{\delta^2\Gamma^{\text{2PI}}_{\text{int}}}{\delta \phi(x_1)\delta D(z_1,z_2)}\bigg|_{\overline{\phi},\overline{G},\overline{D}}\overline{D}(z_1,z_3)\frac{\delta^{n-1} \overline{\Sigma}(z_3,z_4)}{\delta\phi(x_2)\dots\delta\phi(x_n)}\overline{D}(z_4,z_2)\bigg\}
\\
\frac{\delta^n \overline{\Pi}(y_1,y_2)}{\delta \phi(x_1) \dots \delta \phi(x_n)} = &
{\cal B}^{(n)}(y_1,y_2,x_1,\dots,x_n) \nonumber \\
+& \int_{z_1\dots z_4} \frac{\delta^2 2\ii \Gamma^{\text{2PI}}_{\text{int}}}{\delta G(y_1,y_2) \,\delta G(z_1,z_2)}\bigg|_{\op,\oG,\oD} 
\oG(z_1,z_3) \frac{\delta^{n-1} \oPi(z_3,z_4)}{\delta \phi(x_1) \dots \delta \phi(x_n)}
\oG(z_4,z_2)  \nonumber \\
+& \int_{z_1\dots z_4}\text{tr}\bigg\{ \frac{\delta^2 2\ii \Gamma^{\text{2PI}}_{\text{int}}}{\delta G(y_1,y_2) \,\delta D(z_1,z_2)}\bigg|_{\op,\oG,\oD} 
\oD(z_1,z_3) \frac{\delta^{n-1} \overline{\Sigma}(z_3,z_4)}{\delta \phi(x_1) \dots \delta \phi(x_n)}
\oD(z_4,z_2) \bigg\}
\label{eq:Pi_derivatives}
\\
\frac{\delta^n \overline{\Sigma}(y_1,y_2)}{\delta \phi(x_1) \dots \delta \phi(x_n)} = &
{\cal C}^{(n)}(y_1,y_2,x_1,\dots,x_n) \nonumber \\
+& \int_{z_1\dots z_4} \frac{\delta^2 \ii \Gamma^{\text{2PI}}_{\text{int}}}{\delta D(y_1,y_2) \,\delta G(z_1,z_2)}\bigg|_{\op,\oG,\oD} 
\oG(z_1,z_3) \frac{\delta^{n-1} \oPi(z_3,z_4)}{\delta \phi(x_1) \dots \delta \phi(x_n)}
\oG(z_4,z_2)  \nonumber \\
+& \int_{z_1\dots z_4} \frac{\delta^2 \ii \Gamma^{\text{2PI}}_{\text{int}}}{\delta D(y_1,y_2) \,\delta D(z_1,z_2)}\bigg|_{\op,\oG,\oD} 
\oD(z_1,z_3) \frac{\delta^{n-1} \overline{\Sigma}(z_3,z_4)}{\delta \phi(x_1) \dots \delta \phi(x_n)}
\oD(z_4,z_2) 
\label{eq:Sigma_derivatives}
\end{align}
The functions ${\cal A}^{(n)}$, ${\cal B}^{(n)}$ and ${\cal C}^{(n)}$ contain both various derivatives of 
$\Gamma^{\text{2PI}}_{\text{int}}$ with respect to $\phi$, $G$  and $D$ at the stationarity point $(\op,\oG,\oD)$ \cite{Berges:2005hc}.
Moreover, they contain lower derivatives $\delta^k \oPi/\delta \phi^k$ and 
$\delta^k \overline{\Sigma}/\delta \phi^k$ ($k=1,\dots,n-1$). 
Consequently, $n$-point functions are expressed solely via parts of the 2PI generating functional and their derivatives. This comes at the expense of infinite resummations of self-energy diagrams and their derivatives. A formal solution to the self-energy equations can be given in terms of the various vertex functions.
Using these, the solutions to \eqref{eq:Pi_derivatives} and \eqref{eq:Sigma_derivatives} are given by
\begin{align}
\frac{\delta^n \overline{\Pi}(y_1,y_2)}{\delta \phi(x_1) \dots \delta \phi(x_n)} = &
{\cal B}^{(n)}(y_1,y_2,x_1,\dots,x_n) \nonumber \\
+& \frac{\ii}{2} \int_{z_1\dots z_4} \oV^{(4)}(x_1,x_2,z_1,z_2) \oG(z_1,z_3)
 {\cal B}^{(n)}(z_3,z_4,x_1,\dots,x_n)
\oG(z_4,z_2) \nonumber \\
+& \ii \int_{z_1\dots z_4} \text{tr}\left\{ V^{(4)}_{\phi\psi}(x_1,x_2,z_1,z_2) \oD(z_1,z_3)
 {\cal C}^{(n)}(z_3,z_4,x_1,\dots,x_n)
\oD(z_4,z_2)\right\}\,.
\label{eq:Pi_derivatives_sol} 
\end{align}

\begin{align}
\frac{\delta^n \overline{\Sigma}(y_1,y_2)}{\delta \phi(x_1) \dots \delta \phi(x_n)} = &
{\cal C}^{(n)}(y_1,y_2,x_1,\dots,x_n) \nonumber \\
+& \ii \int_{z_1\dots z_4}  V_{\psi\phi}^{(4)}(x_1,x_2,z_1,z_2) 
 {\cal B}^{(n)}(z_3,z_4,x_1,\dots,x_n)\oG(z_1,z_3)
\oG(z_4,z_2) \nonumber \\
+& \ii \int_{z_1\dots z_4}  V_{\psi\psi}(x_1,x_2,z_1,z_2) \oD(z_1,z_3)
 {\cal C}^{(n)}(z_3,z_4,x_1,\dots,x_n)
\oD(z_4,z_2) \,.
\label{eq:Sigma_derivatives_sol} 
\end{align}
From these equations, one can then obtain solutions to \eqref{eq:npoint2}.
In addition, we would require an auxiliary scalar vertex function 
\begin{align}
V^{(4)}(x_1,x_2,x_3,x_4)  
&= \Lambda^{(4)}(x_1,x_2,x_3,x_4) \nonumber
\\
&\qquad + \frac{\ii}{2} \int_{y_1\dots y_4}
\Lambda^{(4)}(x_1,x_2,y_1,y_2) \oG(y_1,y_3) \oG(y_2,y_4) 
\oV^{(4)}(y_3,y_4,x_3,x_4) \,,
\label{eq:BSEV}
\end{align}
with
\begin{align}
\Lambda^{(4)}(x_1,x_2,x_3,x_4) &\equiv 2 \frac{\delta^3 \Gamma^{\text{2PI}}_{\text{int}}}{\delta G(x_1,x_2) \delta \phi(x_3) \delta \phi(x_4)} \bigg|_{\op,\oG,\oD} \,.
\end{align}

It was mentioned in the previous section that actual calculations require an approximation of the 2PI generating functional. In terms of renormalizability, there are certain requirements to be met when choosing the diagrams for $\Gamma^{\text{2PI}}_{\text{int}}$, which becomes apparent when using the BPHZ analysis for the identification of necessary subtractions. Therein, all diagrams are examined for possible sub-divergences, by employing Weinberg’s theorem \cite{Weinberg:1959nj}. Whenever a part of a diagram is potentially divergent, this part is shrunk to an effective vertex. The new topology created this way must be part of the renormalized $\Gamma^{\text{2PI}}_{\text{int}}$, with the effective vertex replaced by counterterms in order to cancel divergences. 

Effectively, this means that we are dealing with a truncation in perturbation theory where some parts have been resummed, implying that one has to take different mass and coupling constant counterterms, depending on the combination of one- and two-point functions connecting to a `vertex'. According to \cite{Berges:2005hc}, renormalization may be carried out in the vacuum at temperature $T=0$. It is most convenient to work in momentum space, which allows us to employ the usual techniques to determine the various counterterms in an on-shell scheme. 
We extend the usual on-shell conditions of standard QFT, see for e.g.~\cite{Denner:1991kt},
to the 2PI formalism as follows
\begin{align}
\label{eq:renor_cond_scalar}
Z_{\phi,0} G^{-1}_R(p) \bigg|_{p^2=m^2_R} &= Z_{\phi,2} \Gamma^{(2)}_\phi(p) \bigg|_{p^2=m^2_R} = 0 \,,\\ 
Z_{\phi,0}\frac{\partial}{\partial p^2} G^{-1}_R(p) \bigg|_{p^2=m^2_R} &= Z_{\phi,2} \frac{\partial}{\partial p^2}\Gamma^{(2)}_\phi(p) \bigg|_{p^2=m^2_R} = 1 \,,
\end{align}
\begin{align}
\label{eq:renor_cond_fermion}
Z_{\psi,0} \,D^{-1}_R(p) u(p) \bigg|_{p^2=M^2_R} = 0 \,,\quad 
\lim_{p^2\to M^2_R} Z^{-1}_{\psi,0} \left(\frac{\myslash{p} + M_R}{p^2-M^2_R}\right)D^{-1}_R(p)u(p) =  u(p) \,,
\end{align}
\begin{align}
\label{eq:renor_cond_fourpt}
Z_{\phi,0}^{2}\,\overline{V}^{(4)}(p_1,p_2,p_3,p_4) \bigg|_{p^2_i=m^2_R}  &= 
Z_{\phi,2}\,Z_{\phi,0}\,  V^{(4)}(p_1,p_2,p_3,p_4) \bigg|_{p^2_i=m^2_R} \nonumber \\
&= Z_{\phi,2}^{2}\,\Gamma^{(4)}(p_1,p_2,p_3,p_4) \bigg|_{p^2_i=m^2_R} =
-\lambda_R    \,,
\end{align}
\begin{equation}
\label{eq:renor_cond_threept}
Z_{\phi,0}\, Z_{\phi,2}^{\frac{1}{2}}\,V^{(3)}(p_1,p_2,p_3) \bigg|_{p^2_2= p^2_3 = m^2_R} = Z_{\phi,2}^{\frac{3}{2}}\, \Gamma^{(3)}(p_1,p_2,p_3) \bigg|_{p^2_2= p^2_3 = m^2_R}  = -\alpha_R \,,
\end{equation}
\begin{equation}
\label{eq:renor_cond_yukawa}
Z_{\psi,2} Z_{\phi,2}^{\frac{1}{2}}\, \bar{u}(p_1) V^{(3)}_{\psi\phi}(p_1,p_2,p_3) u(p_2) \bigg|_{p^2_2 = M^2_R,\,p^2_3 =m^2_R} = -g_R
\end{equation}
where $u(p)$ is an on-shell spinor and $\bar{u}(p)$ is its Dirac conjugate. $V^{(3)}$ and $V^{(3)}_{\psi\phi}$ denote pure scalar and fermion-fermion-scalar three-point vertices respectively, that will be discussed in the relevant section.
Note that we use the same symbols to denote the $n$-point functions in configuration and momentum space indicated only by the arguments. 
The momentum arguments are not independent, but related via momentum conservation, for e.g. $p_1 + p_2 = p_3 + p_4$ for the four-point functions and $p_1 = p_2 +p_3$ for the three-point functions.

Note that we have made a choice to implement the same renormalization conditions for the physical two-point functions of the fields and the propagators, as in \cite{Pilaftsis:2013xna, Pilaftsis:2017enx}. We do the same for the physical $n$-point functions of the scalar fields, and the corresponding vertex functions. One may also choose to renormalize these functions to different coupling constants, such as in \cite{Kainulainen:2021eki}, but these correspond to shifts in the various counterterms.

\section{Renormalization with only Scalars}
\label{sec:scalars}

We begin with scalars, which can be considered to be a limit where the fermions are so heavy that they need to be integrated out. This allows us to detail the intricacies of the renormalization procedure, without the complication of additional Lorentz structures. As a novel aspect, we present here not only the counterterms for the symmetric phase, but also in the broken one.

\subsection{Revisiting the Hartree Approximation}
\label{sec:hartree}
This case has been treated in literature already several times \cite{Blaizot:2003br,Blaizot:2003an,Pilaftsis:2017enx} and is summarized here to exemplify some of the details involved. Moreover, we have obtained new results, which to our knowledge have not been presented in literature so far, such as finite contributions to the wave function renormalization.  

In this approximation, one sets $\alpha=g=0$ and takes only the leading
contribution to $\Gamma_2$ into account. Expressing $\Gamma^{\text{2PI}}_{\text{int}}$ in terms of renormalized quantities, we get
\begin{align}
&\Gamma^{\text{2PI}}_{\text{int}}[\phi_R,G_R] =
- \frac{1}{2}\int_{x} (\delta Z_{\phi,0} \square_x + \delta m^2_0) G_R(x,y) \big|_{x=y} 
- \frac{1}{2}\int_x \phi_R(x)  (\delta Z_{\phi,2} \square_x + \delta m^2_2) \phi_R(x)
  \nonumber \\ 
& \qquad - \int_x \left[ \frac{1}{8}(\lambda_R + \delta \lambda_0) G^2_R(x,x)
   + \frac{1}{4} (\lambda_R + \delta \lambda_2) G_R(x,x) \phi^2_R(x)
   + \frac{1}{4!} (\lambda_R + \delta \lambda_4) \phi^4_R(x)\right] \,.
\end{align}
This gives the four-point kernel 
\begin{equation}
    \overline{\Lambda}^{(4)} = 4\, \frac{\delta^2 \Gamma^{\text{2PI}}_{\text{int}}}{\delta G_R(k)\delta G_R(q)} = -(\lambda_R + \delta \lambda_0) \,,
    \label{kernelhartree}
\end{equation}
which is evidently independent of the (external) momentum. Thus, the BSE~\eqref{eq:BSEVbar} simplifies considerably within the Hartree approximation. 
As the vertex function is an infinite resummation of iterations of this kernel, stitched together by loops with two propagators (see Fig. \ref{fig:bseqn_ht}), the only origin of a momentum dependence in $\overline{V}^{(4)}$ is from the loop function depending on the sum of external momenta $p = p_1 + p_2$. For brevity, we will thus use the notation $\overline{V}^{(4)}(p)\equiv \overline{V}^{(4)}(p_1,p_2,p_3,p_4)$ throughout this section. Accordingly, the BSE is expressed as
\begin{align}
    \overline{V}^{(4)}(p) = -(\lambda_R + \delta \lambda_0) \, -\, \frac{i}{2}(\lambda_R+\delta \lambda_0)\,\overline{V}^{(4)}(p)\int_q G_R(q)  G_R(p+q) \,.
\label{eq:VRp_hartree}    
\end{align} 
At this point, it is convenient to calculate $p^2$ in the center of mass (COM) system. This gives the familiar result
\begin{equation}
    p^2 = (p_1 + p_2)^2 = 4E_{\ast}^2 \,,
\end{equation}
where $2E_{\ast}$ is the COM energy. Let $p^2_{\ast} = 4m^2_R $ be the renormalization point corresponding to the COM three-momentum $|\vec{p_{\ast}}| = 0$ and let
\begin{equation}
	\oV^{(4)}(p_{\ast}) = - \lambda_R\,.
	\label{eqn:oV4_rc_ht}
\end{equation}
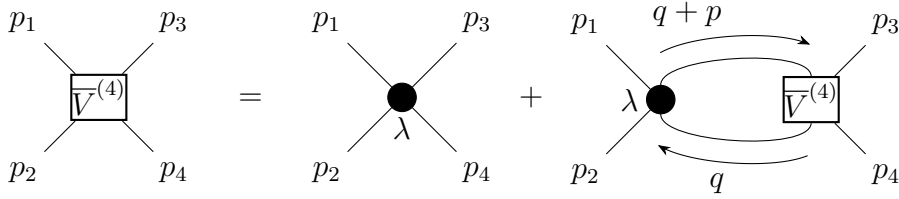
\begin{figure}[t]
\begin{equation*}
\vcenter{\hbox{\begin{tikzpicture}
	\begin{feynman}
	\vertex (v1) at (-1,1){\(p_1\)};
	\vertex (v2) at (-1,-1){\(p_2\)};
	\vertex (v3) at (1,1){\(p_3\)};
	\vertex (v4) at (1,-1){\(p_4\)};
	\vertex[large, VR] (x) at (0,0) {$\oV^{(4)}$};
	\diagram*{
	(v1) -- (x),
	(v2) --(x),
	(x) --  (v3),
	(x) -- (v4)		
	};
	\end{feynman}
\end{tikzpicture}}}
	\quad = \quad
\vcenter{\hbox{\begin{tikzpicture}
	\begin{feynman}
	\vertex (v1) at (-1,1){\(p_1\)};
	\vertex (v2) at (-1,-1){\(p_2\)};
	\vertex (v3) at (1,1){\(p_3\)};
	\vertex (v4) at (1,-1){\(p_4\)};
	\vertex[small, dot] (x) at (0,0) {\(\lambda\)};
	\diagram*{
	(v1) -- (x),
	(v2) -- (x),
	(x) --   (v3),
	(x) --  (v4)		
	};
	\vertex[below=1em of x] {\(\lambda\)};
	\end{feynman}
\end{tikzpicture}}} 
	\,\, + \,\,
	\vcenter{\hbox{\begin{tikzpicture}
	\begin{feynman}
	\vertex (v1) at (-2,1){\(p_1\)};
	\vertex (v2) at (-2,-1){\(p_2\)};
	\vertex (v3) at (2,1){\(p_3\)};
	\vertex (v4) at (2,-1){\(p_4\)};
	\vertex[small, dot] (x1) at (-1,0) {\(\lambda\)};
	\vertex[large, VR] (x2) at (1,0) {$\oV^{(4)}$};
	\diagram*{
	(v1) -- (x1),
	(v2) -- (x1),
	(x1) -- [momentum =\(q+p\),half left, looseness = 0.5] (x2) -- [momentum =\(q\), half left, looseness = 0.5](x1),
	(x2) --  (v3),
	(x2) --  (v4)		
	};
	\vertex[left=1em of x1] {\(\lambda\)};
	\end{feynman}
\end{tikzpicture}}}
\end{equation*} 
\caption{Illustration of the BSE in the Hartree approximation. Here, the convention is $p_{1,2}$ are incoming four-momenta and $p_{3,4}$ are outgoing, with $p = p_1 + p_2$.}
\label{fig:bseqn_ht}
\end{figure}
This allows us to solve exactly for the counterterm
\begin{align}
\delta \lambda_0 = -\lambda_R + \frac{\lambda_R}{1-\frac{1}{2}\lambda_R \mathcal{I}(p^2_{\ast})}\quad \text{with} \quad \mathcal{I}(p^2) = i\int_q G_R(q)G_R(p+q) \,,
\label{ctlambda0_hartree}
\end{align}
which we can insert in \eqref{eq:VRp_hartree} to obtain the vertex function
\begin{equation}
    \overline{V}^{(4)}(p) =-\frac{\lambda_R}{1-\frac{\lambda_R}{2}\left(\mathcal{I}(p^2_{\ast})-\mathcal{I}(p^2)\right)}\,.
    \label{eqn:vbar4_ht}
\end{equation}
From this, we can immediately observe that $\overline{V}^{(4)}(p)$ is finite as any potential divergences present in the loop integral $\mathcal{I}(p^2)$ cancels in the difference $\mathcal{I}(p^2_{\ast})-\mathcal{I}(p^2)$. Furthermore, the renormalization of the auxiliary vertex $V(p)$ proceeds along similar lines as 
\begin{align}
\Lambda^{(4)} = 2 \,\frac{\delta^3 \Gamma^{\text{2PI}}_{\text{int}}}{\delta^2 \phi_R\,\delta G_R(q)}
= - (\lambda_R + \delta \lambda_2) \,.
\end{align}

\noindent Using this in \eqref{eq:BSEVbar} and the result \eqref{eqn:vbar4_ht}, along with the renormalization condition \eqref{eqn:oV4_rc_ht}, we obtain
\begin{align}
    \delta \lambda_2 = \delta \lambda_0 \quad \text{ and } \quad
    V^{(4)}(p) = \overline{V}^{(4)}(p)  \,.
    \label{eqn:ctl2_ht}
\end{align}
We have made a choice to implement the same renormalization conditions for both vertex functions. We will later do the same for the propagator and the two-point function of the scalar field, as in \cite{Pilaftsis:2013xna, Pilaftsis:2017enx}, as well for the physical $n$-point functions of the scalar fields corresponding to the vertex functions. One may also choose to renormalise these functions to different parameters, such as in \cite{Kainulainen:2021eki}, but these correspond to shifts in the various counterterms.  

The determination of these counterterms via the BSEs allows us to treat the sub-divergences, which are not accounted for by the usual BHPZ procedure, that would appear in the renormalization of the two-point function \cite{Blaizot:2003an,Berges:2005hc}, for which we now turn to the gap equation
\begin{align}
    i G^{-1}_R(p) &=  (p^2 - m^2_R)  - i\overline{\Pi}(p^2) \nonumber \\[2mm]
    &=   (p^2 - m^2_R) + (\delta Z_{\phi,0} \, p^2 - \delta m^2_0) - \frac{(\lambda_R + \delta \lambda_2)}{2} \phi_R^2  
    -\frac{(\lambda_R + \delta \lambda_0)}{2} \int_q G_R(q)  \,,
\label{eq:GR_Hartree}    
\end{align}
where $p$ here is the external momentum. Having determined $\delta \lambda_0$ and $\delta \lambda_2$, we seek for the counterterms $\delta Z_{\phi,0}$ and $\delta m_0^2$ which can be used to treat the divergences that can be accounted for by the BHPZ procedure. Enforcing the on-shell renormalization conditions \eqref{eq:renor_cond_scalar}, we have 
\begin{align}
\label{eq:hartree_dm0}
i G^{-1}_R(p) \bigg|_{p^2=m^2_R} &= - i \overline{\Pi}(p^2)  \bigg|_{p^2=m^2_R} \stackrel{!}{=} 0 \,,\\[3mm]
i\frac{\partial}{\partial p^2 } G^{-1}_R(p) \big|_{p^2=m^2_R} &=   1 - i\frac{\partial}{\partial p^2 } \overline{\Pi}(p^2)  \big|_{p^2=m^2_R} \stackrel{!}{=} 1 \,.
\label{eq:hartree_dZ0}
\end{align}
From \eqref{eq:hartree_dZ0}, we have
\begin{equation}
	\delta Z_{\phi,0} = 0 
\end{equation}
and then \eqref{eq:hartree_dm0} gives
\begin{equation}
	\delta m^2_0 = - \frac{\lambda_R + \delta \lambda_2 }{2} \phi^2_R - \frac{\lambda_R + \delta \lambda_0}{2}\int_q G_R(q)\,.
\end{equation}
If we substitute this back into \eqref{eq:GR_Hartree}, we obtain 
\begin{align}
	i G_R^{-1}(p^2) = p^2 - m^2_R\,,
	\label{eqn:prop_ht}
\end{align}
or, in other words, the full propagator is identically the bare one. Consequently, all integrals over propagators can be expressed in terms of the well-known Passarino-Veltman functions \cite{Denner:1991kt}
\begin{align}
    &\int_q G_R(q) = \frac{1}{16\pi^2}A_0(m^2_R) \\
    &\mathcal{I}(p) \equiv i\int_q G_R(q) G_R(p+q) = \frac{1}{16\pi^2}B_0(p^2,m^2_R,m^2_R) \,,
\end{align}
For the counterterms, we thus find
\begin{align}
\label{eqn:os_dl0_exp}
&\delta \lambda_0 = -\lambda_R - 32\pi^2 \epsilon + \mathcal{O}(\epsilon^2) = \delta \lambda_2 \,, \\
&\delta Z_0 = 0 + O(\epsilon)\,, \quad \text{and} \quad 
\delta m^2_0 = -m^2_R  + O(\epsilon) \,,
\label{eq:Hartree_dZ0_dm20}
\end{align}
and also the expression for the vertex function,
\begin{equation}
	\oV^{(4)}(p) = -\frac{\lambda_R}{1-\frac{\lambda_R}{32\pi^2}\left[B_0(4m^2_R,m^2_R,m^2_R) - B_0(p^2,m^2_R,m^2_R)\right]}\,.
	\label{eqn:ov4_os_ht}
\end{equation}
We mention, for completeness, that we did not assume $\phi_R=0$, i.e. these relations hold true in the unbroken and broken phases. Furthermore, in the unbroken phase we can exploit the $\mathbb{Z}_2$ symmetry of the Hartree approximation and make use of the identity \cite{Berges:2005hc}
\begin{align}
    \frac{\delta^2 \Gamma^{\text{2PI}}_{\text{int}}}{\delta \phi_R^2}
    \bigg|_{\phi_R=0} + (\delta Z_2 \, p^2 - \delta m^2_2)
=2\frac{\delta \Gamma^{\text{2PI}}_{\text{int}}}{\delta G_R}
    \bigg|_{\phi_R=0} + (\delta Z_0 \, p^2 - \delta m^2_0)
\label{eq:symphase_ct}
\end{align}
to relate the counterterms for the propagator and the field. Due to the fact that we have imposed the same renormalization conditions for $\Gamma^{(2)}$ and $G^{-1}_R$  (c.f. \eqref{eq:renor_cond_scalar}), we find
\begin{align}
    \delta Z_{\phi,2} = \delta Z_{\phi,0} \quad \text{ and } \quad
    \delta m^2_0 =  \delta m^2_2 \,.
\label{eq:equality_dZ_dm2}    
\end{align}

We now demonstrate explicitly that these relations do not hold for $\phi_R\ne 0$. It is useful at this point to introduce a diagrammatic representation of the various integral equations that we will encounter, which is based on the one in \cite{Berges:2005hc}. The basic building blocks are given in Fig. \ref{fig:graphics}.
\begin{figure}[t]
\begin{tikzpicture}[scale=2]
\node (action) at (0,0) [Gamma] {}; 
\node [left] at (action.west) {$i \Gamma^{\text{2PI}}_{\text{int}} =$};
\draw[-] (3.0,0) node[left] {$G_R(x_1,x_2) =$ \,\,} node[left]{1} -- (3.7,0)node[right]{2};
\node (action2) at (0,-1) [Gamma] {}; 
\node (l1) at (-0.5,-1) [VEV] {}; 
\node (l2) at (0.5,-1) [VEV] {}; 
\draw[-] (l1) node[left] {$\frac{\ii \delta^2 \Gamma^{\text{2PI}}_{\text{int}}}{\delta \phi_R(x_1) \delta \phi_R(x_2)} = \ii \Pi(x_1,x_2) = $ \,\,\,}
node[left] {1} -- (action2);
\draw[-] (action2) -- (l2) node[right] {2} ;
\node (twopoint) at (3.5,-1) [Pi] {}; 
\coordinate (ta) at (3.1,-1);
\node [left] at (ta.west) {$\ii \overline{\Pi}(x_1,x_2) =$\,\,\,};
\draw[double distance=4mm,thick] (twopoint) -- (ta) node[right,above]{\hspace*{-2mm}1\,}
node[right,below]{\hspace*{-2mm}2\,};
\node (fourpoint) at (0,-2) [VR] {$\,\ii\overline{V}^{(4)}$}; 
\coordinate (va) at (-0.4,-2);
\node [left] at (va.west) {$\ii \overline{V}^{(4)}(x_1,x_2,x_3,x_4) =$\,\,\,};
\draw[double distance=4mm,thick] (fourpoint) -- (va) node[right,above]{\hspace*{-2mm}1\,}
node[right,below]{\hspace*{-2mm}2\,};
\coordinate (vb) at (0.4,-2);
\draw[double distance=4mm,thick] (fourpoint) -- (vb) node[right,above]{\hspace*{2mm}3\,}
node[right,below]{\hspace*{2mm}4\,};
\node (threepoint) at (3,-2) [Pi] {}; 
\coordinate (tra) at (2.6,-2);
\node [left] at (tra.west) {$ \frac{\ii \delta\overline{\Pi}(x_1,x_2)}{ \delta \phi_R(x_3)} =$\,\,\,};
\draw[double distance=4mm,thick] (threepoint) -- (tra) node[right,above]{\hspace*{-2mm}1\,}
node[right,below]{\hspace*{-2mm}2\,};
\node (l3) at (3.4,-2) [VEV] {}; 
\draw[-] (threepoint) -- (l3) node[right] {3} ;
\end{tikzpicture}
\caption{Graphical representation of basic building blocks}
\label{fig:graphics}
\end{figure}

We first examine the physical two-point function, for which we have the following integral equation
\begin{align}
\Gamma^{(2)}(x_1,x_2) &\equiv \frac{\delta^2 \Gamma^{\text{2PI}}}{\delta \phi_R(x_1) \delta \phi_R(x_2) } \nonumber \\
&=   \ii G^{-1}_{0,R}(x_1,x_2) +  
\frac{\delta^2 \Gamma^{\text{2PI}}_{\text{int}}}{\delta \phi_R(x_1) \delta \phi_R(x_2) } \bigg|_{G_R} \nonumber \\
&+ \int_{y_1, y_2, y_3, y_4} 
\frac{\delta^2 \Gamma^{\text{2PI}}_{\text{int}}}{\delta \phi_R(x_1) \delta G_R(y_1,y_2) } \bigg|_{G_R} G_R(y_1,y_3) 
\frac{\delta \overline{\Pi}_R(y_3,y_4)}{ \delta \phi_R(x_2) }G_R(y_4,y_2) \,.
\label{eqn:phys2pta}
\end{align}
Using the elements presented in Fig. \ref{fig:graphics}, this equation reads in graphical form
\begin{equation}
\frac{\delta^2 \Gamma^{\text{2PI}}}{\delta \phi_R(x_1) \delta \phi_R(x_2) }   =  \ii G^{-1}_{0,R}(x_1,x_2) +  
 \begin{tikzpicture}[scale=2,baseline={([yshift=-.4ex]current bounding box.center)}]
\node (action2) at (0,-2) [Gamma] {}; 
\node (l1) at (-0.5,-2) [VEV] {}; 
\node (l2) at (0.5,-2) [VEV] {}; 
\draw[-] (l1) node[left] {1} -- (action2);
\draw[-] (action2) -- (l2) node[right] {2} ;
 \end{tikzpicture}
 + \frac{1}{2}
 \begin{tikzpicture}[scale=2,baseline={([yshift=-.4ex]current bounding box.center)}]
\node (action2) at (0,0) [Gamma] {}; 
\node (l1) at (-0.5,0) [VEV] {}; 
\node (threepoint) at (0.6,0) [Pi] {}; 
\node (l2) at (1.1,0) [VEV] {}; 
\draw[-] (l1) node[left] {1} -- (action2);
\draw[-] (threepoint) -- (l2) node[right] {2} ;
\draw[double distance=4mm,thick] (action2) -- (threepoint);
 \end{tikzpicture}
\label{phys2pta}
\end{equation}
\noindent We can replace the derivative of the self-energy w.r.t. a field expectation value $\delta \overline{\Pi}/\delta \phi_R$ using the following graphical representation (c.f. \eqref{eq:Pi_derivatives_sol})
\begin{equation}
\label{eq:dPidphi}
 \begin{tikzpicture}[scale=2,baseline={([yshift=-.4ex]current bounding box.center)}]
\node (threepoint) at (3,0) [Pi] {}; 
\coordinate (tra) at (2.6,0);
\draw[double distance=4mm,thick] (threepoint) -- (tra) node[right,above]{\hspace*{-2mm}1\,}
node[right,below]{\hspace*{-2mm}2\,};
\node (l3) at (3.4,0) [VEV] {}; 
\draw[-] (threepoint) -- (l3) node[right] {3} ;
 \end{tikzpicture}
=
 \begin{tikzpicture}[scale=2,baseline={([yshift=-.4ex]current bounding box.center)}]
\node (threepoint) at (3,0) [Gamma] {}; 
\coordinate (tra) at (2.6,0);
\draw[double distance=4mm,thick] (threepoint) -- (tra) node[right,above]{\hspace*{-2mm}1\,}
node[right,below]{\hspace*{-2mm}2\,};
\node (l3) at (3.4,0) [VEV] {}; 
\draw[-] (threepoint) -- (l3) node[right] {3} ;
 \end{tikzpicture}
+ \frac{1}{2} \quad
 \begin{tikzpicture}[scale=2,baseline={([yshift=-.4ex]current bounding box.center)}]
\node (fourpoint) at (3,0) [VR] {\,$\overline{V}^{(4)}$\,}; 
\node (threepoint) at (3.5,0) [Gamma] {}; 
\coordinate (tra) at (2.6,0);
\draw[double distance=4mm,thick] (fourpoint) -- (tra) node[right,above]{\hspace*{-2mm}1\,}
node[right,below]{\hspace*{-2mm}2\,};
\node (l3) at (4,0) [VEV] {}; 
\draw[-] (threepoint) -- (l3) node[right] {3} ;
\draw[double distance=4mm,thick] (fourpoint) -- (threepoint);
\end{tikzpicture}
\end{equation}

\noindent Inserting this into the integral equation \eqref{phys2pta}, we obtain in momentum space
\begin{align}
\Gamma^{(2)}(p) 
&= (p^2-m^2_R) + (\delta Z_{\phi,2}\,p^2- \delta m^2_2)  - \frac{1}{2}(\lambda_R+\delta\lambda_4)\phi^2_R  - \frac{1}{2}(\lambda_R+\delta\lambda_2) \int _q G_R(q)   \nonumber \\[2mm]
	& \qquad -\frac{1}{8}(\lambda_R+\delta\lambda_2)^2 \left[\mathcal{I}(p^2)\right]
	\, \phi^2_R + \frac{1}{16}(\lambda_R+\delta\lambda_2)^2 \left[\mathcal{I}(p^2)\right]^2 \overline{V}^{(4)}(p) \, \phi^2_R \nonumber \\[2mm]
&= \left[(1+\delta Z_{\phi,2})p^2-2 m^2_R - \delta m^2_2\right] -\frac{1}{2}(\lambda_R+\delta\lambda_4)\phi^2_R \nonumber \\[2mm]
& \qquad + \frac{\lambda_R \phi^2_R}{4}
\left\{1- \frac{\lambda_R}{32\pi^2} \left[B_0(p^2_{\ast},m^2_R,m^2_R)- B_0(p^2,m^2_R,m^2_R)\right] \right\}^{-1} + O(\epsilon) \,.
\end{align}
Imposing the same renormalization conditions as for $G_R$, \eqref{eq:renor_cond_scalar}, this yields
\begin{align}
\label{eqn:dZ2_ht}
\delta Z_{\phi,2} &= \frac{\lambda_R^2 \phi^2_R}{128\pi^2}\,\dot{B}_0(m^2_R,m^2_R,m^2_R)\left\{1- \frac{\lambda_R}{32\pi^2} \left[B_0(p^2_{\ast},m^2_R,m^2_R)- B_0(m^2_R,m^2_R,m^2_R)\right]\right\}^{-1}  \\[2mm]
\delta m^2_2 &= - m^2_R  + m^2_R \delta Z_{\phi,2} - \frac{\left(\lambda_R + \delta \lambda_4\right)
\phi^2_R}{2}  \nonumber \\[2mm] 
&\qquad \qquad + \frac{\lambda_R \phi^2_R}{4} \left\{1 - \frac{\lambda_R}{32\pi^2}\left[B_0(p_{\ast}^2,m^2_R,m^2_R)- B_0(m^2_R,m^2_R,m^2_R)\right] \right\}^{-1} \,,
\label{eqn:dm2_ht}
\end{align}
where 
\begin{equation*}
\dot{B}_0(p^2,m^2_R,m^2_R) = \frac{\partial B_0(q^2,m^2_R,m^2_R)}{\partial q^2} \bigg|_{q^2 = p^2}\,.
\end{equation*} 
One recovers the equality with the corresponding counterterms for $G_R$ in \eqref{eq:equality_dZ_dm2} when $\phi_R = 0$, as claimed in \cite{Berges:2005hc}.

Finally, one needs to determine $\delta\lambda_4$. The treatment of the physical four-point function proceeds in the same manner as for the two-point function, but now a lot more diagrams contribute. An easy way to obtain the four-point function is to take the diagrammatical expression of the two-point function and then take two more derivatives with respect to the field $\phi_R$. Application of the chain rule due to the $\phi_R$ dependence of $G_R$ generates a variety of topologies. Here, we will first give the overview over all the topologies that contribute, in order to generalise to additional couplings, and perform truncations appropriate to the Hartree approximation. For the sake of brevity, the four ``legs'' representing the external space-time points, previously denoted in the diagrams with numerical indices, are now suppressed and it is understood that all permutations of $x_2$, $x_3$ and $x_4$ contribute, as long as the resulting diagrams are not equivalent. An example of a diagram, where permutations must be considered, is 
\begin{equation}
 \begin{tikzpicture}[scale=2,baseline={([yshift=-.4ex]current bounding box.center)}]
\node (fourpoint) at (0,0.3) [Gamma] {}; 
\node (threepoint) at (0,-0.3) [Pi] {}; 
\node (l1) at (-0.4,0.3) [VEV] {}; 
\node (l2) at (0.,0.7) [VEV] {}; 
\node (l3) at (0.4,0.3) [VEV] {}; 
\node (l4) at (0.,-0.7) [VEV] {}; 
\draw[double distance=4mm,thick] (fourpoint) -- (threepoint);
\draw[-] (fourpoint) -- (l1);
\draw[-] (fourpoint) -- (l2);
\draw[-] (fourpoint) -- (l3);
\draw[-] (threepoint) -- (l4);
\end{tikzpicture}
=
 \begin{tikzpicture}[scale=2,baseline={([yshift=-.4ex]current bounding box.center)}]
\node (fourpoint) at (0,0.3) [Gamma] {}; 
\node (threepoint) at (0,-0.3) [Pi] {}; 
\node (l1) at (-0.4,0.3) [VEV] {}; 
\node (l2) at (0.,0.7) [VEV] {}; 
\node (l3) at (0.4,0.3) [VEV] {}; 
\node (l4) at (0.,-0.7) [VEV] {}; 
\draw[double distance=4mm,thick] (fourpoint) -- (threepoint);
\draw[-] (fourpoint) -- (l1) node[left] {1};
\draw[-] (fourpoint) -- (l2) node[above] {4};
\draw[-] (fourpoint) -- (l3) node[right] {3};
\draw[-] (threepoint) -- (l4) node[below] {2};
\end{tikzpicture}
+
 \begin{tikzpicture}[scale=2,baseline={([yshift=-.4ex]current bounding box.center)}]
\node (fourpoint) at (0,0.3) [Gamma] {}; 
\node (threepoint) at (0,-0.3) [Pi] {}; 
\node (l1) at (-0.4,0.3) [VEV] {}; 
\node (l2) at (0.,0.7) [VEV] {}; 
\node (l3) at (0.4,0.3) [VEV] {}; 
\node (l4) at (0.,-0.7) [VEV] {}; 
\draw[double distance=4mm,thick] (fourpoint) -- (threepoint);
\draw[-] (fourpoint) -- (l1) node[left] {1};
\draw[-] (fourpoint) -- (l2) node[above] {4};
\draw[-] (fourpoint) -- (l3) node[right] {2};
\draw[-] (threepoint) -- (l4) node[below] {3};
\end{tikzpicture}
+
 \begin{tikzpicture}[scale=2,baseline={([yshift=-.4ex]current bounding box.center)}]
\node (fourpoint) at (0,0.3) [Gamma] {}; 
\node (threepoint) at (0,-0.3) [Pi] {}; 
\node (l1) at (-0.4,0.3) [VEV] {}; 
\node (l2) at (0.,0.7) [VEV] {}; 
\node (l3) at (0.4,0.3) [VEV] {}; 
\node (l4) at (0.,-0.7) [VEV] {}; 
\draw[double distance=4mm,thick] (fourpoint) -- (threepoint);
\draw[-] (fourpoint) -- (l1) node[left] {1};
\draw[-] (fourpoint) -- (l2) node[above] {3};
\draw[-] (fourpoint) -- (l3) node[right] {2};
\draw[-] (threepoint) -- (l4) node[below] {4};
\end{tikzpicture}
\end{equation}
It is important to note that the index 1 is not part of the permutation. This happens because the diagrams are obtained by a subsequent differentiation with respect to the four fields $\phi_R(x_1)$, \dots\,, $\phi_R(x_4)$. The first self-energy box only appears in the second differentiation and thus the 1-index is always attached to the $\Gamma^{\text{2PI}}_{\text{int}}$ blob. In the subsequent discussion, the leg to the left side is always considered to correspond to the 1-index. At first glance, this might seem to not be in line with the symmetry properties of the 4-point function, such as $\Gamma^{(4)}(x_1, x_2 ,x_3, x_4) = \Gamma^{(4)} (x_2, x_1, x_3, x_4)$ and so on. However, these properties are only hidden in the above case and are apparent once the identities of the self-energy boxes are inserted. 

The diagrams that in principal contribute to the 4-point function are
\begin{align}
\Gamma^{(4)} \equiv \frac{\delta^4 \Gamma}{\delta \phi^4} = & \quad
 \begin{tikzpicture}[scale=2,baseline={([yshift=-.4ex]current bounding box.center)}]
\node (fourpoint) at (0,0) [Gamma] {}; 
\node (l1) at (-0.4,0) [VEV] {}; 
\node (l2) at (0.,0.4) [VEV] {}; 
\node (l3) at (0.4,0) [VEV] {}; 
\node (l4) at (0.,-0.4) [VEV] {}; 
\draw[-] (fourpoint) -- (l1);
\draw[-] (fourpoint) -- (l2);
\draw[-] (fourpoint) -- (l3);
\draw[-] (fourpoint) -- (l4);
\end{tikzpicture}
+
\frac{1}{2} \,\,
 \begin{tikzpicture}[scale=2,baseline={([yshift=-.4ex]current bounding box.center)}]
\node (v1) at (0,0.3) [Gamma,fill=red] {}; 
\node (v2) at (0,-0.3) [Pi] {}; 
\node (l1) at (-0.4,0.3) [VEV] {}; 
\node (l2) at (0.,0.7) [VEV] {}; 
\node (l3) at (0.4,0.3) [VEV] {}; 
\node (l4) at (0.,-0.7) [VEV] {}; 
\draw[double distance=4mm,thick] (v1) -- (v2);
\draw[-] (v1) -- (l1);
\draw[-] (v1) -- (l2);
\draw[-] (v1) -- (l3);
\draw[-] (v2) -- (l4);
\end{tikzpicture}
+
\frac{1}{4} \,\,
 \begin{tikzpicture}[scale=2,baseline={([yshift=-.4ex]current bounding box.center)}]
\node (v1) at (0,0.3) [Gamma,fill=red] {}; 
\node (v2) at (0,-0.3) [Pi] {}; 
\node (v3) at (0.6,0.3) [Pi] {}; 
\node (l1) at (-0.4,0.3) [VEV] {}; 
\node (l2) at (0.,0.7) [VEV] {}; 
\node (l3) at (0.9,0.3) [VEV] {}; 
\node (l4) at (0.,-0.7) [VEV] {}; 
\draw[double distance=4mm,thick] (v1) -- (v2);
\draw[double distance=4mm,thick] (v1) -- (v3);
\draw[-] (v1) -- (l1);
\draw[-] (v1) -- (l2);
\draw[-] (v3) -- (l3);
\draw[-] (v2) -- (l4);
\end{tikzpicture}
+
\frac{1}{8} \,\,
 \begin{tikzpicture}[scale=2,baseline={([yshift=-.4ex]current bounding box.center)}]
\node (v1) at (0,0.3) [Gamma,fill=red] {}; 
\node (v2) at (0,-0.3) [Pi] {}; 
\node (v3) at (0.6,0.3) [Pi] {}; 
\node (v4) at (0.0,0.9) [Pi] {}; 
\node (l1) at (-0.4,0.3) [VEV] {}; 
\node (l2) at (0.,1.3) [VEV] {}; 
\node (l3) at (0.9,0.3) [VEV] {}; 
\node (l4) at (0.,-0.7) [VEV] {}; 
\draw[double distance=4mm,thick] (v1) -- (v2);
\draw[double distance=4mm,thick] (v1) -- (v3);
\draw[double distance=4mm,thick] (v1) -- (v4);
\draw[-] (v1) -- (l1);
\draw[-] (v4) -- (l2);
\draw[-] (v3) -- (l3);
\draw[-] (v2) -- (l4);
\end{tikzpicture}
\nonumber \\
& +
 \begin{tikzpicture}[scale=2,baseline={([yshift=-.4ex]current bounding box.center)}]
\node (v1) at (0,0.3) [Gamma] {}; 
\node (v2) at (0,-0.3) [Pi] {}; 
\node (v3) at (0.6,0.3) [Pi] {}; 
\node (l1) at (-0.4,0.3) [VEV] {}; 
\node (l2) at (0.,0.7) [VEV] {}; 
\node (l3) at (0.9,0.3) [VEV] {}; 
\node (l4) at (0.,-0.7) [VEV] {}; 
\draw (v1) -- (v2);
\draw (v1) -- (v3);
\draw (v2) -- (v3);
\draw[-] (v1) -- (l1);
\draw[-] (v1) -- (l2);
\draw[-] (v3) -- (l3);
\draw[-] (v2) -- (l4);
\end{tikzpicture}
+
\frac{1}{2} \,\,
 \begin{tikzpicture}[scale=2,baseline={([yshift=-.4ex]current bounding box.center)}]
\node (v1) at (0,0.3) [Gamma,fill=red] {}; 
\node (v2) at (0,-0.3) [Pi] {}; 
\node (v3) at (0.6,0.3) [Pi] {}; 
\node (v4) at (0.0,0.9) [Pi] {}; 
\node (l1) at (-0.4,0.3) [VEV] {}; 
\node (l2) at (0.,1.3) [VEV] {}; 
\node (l3) at (0.9,0.3) [VEV] {}; 
\node (l4) at (0.,-0.7) [VEV] {}; 
\draw (v1) -- (v2);
\draw (v1) -- (v3);
\draw (v2) -- (v3);
\draw[double distance=4mm,thick] (v1) -- (v4);
\draw[-] (v1) -- (l1);
\draw[-] (v4) -- (l2);
\draw[-] (v3) -- (l3);
\draw[-] (v2) -- (l4);
\end{tikzpicture}
+
 \begin{tikzpicture}[scale=2,baseline={([yshift=-.4ex]current bounding box.center)}]
\node (v1) at (0,0.3) [Gamma] {}; 
\node (v2) at (0,-0.3) [Pi] {}; 
\node (v3) at (0.6,0.3) [Pi] {}; 
\node (v4) at (0.0,0.9) [Pi] {}; 
\node (l1) at (-0.4,0.3) [VEV] {}; 
\node (l2) at (0.,1.3) [VEV] {}; 
\node (l3) at (0.9,0.3) [VEV] {}; 
\node (l4) at (0.,-0.7) [VEV] {}; 
\draw (v1) -- (v2);
\draw (v4) -- (v3);
\draw (v2) -- (v3);
\draw (v1) -- (v4);
\draw[-] (v1) -- (l1);
\draw[-] (v4) -- (l2);
\draw[-] (v3) -- (l3);
\draw[-] (v2) -- (l4);
\end{tikzpicture}
\nonumber \\
& + \frac{1}{2} \,\,
 \begin{tikzpicture}[scale=2,baseline={([yshift=-.4ex]current bounding box.center)}]
\node (v1) at (0,0.3) [Gamma] {}; 
\node (v2) at (0.6,0.3) [Pi] {}; 
\node (l1) at (-0.4,0.3) [VEV] {}; 
\node (l2) at (0.,0.7) [VEV] {}; 
\node (l3) at (0.9,0.3) [VEV] {}; 
\node (l4) at (0.6,-0.1) [VEV] {}; 
\draw[double distance=4mm,thick] (v1) -- (v2);
\draw[-] (v1) -- (l1);
\draw[-] (v1) -- (l2);
\draw[-] (v2) -- (l3);
\draw[-] (v2) -- (l4);
\end{tikzpicture}
+
\frac{1}{4} \,\,
 \begin{tikzpicture}[scale=2,baseline={([yshift=-.4ex]current bounding box.center)}]
\node (v1) at (0,0.3) [Gamma,fill=red] {}; 
\node (v3) at (0.6,0.3) [Pi] {}; 
\node (v4) at (0.0,0.9) [Pi] {}; 
\node (l1) at (-0.4,0.3) [VEV] {}; 
\node (l2) at (0.,1.3) [VEV] {}; 
\node (l3) at (0.9,0.3) [VEV] {}; 
\node (l4) at (0.6,-0.1) [VEV] {}; 
\draw[double distance=4mm,thick] (v1) -- (v4);
\draw[double distance=4mm,thick] (v1) -- (v3);
\draw[-] (v1) -- (l1);
\draw[-] (v4) -- (l2);
\draw[-] (v3) -- (l3);
\draw[-] (v3) -- (l4);
\end{tikzpicture}
+
 \begin{tikzpicture}[scale=2,baseline={([yshift=-.4ex]current bounding box.center)}]
\node (v1) at (0,0.3) [Gamma] {}; 
\node (v3) at (0.6,0.3) [Pi] {}; 
\node (v4) at (0.0,0.9) [Pi] {}; 
\node (l1) at (-0.4,0.3) [VEV] {}; 
\node (l2) at (0.,1.3) [VEV] {}; 
\node (l3) at (0.9,0.3) [VEV] {}; 
\node (l4) at (0.6,-0.1) [VEV] {}; 
\draw (v1) -- (v3);
\draw (v4) -- (v3);
\draw (v1) -- (v4);
\draw[-] (v1) -- (l1);
\draw[-] (v4) -- (l2);
\draw[-] (v3) -- (l3);
\draw[-] (v3) -- (l4);
\end{tikzpicture}
\nonumber \\
& + \frac{1}{2} \,\,
 \begin{tikzpicture}[scale=2,baseline={([yshift=-.4ex]current bounding box.center)}]
\node (v1) at (0,0.3) [Gamma] {}; 
\node (v2) at (0.6,0.3) [Pi] {}; 
\node (l1) at (-0.4,0.3) [VEV] {}; 
\node (l2) at (0.6,0.7) [VEV] {}; 
\node (l3) at (0.9,0.3) [VEV] {}; 
\node (l4) at (0.6,-0.1) [VEV] {}; 
\draw[double distance=4mm,thick] (v1) -- (v2);
\draw[-] (v1) -- (l1);
\draw[-] (v2) -- (l2);
\draw[-] (v2) -- (l3);
\draw[-] (v2) -- (l4);
\end{tikzpicture} \,\,\,.
\label{eq:dGamma4_dphi4}
\end{align}
The diagrams, where the $\Gamma^{\text{2PI}}_{\text{int}}$ blob is marked in red, do not contribute in the Hartree approximation due to the limited number of terms contained in $\Gamma^{\text{2PI}}_{\text{int}}$. This reduces the number of diagrams one has to consider by about half but this only works for the 2PI kernels and not for the self-energy boxes which adhere to their own integral equations. We have seen that derivatives of the self-energy with respect to one, two and three fields $\phi_R$ are present. The single derivative is given in \eqref{eq:dPidphi}. The other two can be expressed using diagrams that contain only the first derivative of the self-energy. 
We will continue to mark $\Gamma^{\text{2PI}}_{\text{int}}$ in red if the contribution vanishes in the Hartree approximation, and permutations of field indices are implied wherever they lead to non-equivalent topologies.
\begin{align}
 \begin{tikzpicture}[scale=2,baseline={([yshift=-.4ex]current bounding box.center)}]
\node (v1) at (0.0,0.) [Pi] {}; 
\node (l1) at (-0.4,0) {}; 
\node (l2) at (0.4,0.) [VEV] {}; 
\node (l3) at (0,-0.4) [VEV] {}; 
\draw[double distance=4mm,thick] (l1) -- (v1);
\draw[-] (v1) -- (l2);
\draw[-] (v1) -- (l3);
\end{tikzpicture}
= & \,\,
 \begin{tikzpicture}[scale=2,baseline={([yshift=-.4ex]current bounding box.center)}]
\node (v1) at (0.0,0.) [Gamma] {}; 
\node (l1) at (-0.4,0) {}; 
\node (l2) at (0.4,0.) [VEV] {}; 
\node (l3) at (0,-0.4) [VEV] {}; 
\draw[double distance=4mm,thick] (l1) -- (v1);
\draw[-] (v1) -- (l2);
\draw[-] (v1) -- (l3);
\end{tikzpicture}
+ \frac{1}{2} \,
 \begin{tikzpicture}[scale=2,baseline={([yshift=-.4ex]current bounding box.center)}]
\node (v1) at (0.0,0.) [Gamma] {}; 
\node (v2) at (-0.6,0.) [VR] {\,$\overline{V}^{(4)}$\,}; 
\node (l1) at (-1.,0) {}; 
\node (l2) at (0.4,0.) [VEV] {}; 
\node (l3) at (0,-0.4) [VEV] {}; 
\draw[double distance=4mm,thick] (l1) -- (v2);
\draw[double distance=4mm,thick] (v1) -- (v2);
\draw[-] (v1) -- (l2);
\draw[-] (v1) -- (l3);
\end{tikzpicture}
\nonumber \\ & 
+ \frac{1}{2} \,
 \begin{tikzpicture}[scale=2,baseline={([yshift=-.4ex]current bounding box.center)}]
\node (v1) at (0.0,0.) [Gamma,fill=red] {}; 
\node (v3) at (0.0,-0.6) [Pi] {}; 
\node (l1) at (-0.4,0) {}; 
\node (l2) at (0.4,0.) [VEV] {}; 
\node (l3) at (0,-1.0) [VEV] {}; 
\draw[double distance=4mm,thick] (l1) -- (v1);
\draw[double distance=4mm,thick] (v1) -- (v3);
\draw[-] (v1) -- (l2);
\draw[-] (v3) -- (l3);
\end{tikzpicture}
+ \frac{1}{4} \,
 \begin{tikzpicture}[scale=2,baseline={([yshift=-.4ex]current bounding box.center)}]
\node (v1) at (0.0,0.) [Gamma,fill=red] {}; 
\node (v2) at (-0.6,0.) [VR] {\,$\overline{V}^{(4)}$\,}; 
\node (v3) at (0.0,-0.6) [Pi] {}; 
\node (l1) at (-1.,0) {}; 
\node (l2) at (0.4,0.) [VEV] {}; 
\node (l3) at (0,-1.0) [VEV] {}; 
\draw[double distance=4mm,thick] (l1) -- (v2);
\draw[double distance=4mm,thick] (v1) -- (v2);
\draw[double distance=4mm,thick] (v1) -- (v3);
\draw[-] (v1) -- (l2);
\draw[-] (v3) -- (l3);
\end{tikzpicture}
\nonumber \\ & 
+ \frac{1}{4} \,
 \begin{tikzpicture}[scale=2,baseline={([yshift=-.4ex]current bounding box.center)}]
\node (v1) at (0.0,0.) [Gamma,fill=red] {}; 
\node (v3) at (0.0,-0.6) [Pi] {}; 
\node (v4) at (0.6,0.) [Pi] {}; 
\node (l1) at (-0.4,0) {}; 
\node (l2) at (1.,0.) [VEV] {}; 
\node (l3) at (0,-1.0) [VEV] {}; 
\draw[double distance=4mm,thick] (l1) -- (v1);
\draw[double distance=4mm,thick] (v1) -- (v3);
\draw[double distance=4mm,thick] (v1) -- (v4);
\draw[-] (v4) -- (l2);
\draw[-] (v3) -- (l3);
\end{tikzpicture}
+ \frac{1}{8} \,
 \begin{tikzpicture}[scale=2,baseline={([yshift=-.4ex]current bounding box.center)}]
\node (v1) at (0.0,0.) [Gamma,fill=red] {}; 
\node (v2) at (-0.6,0.) [VR] {\,$\overline{V}^{(4)}$\,}; 
\node (v3) at (0.0,-0.6) [Pi] {}; 
\node (v4) at (0.6,0.) [Pi] {}; 
\node (l1) at (-1.,0) {}; 
\node (l2) at (1.,0.) [VEV] {}; 
\node (l3) at (0,-1.0) [VEV] {}; 
\draw[double distance=4mm,thick] (l1) -- (v2);
\draw[double distance=4mm,thick] (v1) -- (v2);
\draw[double distance=4mm,thick] (v1) -- (v3);
\draw[double distance=4mm,thick] (v1) -- (v4);
\draw[-] (v4) -- (l2);
\draw[-] (v3) -- (l3);
\end{tikzpicture}
\nonumber \\ & 
+  \,
 \begin{tikzpicture}[scale=2,baseline={([yshift=-.4ex]current bounding box.center)}]
\node (v1) at (0.0,0.) [Gamma] {}; 
\node (v3) at (0.0,-0.6) [Pi] {}; 
\node (v4) at (0.6,0.) [Pi] {}; 
\node (l1) at (-0.4,0) {}; 
\node (l2) at (1.,0.) [VEV] {}; 
\node (l3) at (0,-1.0) [VEV] {}; 
\draw[double distance=4mm,thick] (l1) -- (v1);
\draw (v1) -- (v3);
\draw (v1) -- (v4);
\draw (v3) -- (v4);
\draw[-] (v4) -- (l2);
\draw[-] (v3) -- (l3);
\end{tikzpicture}
+ \frac{1}{2} \,
 \begin{tikzpicture}[scale=2,baseline={([yshift=-.4ex]current bounding box.center)}]
\node (v1) at (0.0,0.) [Gamma] {}; 
\node (v2) at (-0.6,0.) [VR] {\,$\overline{V}^{(4)}$\,}; 
\node (v3) at (0.0,-0.6) [Pi] {}; 
\node (v4) at (0.6,0.) [Pi] {}; 
\node (l1) at (-1.,0) {}; 
\node (l2) at (1.,0.) [VEV] {}; 
\node (l3) at (0,-1.0) [VEV] {}; 
\draw[double distance=4mm,thick] (l1) -- (v2);
\draw[double distance=4mm,thick] (v1) -- (v2);
\draw (v1) -- (v3);
\draw (v1) -- (v4);
\draw (v3) -- (v4);
\draw[-] (v4) -- (l2);
\draw[-] (v3) -- (l3);
\end{tikzpicture}
\label{eq:d2_Pi_dphi2}
\end{align}
For the triple derivative $d^3 \overline{\Pi} / d \phi^3_R$, it is more convenient to not insert instances of the double derivative $d^3 \overline{\Pi} / d \phi^2_R$ as the expressions get too lengthy otherwise. One simply needs to substitute the corresponding diagrams from \eqref{eq:d2_Pi_dphi2} at places where boxes with two field derivatives appear. One then finds
\begin{align}
 \begin{tikzpicture}[scale=2,baseline={([yshift=-.4ex]current bounding box.center)}]
\node (v1) at (0.0,0.) [Pi] {}; 
\node (l1) at (-0.4,0) {}; 
\node (l2) at (0.4,0.) [VEV] {}; 
\node (l3) at (0,-0.4) [VEV] {}; 
\node (l4) at (0,0.4) [VEV] {}; 
\draw[double distance=4mm,thick] (l1) -- (v1);
\draw[-] (v1) -- (l2);
\draw[-] (v1) -- (l3);
\draw[-] (v1) -- (l4);
\end{tikzpicture}
= & \,\,
 \begin{tikzpicture}[scale=2,baseline={([yshift=-.4ex]current bounding box.center)}]
\node (v1) at (0.0,0.) [Gamma,fill=red] {}; 
\node (l1) at (-0.4,0) {}; 
\node (l2) at (0.4,0.) [VEV] {}; 
\node (l3) at (0,-0.4) [VEV] {}; 
\node (l4) at (0,0.4) [VEV] {}; 
\draw[double distance=4mm,thick] (l1) -- (v1);
\draw[-] (v1) -- (l2);
\draw[-] (v1) -- (l3);
\draw[-] (v1) -- (l4);
\end{tikzpicture}
+ \frac{1}{2} \,
 \begin{tikzpicture}[scale=2,baseline={([yshift=-.4ex]current bounding box.center)}]
\node (v1) at (0.0,0.) [Gamma,fill=red] {}; 
\node (v2) at (-0.6,0.) [VR] {\,$\overline{V}^{(4)}$\,}; 
\node (l1) at (-1.,0) {}; 
\node (l2) at (0.4,0.) [VEV] {}; 
\node (l3) at (0,-0.4) [VEV] {}; 
\node (l4) at (0,0.4) [VEV] {}; 
\draw[double distance=4mm,thick] (l1) -- (v2);
\draw[double distance=4mm,thick] (v1) -- (v2);
\draw[-] (v1) -- (l2);
\draw[-] (v1) -- (l3);
\draw[-] (v1) -- (l4);
\end{tikzpicture}
\nonumber \\ & 
+ \frac{1}{2} \,
 \begin{tikzpicture}[scale=2,baseline={([yshift=-.4ex]current bounding box.center)}]
\node (v1) at (0.0,0.) [Gamma,fill=red] {}; 
\node (v3) at (0.0,-0.6) [Pi] {}; 
\node (l1) at (-0.4,0) {}; 
\node (l2) at (0.4,0.) [VEV] {}; 
\node (l3) at (0,-1.0) [VEV] {}; 
\node (l4) at (0,0.4) [VEV] {}; 
\draw[double distance=4mm,thick] (l1) -- (v1);
\draw[double distance=4mm,thick] (v1) -- (v3);
\draw[-] (v1) -- (l2);
\draw[-] (v3) -- (l3);
\draw[-] (v1) -- (l4);
\end{tikzpicture}
+ \frac{1}{4} \,
 \begin{tikzpicture}[scale=2,baseline={([yshift=-.4ex]current bounding box.center)}]
\node (v1) at (0.0,0.) [Gamma,fill=red] {}; 
\node (v2) at (-0.6,0.) [VR] {\,$\overline{V}^{(4)}$\,}; 
\node (v3) at (0.0,-0.6) [Pi] {}; 
\node (l1) at (-1.,0) {}; 
\node (l2) at (0.4,0.) [VEV] {}; 
\node (l3) at (0,-1.0) [VEV] {}; 
\node (l4) at (0,0.4) [VEV] {}; 
\draw[double distance=4mm,thick] (l1) -- (v2);
\draw[double distance=4mm,thick] (v1) -- (v2);
\draw[double distance=4mm,thick] (v1) -- (v3);
\draw[-] (v1) -- (l2);
\draw[-] (v3) -- (l3);
\draw[-] (v1) -- (l4);
\end{tikzpicture}
\nonumber \\ & 
+ \frac{1}{4} \,
 \begin{tikzpicture}[scale=2,baseline={([yshift=-.4ex]current bounding box.center)}]
\node (v1) at (0.0,0.) [Gamma,fill=red] {}; 
\node (v3) at (0.0,-0.6) [Pi] {}; 
\node (v4) at (0.6,0.) [Pi] {}; 
\node (l1) at (-0.4,0) {}; 
\node (l2) at (1.,0.) [VEV] {}; 
\node (l3) at (0,-1.0) [VEV] {}; 
\node (l4) at (0,0.4) [VEV] {}; 
\draw[double distance=4mm,thick] (l1) -- (v1);
\draw[double distance=4mm,thick] (v1) -- (v3);
\draw[double distance=4mm,thick] (v1) -- (v4);
\draw[-] (v4) -- (l2);
\draw[-] (v3) -- (l3);
\draw[-] (v1) -- (l4);
\end{tikzpicture}
+ \frac{1}{8} \,
 \begin{tikzpicture}[scale=2,baseline={([yshift=-.4ex]current bounding box.center)}]
\node (v1) at (0.0,0.) [Gamma,fill=red] {}; 
\node (v2) at (-0.6,0.) [VR] {\,$\overline{V}^{(4)}$\,}; 
\node (v3) at (0.0,-0.6) [Pi] {}; 
\node (v4) at (0.6,0.) [Pi] {}; 
\node (l1) at (-1.,0) {}; 
\node (l2) at (1.,0.) [VEV] {}; 
\node (l3) at (0,-1.0) [VEV] {}; 
\node (l4) at (0,0.4) [VEV] {}; 
\draw[double distance=4mm,thick] (l1) -- (v2);
\draw[double distance=4mm,thick] (v1) -- (v2);
\draw[double distance=4mm,thick] (v1) -- (v3);
\draw[double distance=4mm,thick] (v1) -- (v4);
\draw[-] (v4) -- (l2);
\draw[-] (v3) -- (l3);
\draw[-] (v1) -- (l4);
\end{tikzpicture}
\nonumber \\ & 
+ \frac{1}{8} \,
 \begin{tikzpicture}[scale=2,baseline={([yshift=-.4ex]current bounding box.center)}]
\node (v1) at (0.0,0.) [Gamma,fill=red] {}; 
\node (v3) at (0.0,-0.6) [Pi] {}; 
\node (v4) at (0.6,0.) [Pi] {}; 
\node (v5) at (0.,0.6) [Pi] {}; 
\node (l1) at (-0.4,0) {}; 
\node (l2) at (1.,0.) [VEV] {}; 
\node (l3) at (0,-1.0) [VEV] {}; 
\node (l4) at (0,1.) [VEV] {}; 
\draw[double distance=4mm,thick] (l1) -- (v1);
\draw[double distance=4mm,thick] (v1) -- (v3);
\draw[double distance=4mm,thick] (v1) -- (v4);
\draw[double distance=4mm,thick] (v1) -- (v5);
\draw[-] (v4) -- (l2);
\draw[-] (v3) -- (l3);
\draw[-] (v5) -- (l4);
\end{tikzpicture}
+ \frac{1}{16} \,
 \begin{tikzpicture}[scale=2,baseline={([yshift=-.4ex]current bounding box.center)}]
\node (v1) at (0.0,0.) [Gamma,fill=red] {}; 
\node (v2) at (-0.6,0.) [VR] {\,$\overline{V}^{(4)}$\,}; 
\node (v3) at (0.0,-0.6) [Pi] {}; 
\node (v4) at (0.6,0.) [Pi] {}; 
\node (v5) at (0.,0.6) [Pi] {}; 
\node (l1) at (-1.,0) {}; 
\node (l2) at (1.,0.) [VEV] {}; 
\node (l3) at (0,-1.0) [VEV] {}; 
\node (l4) at (0,1.) [VEV] {}; 
\draw[double distance=4mm,thick] (l1) -- (v2);
\draw[double distance=4mm,thick] (v1) -- (v2);
\draw[double distance=4mm,thick] (v1) -- (v3);
\draw[double distance=4mm,thick] (v1) -- (v4);
\draw[double distance=4mm,thick] (v1) -- (v5);
\draw[-] (v4) -- (l2);
\draw[-] (v3) -- (l3);
\draw[-] (v5) -- (l4);
\end{tikzpicture}
\nonumber \\ & 
+  \,
 \begin{tikzpicture}[scale=2,baseline={([yshift=-.4ex]current bounding box.center)}]
\node (v1) at (0.0,0.) [Gamma,fill=red] {}; 
\node (v3) at (0.0,-0.6) [Pi] {}; 
\node (v4) at (0.6,0.) [Pi] {}; 
\node (l1) at (-0.4,0) {}; 
\node (l2) at (1.,0.) [VEV] {}; 
\node (l3) at (0,-1.0) [VEV] {}; 
\node (l4) at (0,0.4) [VEV] {}; 
\draw[double distance=4mm,thick] (l1) -- (v1);
\draw (v1) -- (v3);
\draw (v1) -- (v4);
\draw (v3) -- (v4);
\draw[-] (v4) -- (l2);
\draw[-] (v3) -- (l3);
\draw[-] (v1) -- (l4);
\end{tikzpicture}
+ \frac{1}{2} \,
 \begin{tikzpicture}[scale=2,baseline={([yshift=-.4ex]current bounding box.center)}]
\node (v1) at (0.0,0.) [Gamma,fill=red] {}; 
\node (v2) at (-0.6,0.) [VR] {\,$\overline{V}^{(4)}$\,}; 
\node (v3) at (0.0,-0.6) [Pi] {}; 
\node (v4) at (0.6,0.) [Pi] {}; 
\node (l1) at (-1.,0) {}; 
\node (l2) at (1.,0.) [VEV] {}; 
\node (l3) at (0,-1.0) [VEV] {}; 
\node (l4) at (0,0.4) [VEV] {}; 
\draw[double distance=4mm,thick] (l1) -- (v2);
\draw[double distance=4mm,thick] (v1) -- (v2);
\draw (v1) -- (v3);
\draw (v1) -- (v4);
\draw (v3) -- (v4);
\draw[-] (v4) -- (l2);
\draw[-] (v3) -- (l3);
\draw[-] (v1) -- (l4);
\end{tikzpicture}
\nonumber \\ &
+  \frac{1}{2} \,
 \begin{tikzpicture}[scale=2,baseline={([yshift=-.4ex]current bounding box.center)}]
\node (v1) at (0.0,0.) [Gamma,fill=red] {}; 
\node (v3) at (0.0,-0.6) [Pi] {}; 
\node (v4) at (0.6,0.) [Pi] {}; 
\node (v5) at (0.,0.6) [Pi] {}; 
\node (l1) at (-0.4,0) {}; 
\node (l2) at (1.,0.) [VEV] {}; 
\node (l3) at (0,-1.0) [VEV] {}; 
\node (l4) at (0,1.) [VEV] {}; 
\draw[double distance=4mm,thick] (l1) -- (v1);
\draw[double distance=4mm,thick] (v1) -- (v5);
\draw (v1) -- (v3);
\draw (v1) -- (v4);
\draw (v3) -- (v4);
\draw[-] (v4) -- (l2);
\draw[-] (v3) -- (l3);
\draw[-] (v5) -- (l4);
\end{tikzpicture}
+ \frac{1}{4} \,
 \begin{tikzpicture}[scale=2,baseline={([yshift=-.4ex]current bounding box.center)}]
\node (v1) at (0.0,0.) [Gamma,fill=red] {}; 
\node (v2) at (-0.6,0.) [VR] {\,$\overline{V}^{(4)}$\,}; 
\node (v3) at (0.0,-0.6) [Pi] {}; 
\node (v4) at (0.6,0.) [Pi] {}; 
\node (v5) at (0.,0.6) [Pi] {}; 
\node (l1) at (-1.,0) {}; 
\node (l2) at (1.,0.) [VEV] {}; 
\node (l3) at (0,-1.0) [VEV] {}; 
\node (l4) at (0,1.) [VEV] {}; 
\draw[double distance=4mm,thick] (l1) -- (v2);
\draw[double distance=4mm,thick] (v1) -- (v2);
\draw[double distance=4mm,thick] (v1) -- (v5);
\draw (v1) -- (v3);
\draw (v1) -- (v4);
\draw (v3) -- (v4);
\draw[-] (v4) -- (l2);
\draw[-] (v3) -- (l3);
\draw[-] (v5) -- (l4);
\end{tikzpicture}
\nonumber 
\end{align}
\begin{align}
& +   \,
 \begin{tikzpicture}[scale=2,baseline={([yshift=-.4ex]current bounding box.center)}]
\node (v1) at (0.0,0.) [Gamma] {}; 
\node (v3) at (0.0,-0.6) [Pi] {}; 
\node (v4) at (0.6,0.) [Pi] {}; 
\node (v5) at (0.,0.6) [Pi] {}; 
\node (l1) at (-0.4,0) {}; 
\node (l2) at (1.,0.) [VEV] {}; 
\node (l3) at (0,-1.0) [VEV] {}; 
\node (l4) at (0,1.) [VEV] {}; 
\draw[double distance=4mm,thick] (l1) -- (v1);
\draw (v1) -- (v5);
\draw (v1) -- (v3);
\draw (v5) -- (v4);
\draw (v3) -- (v4);
\draw[-] (v4) -- (l2);
\draw[-] (v3) -- (l3);
\draw[-] (v5) -- (l4);
\end{tikzpicture}
+ \frac{1}{2} \,
 \begin{tikzpicture}[scale=2,baseline={([yshift=-.4ex]current bounding box.center)}]
\node (v1) at (0.0,0.) [Gamma] {}; 
\node (v2) at (-0.6,0.) [VR] {\,$\overline{V}^{(4)}$\,}; 
\node (v3) at (0.0,-0.6) [Pi] {}; 
\node (v4) at (0.6,0.) [Pi] {}; 
\node (v5) at (0.,0.6) [Pi] {}; 
\node (l1) at (-1.,0) {}; 
\node (l2) at (1.,0.) [VEV] {}; 
\node (l3) at (0,-1.0) [VEV] {}; 
\node (l4) at (0,1.) [VEV] {}; 
\draw[double distance=4mm,thick] (l1) -- (v2);
\draw[double distance=4mm,thick] (v1) -- (v2);
\draw (v1) -- (v5);
\draw (v1) -- (v3);
\draw (v5) -- (v4);
\draw (v3) -- (v4);
\draw[-] (v4) -- (l2);
\draw[-] (v3) -- (l3);
\draw[-] (v5) -- (l4);
\end{tikzpicture}
\nonumber \\ & 
+ \frac{1}{2} \,
 \begin{tikzpicture}[scale=2,baseline={([yshift=-.4ex]current bounding box.center)}]
\node (v1) at (0.0,0.) [Gamma,fill=red] {}; 
\node (v3) at (0.0,-0.6) [Pi] {}; 
\node (l1) at (-0.4,0) {}; 
\node (l2) at (0.4,-0.6) [VEV] {}; 
\node (l3) at (0,-1.0) [VEV] {}; 
\node (l4) at (0,0.4) [VEV] {}; 
\draw[double distance=4mm,thick] (l1) -- (v1);
\draw[double distance=4mm,thick] (v1) -- (v3);
\draw[-] (v3) -- (l2);
\draw[-] (v3) -- (l3);
\draw[-] (v1) -- (l4);
\end{tikzpicture}
+ \frac{1}{4} \,
 \begin{tikzpicture}[scale=2,baseline={([yshift=-.4ex]current bounding box.center)}]
\node (v1) at (0.0,0.) [Gamma,fill=red] {}; 
\node (v2) at (-0.6,0.) [VR] {\,$\overline{V}^{(4)}$\,}; 
\node (v3) at (0.0,-0.6) [Pi] {}; 
\node (l1) at (-1.,0) {}; 
\node (l2) at (0.4,-0.6) [VEV] {}; 
\node (l3) at (0,-1.0) [VEV] {}; 
\node (l4) at (0,0.4) [VEV] {}; 
\draw[double distance=4mm,thick] (l1) -- (v2);
\draw[double distance=4mm,thick] (v1) -- (v2);
\draw[double distance=4mm,thick] (v1) -- (v3);
\draw[-] (v3) -- (l2);
\draw[-] (v3) -- (l3);
\draw[-] (v1) -- (l4);
\end{tikzpicture}
\nonumber \\ & 
+ \frac{1}{4} \,
 \begin{tikzpicture}[scale=2,baseline={([yshift=-.4ex]current bounding box.center)}]
\node (v1) at (0.0,0.) [Gamma,fill=red] {}; 
\node (v3) at (0.0,-0.6) [Pi] {}; 
\node (v5) at (0.0,0.6) [Pi] {}; 
\node (l1) at (-0.4,0) {}; 
\node (l2) at (0.4,-0.6) [VEV] {}; 
\node (l3) at (0,-1.0) [VEV] {}; 
\node (l4) at (0,1.) [VEV] {}; 
\draw[double distance=4mm,thick] (l1) -- (v1);
\draw[double distance=4mm,thick] (v1) -- (v3);
\draw[double distance=4mm,thick] (v1) -- (v5);
\draw[-] (v3) -- (l2);
\draw[-] (v3) -- (l3);
\draw[-] (v5) -- (l4);
\end{tikzpicture}
+ \frac{1}{8} \,
 \begin{tikzpicture}[scale=2,baseline={([yshift=-.4ex]current bounding box.center)}]
\node (v1) at (0.0,0.) [Gamma,fill=red] {}; 
\node (v2) at (-0.6,0.) [VR] {\,$\overline{V}^{(4)}$\,}; 
\node (v3) at (0.0,-0.6) [Pi] {}; 
\node (v5) at (0.0,0.6) [Pi] {}; 
\node (l1) at (-1.,0) {}; 
\node (l2) at (0.4,-0.6) [VEV] {}; 
\node (l3) at (0,-1.0) [VEV] {}; 
\node (l4) at (0,1.) [VEV] {}; 
\draw[double distance=4mm,thick] (l1) -- (v2);
\draw[double distance=4mm,thick] (v1) -- (v2);
\draw[double distance=4mm,thick] (v1) -- (v3);
\draw[double distance=4mm,thick] (v1) -- (v5);
\draw[-] (v3) -- (l2);
\draw[-] (v3) -- (l3);
\draw[-] (v5) -- (l4);
\end{tikzpicture}
\nonumber \\ & 
+  \,
 \begin{tikzpicture}[scale=2,baseline={([yshift=-.4ex]current bounding box.center)}]
\node (v1) at (0.0,0.) [Gamma] {}; 
\node (v3) at (0.0,-0.6) [Pi] {}; 
\node (v4) at (0.6,0.) [Pi] {}; 
\node (l1) at (-0.4,0) {}; 
\node (l2) at (1.,0.) [VEV] {}; 
\node (l3) at (0,-1.0) [VEV] {}; 
\node (l4) at (0.6,0.4) [VEV] {}; 
\draw[double distance=4mm,thick] (l1) -- (v1);
\draw (v1) -- (v3);
\draw (v1) -- (v4);
\draw (v3) -- (v4);
\draw[-] (v4) -- (l2);
\draw[-] (v3) -- (l3);
\draw[-] (v4) -- (l4);
\end{tikzpicture}
+ \frac{1}{2} \,
 \begin{tikzpicture}[scale=2,baseline={([yshift=-.4ex]current bounding box.center)}]
\node (v1) at (0.0,0.) [Gamma] {}; 
\node (v2) at (-0.6,0.) [VR] {\,$\overline{V}^{(4)}$\,}; 
\node (v3) at (0.0,-0.6) [Pi] {}; 
\node (v4) at (0.6,0.) [Pi] {}; 
\node (l1) at (-1.,0) {}; 
\node (l2) at (1.,0.) [VEV] {}; 
\node (l3) at (0,-1.0) [VEV] {}; 
\node (l4) at (0.6,0.4) [VEV] {}; 
\draw[double distance=4mm,thick] (l1) -- (v2);
\draw[double distance=4mm,thick] (v1) -- (v2);
\draw (v1) -- (v3);
\draw (v1) -- (v4);
\draw (v3) -- (v4);
\draw[-] (v4) -- (l2);
\draw[-] (v3) -- (l3);
\draw[-] (v4) -- (l4);
\end{tikzpicture} \,\,.
\label{eq:d3_Pi_dphi3}
\end{align}
Even though many of the diagrams vanish in the Hartree approximation, there are still quite a few to evaluate. However, before doing so explicitly, one can show that some of them only contribute at $O(\epsilon)$. This can be seen as follows: potential topologies that can contribute in any of the $\Gamma^{\text{2PI}}_{\text{int}}$ blobs that appear in loops and do not vanish is $\sim (\lambda_R + \delta\lambda_n)$, $n = 0, 2$. Due to the shifts \eqref{eqn:os_dl0_exp}, this gives a factor that is always $O(\epsilon)$. Other factors of $\epsilon$ come from the loop integrals: one- and two-point integrals are $O(\epsilon^{-1})$ and $n$-point functions with $n \ge 3$ are $O(1)$. Finally, the vertex function $\overline{V}^{(4)}$ is of course $O(1)$. Thus, one can count the powers of $\epsilon$ for each diagram and, as it turns out, many diagrams only contribute at $O(\epsilon)$ to the total four-point function. 

This analysis can be first applied to the self-energy boxes, revealing that $\delta^n \overline{\Pi} / \delta \phi^n_R$ $(n = 1, 2, 3)$ are all $O(1)$. In \eqref{eq:d2_Pi_dphi2}, one finds that only the second and last terms are finite and the other two non-vanishing ones give $O(\epsilon)$ contributions. In the case of \eqref{eq:d3_Pi_dphi3}, an analogous analysis reveals that only two contribute: the fourteenth and the last one. Finally, we return to the four-point function \eqref{eq:dGamma4_dphi4} and applying the same analysis, we find that the first, eighth and last diagrams contribute at $O(1)$. 

There are 3- and 4-point loop integrals among these diagrams which, in dimensional regularisation, are the well-known $C_0$ and $D_0$ functions
\begin{align}
C_0(p_1,p_2) &=  (16\pi^2) \int_q G_R(q) G_R(q+p_1)  G_R(q+p_2) \\ 
D_0(p_1,p_2,p_3) &= -i (16\pi^2) \int_q G_R(q) G_R(q+p_1)  G_R(q+p_2) G_R(q+p_4)  \,.
\end{align}
Note that both functions are ultraviolet finite and explicit expressions can be found, for example in \cite{Denner:1991kt}. 

Assembling the different contributions, with the momentum assignments $p_{1,2}$ incoming and $p_{3,4}$ outgoing, we obtain
\begin{align}
\label{eqn:gamma4_ht}
&\Gamma^{(4)}(p_1,p_2,p_3,p_4) = -(\lambda_R + \delta \lambda_4)
+ \lambda_R \left[J(p_1+p_2) +J(p_1-p_3) +J(p_1-p_4) \right] \nonumber \\[2mm]
& + \frac{\lambda^3_R \phi^2_R}{16\pi^2} 
\big[J(p_1+p_2) J(p_3)J(p_4)C_0(p_1+p_2,p_4) + J(p_1-p_3)J(p_2)J(p_4)C_0(p_1-p_3,p_4) \nonumber \\[2mm]
& \hspace{13mm}+ J(p_1-p_4) J(p_2) J(p_3) C_0(p_1-p_4,p_3) + J(p_1) J(p_2) J(p_3+p_4) C_0(p_1,p_3+p_4)  \nonumber \\[2mm]
& \hspace{13mm}+ J(p_1) J(p_3) J(p_2-p_4)C_0(p_1,p_2-p_4) + J(p_1)J(p_4)J(p_2-p_3)C_0(p_1,p_2-p_3)  \big]\nonumber \\[2mm]
& - \frac{\lambda^4_R \phi^4_R}{16\pi^2} \,J(p_1) J(p_2) J(p_3) J(p_4) \big[D_0(p_2,p_1+p_2,p_3) + D_0(p_2,p_2-p_3,p_3)+ D_0(p_2,p_2-p_4,p_4)\big] \nonumber \\[2mm]
& + \frac{\lambda^5_R \phi^4_R}{(16\pi^2)^2} \,J(p_1) J(p_2) J(p_3) J(p_4) \big[ J(p_1+p_2) C_0(p_1,p_1+p_2) C_0(p_3+p_4,p_4)  \nonumber \\[2mm]
& \hspace{5mm} +  J(p_1-p_3) C_0(p_1,p_1-p_3) C_0(p_4-p_2,p_4) +  J(p_1-p_4) C_0(p_1,p_1-p_4) C_0(p_3-p_2,p_3)  \big] \,,
\end{align} 
where we have introduced
\begin{align}
J(p) \equiv \left[ 1- \frac{\lambda_R}{32\pi^2} \left( B_0(p^2_{\ast},m^2_R,m^2_R) - B_0(p^2,m^2_R,m^2_R)
\right) \right]^{-1}
\end{align}
which originates from the vertex function. The counterterm, $\delta \lambda_4$ can then easily be obtained by the renormalization condition \eqref{eq:renor_cond_fourpt}.

For completeness, one should also note that only in the broken phase $(\phi_R \neq 0)$, there exists a physical three-point function, for which we have the following topologies
\begin{align}
\Gamma^{(3)} \equiv \frac{\delta^3 \Gamma}{\delta \phi^3} = & \quad
 \begin{tikzpicture}[scale=2,baseline={([yshift=-.4ex]current bounding box.center)}]
\node (fourpoint) at (0,0) [Gamma] {}; 
\node (l1) at (-0.4,0) [VEV] {}; 
\node (l2) at (0.,0.4) [VEV] {}; 
\node (l3) at (0.4,0) [VEV] {}; 
\draw[-] (fourpoint) -- (l1);
\draw[-] (fourpoint) -- (l2);
\draw[-] (fourpoint) -- (l3);
\end{tikzpicture}
+
\frac{1}{2} \,\,
 \begin{tikzpicture}[scale=2,baseline={([yshift=-.4ex]current bounding box.center)}]
\node (v1) at (0,0.3) [Gamma] {}; 
\node (v3) at (0.6,0.3) [Pi] {}; 
\node (l1) at (-0.4,0.3) [VEV] {}; 
\node (l2) at (0.,0.7) [VEV] {}; 
\node (l3) at (0.9,0.3) [VEV] {}; 
\draw[double distance=4mm,thick] (v1) -- (v3);
\draw[-] (v1) -- (l1);
\draw[-] (v1) -- (l2);
\draw[-] (v3) -- (l3);
\end{tikzpicture}
\nonumber \\[8mm]
&+
\frac{1}{4} \,\,
 \begin{tikzpicture}[scale=2,baseline={([yshift=-.4ex]current bounding box.center)}]
\node (v1) at (0,0.3) [Gamma, fill=red] {}; 
\node (v2) at (0.6,0.3) [Pi] {}; 
\node (v3) at (0.0,0.9) [Pi] {}; 
\node (l1) at (-0.4,0.3) [VEV] {}; 
\node (l2) at (0.,1.3) [VEV] {}; 
\node (l3) at (0.9,0.3) [VEV] {}; 
\draw[double distance=4mm,thick] (v1) -- (v2);
\draw[double distance=4mm,thick] (v1) -- (v2);
\draw[double distance=4mm,thick] (v1) -- (v3);
\draw[-] (v1) -- (l1);
\draw[-] (v3) -- (l2);
\draw[-] (v2) -- (l3);
\end{tikzpicture}
+
 \begin{tikzpicture}[scale=2,baseline={([yshift=-.4ex]current bounding box.center)}]
\node (v1) at (0,0.3) [Gamma] {}; 
\node (v2) at (0,-0.3) [Pi] {}; 
\node (v3) at (0.6,0.3) [Pi] {}; 
\node (l1) at (-0.4,0.3) [VEV] {}; 
\node (l2) at (0.9,0.3) [VEV] {}; 
\node (l3) at (0.,-0.7) [VEV] {}; 
\draw (v1) -- (v2);
\draw (v1) -- (v3);
\draw (v2) -- (v3);
\draw[-] (v1) -- (l1);
\draw[-] (v3) -- (l2);
\draw[-] (v2) -- (l3);
\end{tikzpicture}
\nonumber \\[8mm]
& + \frac{1}{2} \,\,
 \begin{tikzpicture}[scale=2,baseline={([yshift=-.4ex]current bounding box.center)}]
\node (v1) at (0,0.3) [Gamma] {}; 
\node (v2) at (0.6,0.3) [Pi] {}; 
\node (l1) at (-0.4,0.3) [VEV] {}; 
\node (l2) at (0.9,0.3) [VEV] {}; 
\node (l3) at (0.6,-0.1) [VEV] {}; 
\draw[double distance=4mm,thick] (v1) -- (v2);
\draw[-] (v1) -- (l1);
\draw[-] (v2) -- (l2);
\draw[-] (v2) -- (l3);
\end{tikzpicture} \,\,\,\,.
\label{eq:dGamma3_dphi3}
\end{align}
In this case, we work with the convention that $p_1$ is the incoming four-momentum and $p_2$ and $p_3$ are outgoing. One then obtains the expression
\begin{align}
    \Gamma^{(3)}(p_1,p_2,p_3) &= -(\lambda_R+\delta\lambda_4)\phi_R + \frac{1}{2} \lambda_R\, \phi_R \left[J(p_1)+J(p_2)+J(p_3)\right]  \nonumber \\ &
    \quad - \frac{1}{2}\lambda^3_R\,\phi^3_R\, J(p_1)J(p_2)J(p_3)\frac{\left[C_0(p_1,p_2)+C_0(p_1,p_3)\right]}{16\pi^2} +\mathcal{O}(\epsilon)\,,
\end{align}
which is fully determined from the counterterm $\delta \lambda_4$.


\subsection{Scalar Sunset Approximation}
Introducing a trilinear coupling, the effective action is now given by
\begin{align}
&\Gamma^{\text{2PI}}_{\text{int}}[\phi_R,G_R] =
- \frac{1}{2}\int_{x} (\delta Z_{\phi,0} \square_x + \delta m^2_0) G_R(x,y) \big|_{x=y}
- \frac{1}{2}\int_x \phi_R(x)  (\delta Z_{\phi,2} \square_x + \delta m^2_2) \phi_R(x)
  \nonumber \\ 
& - \int_x \left[ \frac{1}{8}(\lambda_R + \delta \lambda_0) G^2_R(x,x)
   + \frac{1}{4} (\lambda_R + \delta \lambda_2) G_R(x,x) \phi^2_R(x)
   + \frac{1}{4!} (\lambda_R + \delta \lambda_4) \phi^4_R(x)
\right] \nonumber \\
& - \int_x \delta t_1 \phi_R(x) - \int_x \left[\frac{1}{2}(\alpha_R + \delta \alpha_1) \phi_R(x)G_R(x,x)
   + \frac{1}{3!} (\alpha_R + \delta \alpha_3) \phi^3_R(x)
\right] \nonumber \\
& +\frac{\ii}{12} \int_x \int_y \left[(\alpha_R + \delta \alpha_0)+(\lambda_R + \delta \lambda_1)\phi(x)\right]\left[(\alpha_R + \delta \alpha_0)+(\lambda_R + \delta \lambda_1)\phi(y)\right] \,G^3_R(x,y)
\end{align}
In this case, we carry out the renormalization at non-vanishing field expectation value $\phi_R \neq 0$, as there is no symmetry in the model. 

We may set the counterterms, $\delta \alpha_0 = \delta \lambda_1 = 0$, which is possible as they amount to finite renormalizations at this level of the 2PI truncation; more specifically, the first non-trivial contribution to these are obtained when one includes the basketball diagram \cite{Patkos:2008ik,Pilaftsis:2013xna, Pilaftsis:2015cka}. We then obtain the following four-point kernels which are now momentum dependent
\begin{align}
    &\oL^{(4)}(p_1,p_2,p_3,p_4) = -(\lambda_R + \delta \lambda_0) + 2i(\alpha_R + \lambda_R \phi_R)^2 G_R(p_3-p_1) \,,
\label{eq:4ptkalpha1} \\[4mm]
    &\Lambda^{(4)}(p_1,p_2,p_3,p_4) = -(\lambda_R + \delta \lambda_2) + i \lambda_R^2 \int_q G_R(q)G_R(p+q) \nonumber \\[2mm]
    &\qquad \qquad \equiv -(\lambda_R + \delta \lambda_2) + \lambda_R^2\,\mathcal{I}(p)\,,
    \label{eq:4ptkalpha2}
\end{align}
where $p =p_1 +p_2$. In addition, we also define the following three-point kernel 
\begin{align}
    \Lambda^{(3)}(p_1,p_2,p_3) = 2 \frac{\delta^2 \Gamma^{\text{2PI}}_{\text{int}}}{\delta \phi_R\, \delta G_R(p_1)} = -(\alpha_R+\delta \alpha_1) +  \lambda_R (\alpha_R + \lambda_R\phi_R) \,\mathcal{I}(p_1)\,.
        \label{eq:3ptkalpha}
\end{align}
The corresponding BSE for its resummation is 
\begin{align}
	&V^{(3)}(p_1,p_2,p_3) = \Lambda^{(3)}(p_1,p_2,p_3) \nonumber \\[2mm]
	&\quad +  \frac{i}{2}\int_q \Lambda^{(3)}(p_1,p_2,p_3)  G_R(q) G_R(p_1+q)\,\overline{V}^{(4)}(q+p_1,-q,p_2,p_3)\,.
	\label{eqn:v3_ss}
\end{align}
We now implement appropriate renormalization conditions in order to determine the various coupling constant counterterms. For the four-point vertices, we continue to work with the convention that the we have $p_{1,2}$ as incoming and $p_{3,4}$ as outgoing momenta. We first have from \eqref{eq:BSEVbar},
\begin{align}
    &\overline{V}^{(4)}(p_1,p_2,p_3,p_4) = \overline{\Lambda}^{(4)}(p_1,p_2,p_3,p_4) \nonumber \\[2mm]
    &\qquad \qquad  + \frac{i}{2}\int_q \overline{\Lambda}^{(4)}(p_1,p_2,q+p,-q)\, G_R(q) G_R(p+q)\,\overline{V}^{(4)}(q+p,-q,p_3,p_4)\,.
    \label{4ptvalpha}
\end{align} 
Let us set the renormalization condition,
\begin{equation}
    \overline{V}^{(4)}(p_{1 \ast},p_{2 \ast},p_{3 \ast},p_{4 \ast}) = -\lambda_R + 2i (\alpha_R+\lambda_R\phi_R)^2 \,G_R(p_{3\ast}-p_{1\ast})\,,
    \label{eq:v4alpha_rc}
\end{equation}
which accounts for the fact that four scalars may scatter directly via the quartic coupling, or by a scalar exchange through two trilinear couplings. Like in the Hartree approximation, we work in the COM system, which is characterised by a COM momentum and scattering angle.
In this case, the second term induces a dependence
on the scattering angle and, thus, we set $|\vec{p_{\ast}}| = m_R$ and $\theta_{\ast} = \pi$ as our renormalization point. 
Using \eqref{eq:v4alpha_rc} and solving for $\delta \lambda_0$, we obtain
\begin{equation}
    \delta \lambda_0  = -\lambda_R + \frac{\lambda_R -  (\alpha_R+\lambda_R\phi_R)^2\, I_2(p_{\ast},p_{2 \ast},p_{3 \ast},p_{4 \ast})}{1+\frac{1}{2}I_1(p_{\ast},p_{3 \ast},p_{4 \ast})} \,,
    \label{eq:dllambda0_tri}
\end{equation}
where we have defined the following integrals over the vertex functions:
\begin{align}
    \label{eqn:vertexintegral1}
    I_1(p,k,r) &= i \int_q G_R(q)G_R(p+q)\overline{V}^{(4)}(q+p,-q,k,r) \,,\\[2mm]
     \label{eqn:vertexintegral2}
    I_2(p,k,r,l) &= \int_q G_R(q) G_R(q+p)G_R(q+k)G_R(q+r)\overline{V}^{(4)}(q+p,-q,r,l) \,.
\end{align}
Note that the functions $I_1$ and $I_2$ are at most logarithmically divergent and finite respectively, which we can ascertain by counting the number of propagators involved. The counterterm $\delta \lambda_0$ from \eqref{eq:dllambda0_tri} is hence discerned as finite. Plugging this back into \eqref{eq:4ptkalpha1} and \eqref{4ptvalpha}, we obtain the following expression for the four-point function
\begin{align}
    &\overline{V}^{(4)}(p_1,p_2,p_3,p_4) =-\lambda_R\left(\frac{1+\frac{1}{2}I_1(p,p_{3},p_{4})}{1+\frac{1}{2}I_1(p_{\ast},p_{3 \ast},p_{4 \ast})}\right)
    + 2i(\alpha_R+\lambda_R\phi_R)^2 G_R(p_3 - p_1) \nonumber \\[2mm]
    &+ (\alpha_R+\lambda_R\phi_R)^2 \left[I_2(p,p_2,p_3,p_4) - I_2 (p_{\ast},p_{3 \ast},p_{4 \ast})
    \left(\frac{1+\frac{1}{2}I_1(p,p_{3},p_{4})}{1+\frac{1}{2}I_1(p_{\ast},p_{3 \ast},p_{4 \ast})}\right)\right]\,,
    \label{v4Ralpha}
\end{align}
which is discerned to be finite from the arguments related to the loop integrals presented above.

As our model does not possess the $\mathbb{Z}_2$ symmetry in the Hartree case, our starting point to determine the counterterm $\delta \lambda_2$ is the following BSE 
\begin{align}
      &V^{(4)}(p_1,p_2,p_3,p_4) = \Lambda^{(4)}(p_1,p_2,p_3,p_4) \nonumber \\[2mm]
    &\qquad \qquad  + \frac{i}{2}\int_q \Lambda^{(4)}(p_1,p_2,q+p,-q) G_R(q) G_R(q+p) \overline{V}^{(4)}(q+p,-q,p_3,p_4) \,.
\label{eqn:v4aux_ss}
\end{align}
From the definition \eqref{eq:4ptkalpha2}, $\Lambda^{(4)}$ does not depend on the integrating loop momentum, which simplifies matters. We impose the renormalization condition
\begin{equation}
    V^{(4)}(p_{1\ast},p_{2\ast},p_{3\ast},p_{4\ast}) = -\lambda_R
\end{equation}
and this leads to the following result
\begin{equation}
    \delta \lambda_2  = -\lambda_R + \lambda^2_R\,\mathcal{I}(p_{\ast})+ \frac{\lambda_R}{1+\frac{1}{2}I_1(p_{\ast},p_{3 \ast},p_{4\ast})}\,\,.
    \label{eqn:dellambda2_tri}
\end{equation}
Firstly, we notice that $\delta \lambda_2 \neq \delta \lambda_0$ unlike in the Hartree approximation, even if we set $\alpha = 0$. This is because, in the scalar sunset approximation, the effective trilinear coupling $\sim \lambda_R \phi_R$ gives a contribution which is the second term of \eqref{eqn:dellambda2_tri}. The very same term introduces a divergence in $\delta \lambda_2$ due to the loop integral $\mathcal{I}(p) = \ii\int_q G_R(q) G_R(q+p)$.

We now examine the three-point function \eqref{eqn:v3_ss}, which has a similar structure to $V^{(4)}$. Firstly, let us describe the kinematics: we take two of the scalars with four-m$p_2$ and $p_3$ to be on-shell. By the conservation of four-momentum, we can calculate $p_1$ as 
\begin{equation}
    p^2_1 = (p_2 + p_3)^2 = 4(|\vec{p}|^2 + m^2_R) \,,
\end{equation}
where we work in the COM frame for $p_2$ and $p_3$ so 
that these scalars are produced back-to-back. For our renormalization condition, we fix the four vectors as follows
\begin{equation}
    p_{2\ast} = \left[\sqrt{2}m_R,\,\vec{p_{\ast}}\right]\,,\quad 
    p_{3\ast} = \left[\sqrt{2}m_R,\,-\vec{p_{\ast}}\right]\,\quad
    p_{1\ast} = \left[2\sqrt{2}m_R,0\right]    
\end{equation}
with $|\vec{p_{\ast}}| = m_R$ and require
\begin{equation}
    V^{(3)}(p_{1 \ast},p_{2\ast},p_{3\ast}) = -\alpha_R \,.
\end{equation}
This gives us the following relation
\begin{equation}
    \delta \alpha_1  = -\alpha_R + \lambda_R (\alpha_R + \lambda_R\phi_R)\,\mathcal{I}(p^2_{1 \ast}) + \frac{\alpha_R}{1+\frac{1}{2}I_1(p_{1 \ast},p_{2 \ast},p_{3\ast})}\,.
    \label{eqn:delalpha1}
\end{equation}
This counterterm is also divergent for the same reason as $\delta \lambda_2$. 

Turning now to the gap equation, with $p$ being the external momentum, this reads
\begin{align}
    i G^{-1}_R(p)  
    &= (p^2 - m^2_R) + (\delta Z_{\phi,0} \, p^2 - \delta m^2_0) - (\alpha_R+\delta\alpha_1)\phi_R  - \frac{(\lambda_R + \delta \lambda_2)}{2}  \phi_R^2 \nonumber \\[2mm]
    &\qquad  - \frac{(\lambda_R + \delta \lambda_0)}{2}\, \mathcal{T} + \frac{(\alpha_R + \lambda_R \phi_R)^2}{2}\, \mathcal{I}(p)  \,,
\label{eqn:gap_ss}
\end{align}
where we have the one-point integral
\begin{equation}
	\mathcal{T} = \int_q G_R(q)\,.
\end{equation}
Using the same on-shell renormalization conditions in the Hartree approximation, as defined in \eqref{eq:hartree_dm0} and \eqref{eq:hartree_dZ0}, we first pick up a finite contribution to the wave function renormalization given by
\begin{equation}
    \delta Z_{\phi,0} = -\frac{(\alpha_R +\lambda_R \phi_R)^2}{2}\, \frac{\partial \mathcal{I}(p)}{\partial p^2}\bigg|_{p^2 = m^2_R} \,.
    \label{delZ0_alpha}
\end{equation}
The derivative eliminates any divergence present and therefore, $\delta Z_{\phi,0}$ is finite in the scalar sunset approximation. We can then determine the mass counterterm
\begin{align}
    \delta m^2_0 &=  \delta Z_{\phi,0} m^2_R - (\alpha_R+\delta\alpha_1)\phi_R - \frac{(\lambda_R + \delta \lambda_2)}{2} \phi_R^2  -\frac{(\lambda_R+\delta \lambda_0)}{2}\, \mathcal{T} \nonumber \\[2mm] 
    &\qquad \qquad + \frac{(\alpha_R +\lambda_R \phi_R)^2}{2}\,\mathcal{I}(p)\big|_{p^2 =m^2_R} \,.
\end{align}
This counterterm is no longer finite like in the Hartree approximation: the reason for this is that we have noted that the counterterms $\delta \lambda_2$ and $\delta \alpha_1$ are divergent. We plug back $\delta m^2_0$ and $\delta Z_{\phi,0}$ into \eqref{eqn:gap_ss} to obtain the following integral equation to ascertain the propagator
\begin{align}
    iG^{-1}_R(p) &= (p^2-m^2_R)\left[1 -\frac{(\alpha_R +\lambda_R \phi_R)^2}{2}\, \frac{\partial \mathcal{I}(q)}{\partial q^2}\bigg|_{q^2 = m^2_R}\right] +\frac{(\alpha_R +\lambda_R \phi_R)^2}{2}\left[\mathcal{I}(p) - \mathcal{I}(q)\big|_{q^2 =m^2_R}\right]\,.
    \label{eq:gap_alpha}
\end{align}
In this form, the propagator is manifestly finite due to potential divergences dropping out in the difference of the loop integral $\mathcal{I}$ or in its differentiation. However, $G_R$ cannot be given in an explicit form but needs to be determined by solving \eqref{eq:gap_alpha}, for which we use an iterative approach, described as follows:
\begin{enumerate}
	\item Initialising with the free propagator, we evaluate the loop integrals in \eqref{eq:gap_alpha} to yield the first iteration of the propagator in terms of Passarino-Veltman functions
	\begin{align}
    	i(G^{-1}_R(p))^{(0)} &= (p^2-m^2_R)\left[1 -\frac{(\alpha_R +\lambda_R \phi_R)^2}{32\pi^2}\, \dot{B}_0(m^2_R,m^2_R,m^2_R)\right]\nonumber \\[2mm]
	&\qquad +\frac{(\alpha_R +\lambda_R \phi_R)^2}{32\pi^2}\left[B_0(p^2,m^2_R,m^2_R)-B_0(m^2_R,m^2_R,m^2_R)\right]\,.
	\label{eqn:propss_1st}
	\end{align}
	\item For the next iteration, convert \eqref{eqn:propss_1st} to Euclidean space. Then, use a numerical implementation of the loop integral $\mathcal{I}(p)$ given as \cite{Pilaftsis:2013xna}
\begin{align}
    \int_q G^{\rm{E}}_R(||q||) &G^{\text{E}}_R (||p+q||) = \nonumber \\
    & \frac{1}{8\pi^3 p^2} \int^{\Lambda}_0 dq \,q\,G^{\rm{E}}_R(q) \int^{\text{min}\{|(q+||p||)|,\Lambda\}}_{|(q-||p||)|} du\, u \,\sqrt{-\lambda (u^2, q^2, ||p||^2)}\, G^{\rm{E}}_R(u) \,,
\end{align}
where the superscript E denotes Euclidean propagators and the momenta are in Euclidean space. We denote $\lambda(x,y,z) = x^2 +y^2 +z^2 -2xy -2yz -2zx$ as the K\"{a}ll\'{e}n function and $||x||$ as the Euclidean norm of $x$. Finally, $\Lambda$ denotes the UV cutoff which, in principle, tends to infinity, but for our numerical implementation needs to be sufficiently large as compared to the scalar mass, the only relevant mass scale. 
	\item Generate a set of points for this iteration of the propagator and interpolate these to obtain the propagator at this iteration. 
	\item Repeat now from step (2) until the iteration converges to the desired accuracy. As a criterion, we use the relative difference between successive iterations.
\end{enumerate}

\begin{figure}[ht]
\centering
\includegraphics[width=1.0\textwidth, keepaspectratio]{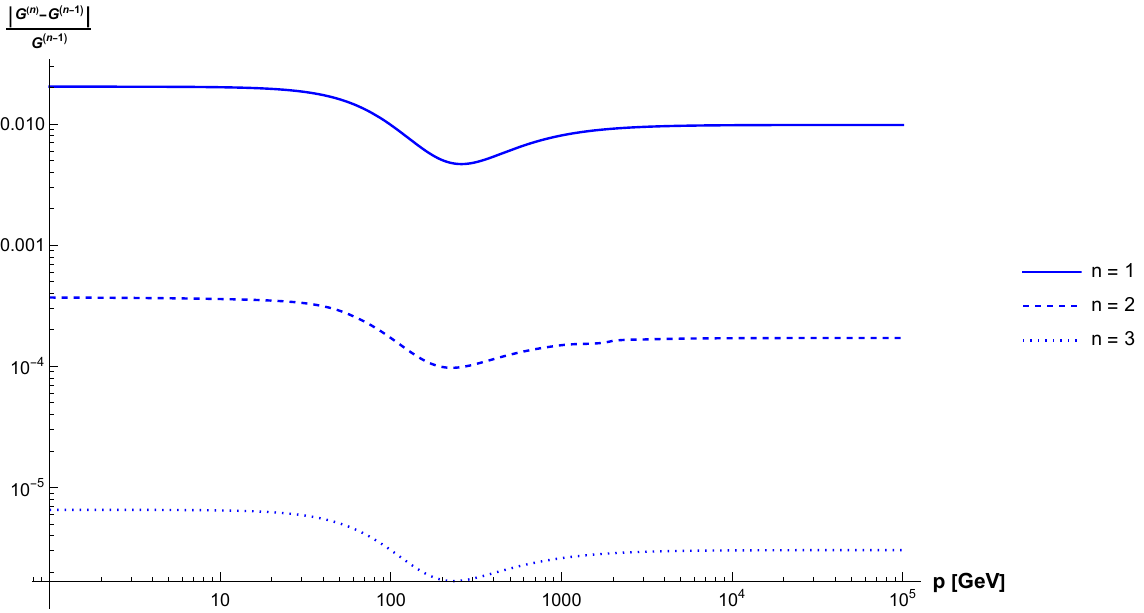}
\caption{The relative difference between successive iterations of the scalar propagator for parameter choices of $m_R = 100 \,\text{GeV}$,  $\alpha_R = 200\,\text{GeV}$ and $\lambda_R = 4$, leading to $\phi_R \approx 70\,\text{GeV}$ for a range of the norm of the Euclidean four-momentum \textbf{p}. We chose the UV cutoff $\Lambda = 10^6\,\text{GeV}$, as explained in the text.}
\label{fig:scalardiff_alpha}
\end{figure}

Our results in Fig. \ref{fig:scalardiff_alpha} show that there is indeed convergence for this iterative procedure as relative differences between successive iterations continue to get smaller for the very large couplings chosen. It would suffice, for smaller couplings hence, to use the first iteration of the propagator for practical purposes. We have also checked the dependence on the UV cutoff, $\Lambda$, and find it to be rather small, with the relative difference being at most $\mathcal{O}(10^{-5})$, even when choosing a cutoff two orders of magnitude higher.

For the determination of the remaining counterterms, we examine various derivatives of the 2PI effective potential. For example, the tadpole counterterm is straightforwardly obtained by taking a single derivative of the effective action w.r.t. $\phi_R$. The stationarity condition then gives
\begin{align}
    \Gamma^{(1)} &= -\delta t_1 - (m^2_R +\delta m^2_2) \phi_R - \frac{(\alpha_R + \delta \alpha_3)}{2}\phi^2_R - \frac{(\alpha_R + \delta \lambda_4)}{6}\phi^2_R \nonumber \\[2mm]
    &\quad - \frac{1}{2}\left[(\alpha_R+\delta \alpha_1) + (\lambda_R + \delta \lambda_4)\phi_R\right]\,\mathcal{T} + \frac{\lambda_R \left(\alpha_R + \lambda_R \phi_R\right)}{6} \,\mathcal{S} \stackrel{!}{=} 0   \,\,,
    \label{eqn:tadpole_ss}
\end{align}
where 
\begin{equation}
	\mathcal{S} = i \int_p \int_q G_R(p)G_R(q)G_R(p+q)\,.
\end{equation}
The tadpole counterterm cannot be determined without first ascertaining the missing field counterterms. The easiest of these to find is $\delta m^2_2$, for which we look at $\Gamma^{(2)}$. Using equations \eqref{eqn:phys2pta} and \eqref{eq:dPidphi} we obtain 
\begin{align}
    &\Gamma^{(2)}(p) = (p^2-m^2_R) + (\delta Z_{\phi,2} p^2  -\delta m^2_2) - (\alpha_R + \delta \alpha_3)\phi_R -\frac{1}{2}(\lambda_R+\delta\lambda_4) \phi^2_R \nonumber \\[2mm]
    & \quad  -\frac{1}{2}(\lambda_R + \delta \lambda_2)\mathcal{T} + \frac{\lambda^2_R}{6} \,\mathcal{S}   \nonumber \\[2mm]
    &\quad - \frac{1}{8}\left[(\alpha_R + \delta\alpha_1) + (\lambda_R+\delta\lambda_2)\phi_R  - \frac{\lambda_R(\alpha_R + \lambda_R\phi_R)}{3}\mathcal{I}(p)\right]^2 \,\mathcal{I}(p)  \nonumber \\[2mm]
    &\quad + \frac{1}{16}\bigg\{\left[(\alpha_R + \delta\alpha_1) + (\lambda_R+\delta\lambda_2)\phi_R  - \frac{\lambda_R(\alpha_R + \lambda_R\phi_R)}{3}\mathcal{I}(p)\right]^2 \nonumber \\[2mm]
    &\quad \quad \quad \quad \quad \quad\int_q \int_k G_R(p+q)G_R(q) \overline{V}^{(4)}(p+q,-q,p+k,-k) G_R(p+k)G_R(k) \bigg\}  \,,
    \label{eqn:scalar2pt_tri}
\end{align}
where the terms in parenthesis for the third and last lines appear from the replacement of the various building blocks in \eqref{eqn:phys2pta} and \eqref{eq:dPidphi}. The earlier analysis in the Hartree approximation does not carry over as  $\delta \lambda_2 \neq \delta \lambda_0$, and moreover, $\delta \lambda_2$ is now contains a divergent part. We can determine the counterterms $\delta Z_{\phi,2}$ and $\delta m^2_2$ with the on-shell renormalization conditions, to give
\begin{align}
    &\delta Z_{\phi,2} =  \frac{1}{8}\left[(\alpha_R + \delta\alpha_1) + (\lambda_R+\delta\lambda_2)\phi_R  - \frac{\lambda_R(\alpha_R + \lambda_R\phi_R)}{3}\mathcal{I}(p)\big|_{p^2=m^2_R}\right]\nonumber \\[2mm]
    &\Bigg\{\bigg\{-\frac{\lambda_R(\alpha_R+\lambda_R\phi_R)}{3}\mathcal{I}(p)\big|_{p^2=m^2_R}\,\frac{\partial \mathcal{I}(p)}{\partial p^2}\bigg|_{p^2 = m^2_R}   \nonumber \\[2mm]
    &\quad + \left[(\alpha_R + \delta\alpha_1)+ (\lambda_R+\delta\lambda_2)\phi_R  - \frac{\lambda_R(\alpha_R + \lambda_R\phi_R)}{3}\mathcal{I}(p)\big|_{p^2=m^2 _R}\right]\frac{\partial \mathcal{I}(p)}{\partial p^2}\bigg|_{p^2 = m^2_R} \bigg\}  \nonumber \\[2mm]
    &+\frac{1}{2}\bigg\{-\frac{\lambda_R(\alpha_R+\lambda_R\phi_R)}{3}\mathcal{I}(p)\big|_{p^2=m^2_R}\,\frac{\partial \mathcal{I}_V(p)}{\partial p^2}\bigg|_{p^2 = m^2_R}  \nonumber \\[2mm]
    &\quad + \left[(\alpha_R + \delta\alpha_1) + (\lambda_R+\delta\lambda_2)\phi_R  - \frac{\lambda_R(\alpha_R + \lambda_R\phi_R)}{3}\mathcal{I}(p)\big|_{p^2=m^2_R}\right]\frac{\partial \mathcal{I}_V(p)}{\partial p^2}\bigg|_{p^2 = m^2_R} \bigg\}\Bigg\}\,,
\end{align}

\begin{align}
    &\delta m^2_2 = m^2_R\,\delta Z_{\phi,2}  - (\alpha_R+\delta\alpha_3)\phi_R - \frac{1}{2}(\lambda_R+\delta\lambda_4)\phi^2_R -\frac{1}{2}(\lambda_R + \delta \lambda_2)\mathcal{T} + \frac{\lambda^2_R}{6} \,\mathcal{S}  \nonumber \\[2mm]
    & - \frac{1}{8}\left[(\alpha_R + \delta\alpha_1) + (\lambda_R+\delta\lambda_2)\phi_R  - \frac{\lambda_R(\alpha_R + \lambda_R\phi_R)}{3}\mathcal{I}(p)\big|_{p^2=m^2_R}\right]^2 \,\mathcal{I}(p)\big|_{p^2=m^2_R}  \nonumber \\[2mm]
    &+ \frac{1}{16}\left[(\alpha_R + \delta\alpha_1) + (\lambda_R+\delta\lambda_2)\phi_R  - \frac{\lambda_R(\alpha_R + \lambda_R\phi_R)}{3}\mathcal{I}(p)\big|_{p^2=m^2_R}  \right]^2\mathcal{I}_V(p)\big|_{p^2=m^2_R} \,,
\end{align}
where 
\begin{equation}
	\mathcal{I}_V(p) = \int_q \int_k G_R(p+q)G_R(q) \overline{V}^{(4)}(p+q,-q,p+k,-k) G_R(p+k)G_R(k) \,.
\end{equation}
On close inspection, we can discern that these counterterms would not be finite like in the Hartree approximation, due to the coupling constant counterterms being $\mathcal{O}(\epsilon^{-1})$. Furthermore, it is also obvious that the equalities $\delta Z_{\phi,2} = \delta Z_{\phi,0}$ and $\delta m^2_2 = \delta m^2_0$ no longer hold true, as we had in the Hartree approximation.

Finally, the only undetermined counterterms are 
$\delta \alpha_3$ and $\delta \lambda_4$, which are needed to completely express $\delta t_1$ and $\delta m^2_2$. 
The corresponding formulas are very lengthy and, thus we give here the corresponding diagrammatic representation.
\begin{align}
\Gamma^{(3)} \equiv \frac{\delta^3 \Gamma}{\delta \phi^3} = & \quad
 \begin{tikzpicture}[scale=2,baseline={([yshift=-.4ex]current bounding box.center)}]
\node (fourpoint) at (0,0) [Gamma] {}; 
\node (l1) at (-0.4,0) [VEV] {}; 
\node (l2) at (0.,0.4) [VEV] {}; 
\node (l3) at (0.4,0) [VEV] {}; 
\draw[-] (fourpoint) -- (l1);
\draw[-] (fourpoint) -- (l2);
\draw[-] (fourpoint) -- (l3);
\end{tikzpicture}
+
\frac{1}{2} \,\,
 \begin{tikzpicture}[scale=2,baseline={([yshift=-.4ex]current bounding box.center)}]
\node (v1) at (0,0.3) [Gamma] {}; 
\node (v3) at (0.6,0.3) [Pi] {}; 
\node (l1) at (-0.4,0.3) [VEV] {}; 
\node (l2) at (0.,0.7) [VEV] {}; 
\node (l3) at (0.9,0.3) [VEV] {}; 
\draw[double distance=4mm,thick] (v1) -- (v3);
\draw[-] (v1) -- (l1);
\draw[-] (v1) -- (l2);
\draw[-] (v3) -- (l3);
\end{tikzpicture}
+
\frac{1}{4} \,\,
 \begin{tikzpicture}[scale=2,baseline={([yshift=-.4ex]current bounding box.center)}]
\node (v1) at (0,0.3) [Gamma] {}; 
\node (v2) at (0.6,0.3) [Pi] {}; 
\node (v3) at (0.0,0.9) [Pi] {}; 
\node (l1) at (-0.4,0.3) [VEV] {}; 
\node (l2) at (0.,1.3) [VEV] {}; 
\node (l3) at (0.9,0.3) [VEV] {}; 
\draw[double distance=4mm,thick] (v1) -- (v2);
\draw[double distance=4mm,thick] (v1) -- (v2);
\draw[double distance=4mm,thick] (v1) -- (v3);
\draw[-] (v1) -- (l1);
\draw[-] (v3) -- (l2);
\draw[-] (v2) -- (l3);
\end{tikzpicture}
+
 \begin{tikzpicture}[scale=2,baseline={([yshift=-.4ex]current bounding box.center)}]
\node (v1) at (0,0.3) [Gamma] {}; 
\node (v2) at (0,-0.3) [Pi] {}; 
\node (v3) at (0.6,0.3) [Pi] {}; 
\node (l1) at (-0.4,0.3) [VEV] {}; 
\node (l2) at (0.9,0.3) [VEV] {}; 
\node (l3) at (0.,-0.7) [VEV] {}; 
\draw (v1) -- (v2);
\draw (v1) -- (v3);
\draw (v2) -- (v3);
\draw[-] (v1) -- (l1);
\draw[-] (v3) -- (l2);
\draw[-] (v2) -- (l3);
\end{tikzpicture}
\nonumber \\[8mm]
& + \frac{1}{2} \,\,
 \begin{tikzpicture}[scale=2,baseline={([yshift=-.4ex]current bounding box.center)}]
\node (v1) at (0,0.3) [Gamma] {}; 
\node (v2) at (0.6,0.3) [Pi] {}; 
\node (l1) at (-0.4,0.3) [VEV] {}; 
\node (l2) at (0.9,0.3) [VEV] {}; 
\node (l3) at (0.6,-0.1) [VEV] {}; 
\draw[double distance=4mm,thick] (v1) -- (v2);
\draw[-] (v1) -- (l1);
\draw[-] (v2) -- (l2);
\draw[-] (v2) -- (l3);
\end{tikzpicture} \,\,\,\,.
\label{eq:Gamma3_alpha}
\end{align}

\begin{align}
\Gamma^{(4)} \equiv \frac{\delta^4 \Gamma}{\delta \phi^4} = & \quad
 \begin{tikzpicture}[scale=2,baseline={([yshift=-.4ex]current bounding box.center)}]
\node (fourpoint) at (0,0) [Gamma] {}; 
\node (l1) at (-0.4,0) [VEV] {}; 
\node (l2) at (0.,0.4) [VEV] {}; 
\node (l3) at (0.4,0) [VEV] {}; 
\node (l4) at (0.,-0.4) [VEV] {}; 
\draw[-] (fourpoint) -- (l1);
\draw[-] (fourpoint) -- (l2);
\draw[-] (fourpoint) -- (l3);
\draw[-] (fourpoint) -- (l4);
\end{tikzpicture}
+
\frac{1}{4} \,\,
 \begin{tikzpicture}[scale=2,baseline={([yshift=-.4ex]current bounding box.center)}]
\node (v1) at (0,0.3) [Gamma] {}; 
\node (v2) at (0,-0.3) [Pi] {}; 
\node (v3) at (0.6,0.3) [Pi] {}; 
\node (l1) at (-0.4,0.3) [VEV] {}; 
\node (l2) at (0.,0.7) [VEV] {}; 
\node (l3) at (0.9,0.3) [VEV] {}; 
\node (l4) at (0.,-0.7) [VEV] {}; 
\draw[double distance=4mm,thick] (v1) -- (v2);
\draw[double distance=4mm,thick] (v1) -- (v3);
\draw[-] (v1) -- (l1);
\draw[-] (v1) -- (l2);
\draw[-] (v3) -- (l3);
\draw[-] (v2) -- (l4);
\end{tikzpicture}
+
\frac{1}{8} \,\,
 \begin{tikzpicture}[scale=2,baseline={([yshift=-.4ex]current bounding box.center)}]
\node (v1) at (0,0.3) [Gamma] {}; 
\node (v2) at (0,-0.3) [Pi] {}; 
\node (v3) at (0.6,0.3) [Pi] {}; 
\node (v4) at (0.0,0.9) [Pi] {}; 
\node (l1) at (-0.4,0.3) [VEV] {}; 
\node (l2) at (0.,1.3) [VEV] {}; 
\node (l3) at (0.9,0.3) [VEV] {}; 
\node (l4) at (0.,-0.7) [VEV] {}; 
\draw[double distance=4mm,thick] (v1) -- (v2);
\draw[double distance=4mm,thick] (v1) -- (v3);
\draw[double distance=4mm,thick] (v1) -- (v4);
\draw[-] (v1) -- (l1);
\draw[-] (v4) -- (l2);
\draw[-] (v3) -- (l3);
\draw[-] (v2) -- (l4);
\end{tikzpicture}
\nonumber \\
& +
 \begin{tikzpicture}[scale=2,baseline={([yshift=-.4ex]current bounding box.center)}]
\node (v1) at (0,0.3) [Gamma] {}; 
\node (v2) at (0,-0.3) [Pi] {}; 
\node (v3) at (0.6,0.3) [Pi] {}; 
\node (l1) at (-0.4,0.3) [VEV] {}; 
\node (l2) at (0.,0.7) [VEV] {}; 
\node (l3) at (0.9,0.3) [VEV] {}; 
\node (l4) at (0.,-0.7) [VEV] {}; 
\draw (v1) -- (v2);
\draw (v1) -- (v3);
\draw (v2) -- (v3);
\draw[-] (v1) -- (l1);
\draw[-] (v1) -- (l2);
\draw[-] (v3) -- (l3);
\draw[-] (v2) -- (l4);
\end{tikzpicture}
+
\frac{1}{2} \,\,
 \begin{tikzpicture}[scale=2,baseline={([yshift=-.4ex]current bounding box.center)}]
\node (v1) at (0,0.3) [Gamma] {}; 
\node (v2) at (0,-0.3) [Pi] {}; 
\node (v3) at (0.6,0.3) [Pi] {}; 
\node (v4) at (0.0,0.9) [Pi] {}; 
\node (l1) at (-0.4,0.3) [VEV] {}; 
\node (l2) at (0.,1.3) [VEV] {}; 
\node (l3) at (0.9,0.3) [VEV] {}; 
\node (l4) at (0.,-0.7) [VEV] {}; 
\draw (v1) -- (v2);
\draw (v1) -- (v3);
\draw (v2) -- (v3);
\draw[double distance=4mm,thick] (v1) -- (v4);
\draw[-] (v1) -- (l1);
\draw[-] (v4) -- (l2);
\draw[-] (v3) -- (l3);
\draw[-] (v2) -- (l4);
\end{tikzpicture}
+
 \begin{tikzpicture}[scale=2,baseline={([yshift=-.4ex]current bounding box.center)}]
\node (v1) at (0,0.3) [Gamma] {}; 
\node (v2) at (0,-0.3) [Pi] {}; 
\node (v3) at (0.6,0.3) [Pi] {}; 
\node (v4) at (0.0,0.9) [Pi] {}; 
\node (l1) at (-0.4,0.3) [VEV] {}; 
\node (l2) at (0.,1.3) [VEV] {}; 
\node (l3) at (0.9,0.3) [VEV] {}; 
\node (l4) at (0.,-0.7) [VEV] {}; 
\draw (v1) -- (v2);
\draw (v4) -- (v3);
\draw (v2) -- (v3);
\draw (v1) -- (v4);
\draw[-] (v1) -- (l1);
\draw[-] (v4) -- (l2);
\draw[-] (v3) -- (l3);
\draw[-] (v2) -- (l4);
\end{tikzpicture}
\nonumber \\
& + \frac{1}{2} \,\,
 \begin{tikzpicture}[scale=2,baseline={([yshift=-.4ex]current bounding box.center)}]
\node (v1) at (0,0.3) [Gamma] {}; 
\node (v2) at (0.6,0.3) [Pi] {}; 
\node (l1) at (-0.4,0.3) [VEV] {}; 
\node (l2) at (0.,0.7) [VEV] {}; 
\node (l3) at (0.9,0.3) [VEV] {}; 
\node (l4) at (0.6,-0.1) [VEV] {}; 
\draw[double distance=4mm,thick] (v1) -- (v2);
\draw[-] (v1) -- (l1);
\draw[-] (v1) -- (l2);
\draw[-] (v2) -- (l3);
\draw[-] (v2) -- (l4);
\end{tikzpicture}
+
\frac{1}{4} \,\,
 \begin{tikzpicture}[scale=2,baseline={([yshift=-.4ex]current bounding box.center)}]
\node (v1) at (0,0.3) [Gamma] {}; 
\node (v3) at (0.6,0.3) [Pi] {}; 
\node (v4) at (0.0,0.9) [Pi] {}; 
\node (l1) at (-0.4,0.3) [VEV] {}; 
\node (l2) at (0.,1.3) [VEV] {}; 
\node (l3) at (0.9,0.3) [VEV] {}; 
\node (l4) at (0.6,-0.1) [VEV] {}; 
\draw[double distance=4mm,thick] (v1) -- (v4);
\draw[double distance=4mm,thick] (v1) -- (v3);
\draw[-] (v1) -- (l1);
\draw[-] (v4) -- (l2);
\draw[-] (v3) -- (l3);
\draw[-] (v3) -- (l4);
\end{tikzpicture}
+
 \begin{tikzpicture}[scale=2,baseline={([yshift=-.4ex]current bounding box.center)}]
\node (v1) at (0,0.3) [Gamma] {}; 
\node (v3) at (0.6,0.3) [Pi] {}; 
\node (v4) at (0.0,0.9) [Pi] {}; 
\node (l1) at (-0.4,0.3) [VEV] {}; 
\node (l2) at (0.,1.3) [VEV] {}; 
\node (l3) at (0.9,0.3) [VEV] {}; 
\node (l4) at (0.6,-0.1) [VEV] {}; 
\draw (v1) -- (v3);
\draw (v4) -- (v3);
\draw (v1) -- (v4);
\draw[-] (v1) -- (l1);
\draw[-] (v4) -- (l2);
\draw[-] (v3) -- (l3);
\draw[-] (v3) -- (l4);
\end{tikzpicture}
\nonumber \\
& + \frac{1}{2} \,\,
 \begin{tikzpicture}[scale=2,baseline={([yshift=-.4ex]current bounding box.center)}]
\node (v1) at (0,0.3) [Gamma] {}; 
\node (v2) at (0.6,0.3) [Pi] {}; 
\node (l1) at (-0.4,0.3) [VEV] {}; 
\node (l2) at (0.6,0.7) [VEV] {}; 
\node (l3) at (0.9,0.3) [VEV] {}; 
\node (l4) at (0.6,-0.1) [VEV] {}; 
\draw[double distance=4mm,thick] (v1) -- (v2);
\draw[-] (v1) -- (l1);
\draw[-] (v2) -- (l2);
\draw[-] (v2) -- (l3);
\draw[-] (v2) -- (l4);
\end{tikzpicture} \,\,\,.
\label{eq:Gamma4_alpha}
\end{align}
Referring now to \eqref{eq:d2_Pi_dphi2} and \eqref{eq:d3_Pi_dphi3}, we first enlist the various non-vanishing contributions of the derivative of the self-energy w.r.t. $\phi_R$ that would be required.
\begin{align}
 \begin{tikzpicture}[scale=2,baseline={([yshift=-.4ex]current bounding box.center)}]
\node (v1) at (0.0,0.) [Pi] {}; 
\node (l1) at (-0.4,0) {}; 
\node (l2) at (0.4,0.) [VEV] {}; 
\node (l3) at (0,-0.4) [VEV] {}; 
\draw[double distance=4mm,thick] (l1) -- (v1);
\draw[-] (v1) -- (l2);
\draw[-] (v1) -- (l3);
\end{tikzpicture}
= & \,\,
 \begin{tikzpicture}[scale=2,baseline={([yshift=-.4ex]current bounding box.center)}]
\node (v1) at (0.0,0.) [Gamma] {}; 
\node (l1) at (-0.4,0) {}; 
\node (l2) at (0.4,0.) [VEV] {}; 
\node (l3) at (0,-0.4) [VEV] {}; 
\draw[double distance=4mm,thick] (l1) -- (v1);
\draw[-] (v1) -- (l2);
\draw[-] (v1) -- (l3);
\end{tikzpicture}
+ \frac{1}{2} \,
 \begin{tikzpicture}[scale=2,baseline={([yshift=-.4ex]current bounding box.center)}]
\node (v1) at (0.0,0.) [Gamma] {}; 
\node (v2) at (-0.6,0.) [VR] {\,$\overline{V}^{(4)}$\,}; 
\node (l1) at (-1.,0) {}; 
\node (l2) at (0.4,0.) [VEV] {}; 
\node (l3) at (0,-0.4) [VEV] {}; 
\draw[double distance=4mm,thick] (l1) -- (v2);
\draw[double distance=4mm,thick] (v1) -- (v2);
\draw[-] (v1) -- (l2);
\draw[-] (v1) -- (l3);
\end{tikzpicture}
\nonumber \\ & 
+ \frac{1}{2} \,
 \begin{tikzpicture}[scale=2,baseline={([yshift=-.4ex]current bounding box.center)}]
\node (v1) at (0.0,0.) [Gamma] {}; 
\node (v3) at (0.0,-0.6) [Pi] {}; 
\node (l1) at (-0.4,0) {}; 
\node (l2) at (0.4,0.) [VEV] {}; 
\node (l3) at (0,-1.0) [VEV] {}; 
\draw[double distance=4mm,thick] (l1) -- (v1);
\draw[double distance=4mm,thick] (v1) -- (v3);
\draw[-] (v1) -- (l2);
\draw[-] (v3) -- (l3);
\end{tikzpicture}
+ \frac{1}{4} \,
 \begin{tikzpicture}[scale=2,baseline={([yshift=-.4ex]current bounding box.center)}]
\node (v1) at (0.0,0.) [Gamma] {}; 
\node (v2) at (-0.6,0.) [VR] {\,$\overline{V}^{(4)}$\,}; 
\node (v3) at (0.0,-0.6) [Pi] {}; 
\node (l1) at (-1.,0) {}; 
\node (l2) at (0.4,0.) [VEV] {}; 
\node (l3) at (0,-1.0) [VEV] {}; 
\draw[double distance=4mm,thick] (l1) -- (v2);
\draw[double distance=4mm,thick] (v1) -- (v2);
\draw[double distance=4mm,thick] (v1) -- (v3);
\draw[-] (v1) -- (l2);
\draw[-] (v3) -- (l3);
\end{tikzpicture}
\nonumber \\ & 
+ \frac{1}{4} \,
 \begin{tikzpicture}[scale=2,baseline={([yshift=-.4ex]current bounding box.center)}]
\node (v1) at (0.0,0.) [Gamma] {}; 
\node (v3) at (0.0,-0.6) [Pi] {}; 
\node (v4) at (0.6,0.) [Pi] {}; 
\node (l1) at (-0.4,0) {}; 
\node (l2) at (1.,0.) [VEV] {}; 
\node (l3) at (0,-1.0) [VEV] {}; 
\draw[double distance=4mm,thick] (l1) -- (v1);
\draw[double distance=4mm,thick] (v1) -- (v3);
\draw[double distance=4mm,thick] (v1) -- (v4);
\draw[-] (v4) -- (l2);
\draw[-] (v3) -- (l3);
\end{tikzpicture}
+ \frac{1}{8} \,
 \begin{tikzpicture}[scale=2,baseline={([yshift=-.4ex]current bounding box.center)}]
\node (v1) at (0.0,0.) [Gamma] {}; 
\node (v2) at (-0.6,0.) [VR] {\,$\overline{V}^{(4)}$\,}; 
\node (v3) at (0.0,-0.6) [Pi] {}; 
\node (v4) at (0.6,0.) [Pi] {}; 
\node (l1) at (-1.,0) {}; 
\node (l2) at (1.,0.) [VEV] {}; 
\node (l3) at (0,-1.0) [VEV] {}; 
\draw[double distance=4mm,thick] (l1) -- (v2);
\draw[double distance=4mm,thick] (v1) -- (v2);
\draw[double distance=4mm,thick] (v1) -- (v3);
\draw[double distance=4mm,thick] (v1) -- (v4);
\draw[-] (v4) -- (l2);
\draw[-] (v3) -- (l3);
\end{tikzpicture}
\nonumber \\ & 
+  \,
 \begin{tikzpicture}[scale=2,baseline={([yshift=-.4ex]current bounding box.center)}]
\node (v1) at (0.0,0.) [Gamma] {}; 
\node (v3) at (0.0,-0.6) [Pi] {}; 
\node (v4) at (0.6,0.) [Pi] {}; 
\node (l1) at (-0.4,0) {}; 
\node (l2) at (1.,0.) [VEV] {}; 
\node (l3) at (0,-1.0) [VEV] {}; 
\draw[double distance=4mm,thick] (l1) -- (v1);
\draw (v1) -- (v3);
\draw (v1) -- (v4);
\draw (v3) -- (v4);
\draw[-] (v4) -- (l2);
\draw[-] (v3) -- (l3);
\end{tikzpicture}
+ \frac{1}{2} \,
 \begin{tikzpicture}[scale=2,baseline={([yshift=-.4ex]current bounding box.center)}]
\node (v1) at (0.0,0.) [Gamma] {}; 
\node (v2) at (-0.6,0.) [VR] {\,$\overline{V}^{(4)}$\,}; 
\node (v3) at (0.0,-0.6) [Pi] {}; 
\node (v4) at (0.6,0.) [Pi] {}; 
\node (l1) at (-1.,0) {}; 
\node (l2) at (1.,0.) [VEV] {}; 
\node (l3) at (0,-1.0) [VEV] {}; 
\draw[double distance=4mm,thick] (l1) -- (v2);
\draw[double distance=4mm,thick] (v1) -- (v2);
\draw (v1) -- (v3);
\draw (v1) -- (v4);
\draw (v3) -- (v4);
\draw[-] (v4) -- (l2);
\draw[-] (v3) -- (l3);
\end{tikzpicture}
\label{eq:d2Pi_dphi2_alpha}
\end{align}

\begin{align}
 \begin{tikzpicture}[scale=2,baseline={([yshift=-.4ex]current bounding box.center)}]
\node (v1) at (0.0,0.) [Pi] {}; 
\node (l1) at (-0.4,0) {}; 
\node (l2) at (0.4,0.) [VEV] {}; 
\node (l3) at (0,-0.4) [VEV] {}; 
\node (l4) at (0,0.4) [VEV] {}; 
\draw[double distance=4mm,thick] (l1) -- (v1);
\draw[-] (v1) -- (l2);
\draw[-] (v1) -- (l3);
\draw[-] (v1) -- (l4);
\end{tikzpicture}
= & \,\,
+ \frac{1}{4} \,
 \begin{tikzpicture}[scale=2,baseline={([yshift=-.4ex]current bounding box.center)}]
\node (v1) at (0.0,0.) [Gamma] {}; 
\node (v3) at (0.0,-0.6) [Pi] {}; 
\node (v4) at (0.6,0.) [Pi] {}; 
\node (l1) at (-0.4,0) {}; 
\node (l2) at (1.,0.) [VEV] {}; 
\node (l3) at (0,-1.0) [VEV] {}; 
\node (l4) at (0,0.4) [VEV] {}; 
\draw[double distance=4mm,thick] (l1) -- (v1);
\draw[double distance=4mm,thick] (v1) -- (v3);
\draw[double distance=4mm,thick] (v1) -- (v4);
\draw[-] (v4) -- (l2);
\draw[-] (v3) -- (l3);
\draw[-] (v1) -- (l4);
\end{tikzpicture}
+ \frac{1}{8} \,
 \begin{tikzpicture}[scale=2,baseline={([yshift=-.4ex]current bounding box.center)}]
\node (v1) at (0.0,0.) [Gamma] {}; 
\node (v2) at (-0.6,0.) [VR] {\,$\overline{V}^{(4)}$\,}; 
\node (v3) at (0.0,-0.6) [Pi] {}; 
\node (v4) at (0.6,0.) [Pi] {}; 
\node (l1) at (-1.,0) {}; 
\node (l2) at (1.,0.) [VEV] {}; 
\node (l3) at (0,-1.0) [VEV] {}; 
\node (l4) at (0,0.4) [VEV] {}; 
\draw[double distance=4mm,thick] (l1) -- (v2);
\draw[double distance=4mm,thick] (v1) -- (v2);
\draw[double distance=4mm,thick] (v1) -- (v3);
\draw[double distance=4mm,thick] (v1) -- (v4);
\draw[-] (v4) -- (l2);
\draw[-] (v3) -- (l3);
\draw[-] (v1) -- (l4);
\end{tikzpicture}
\nonumber \\ & 
+ \frac{1}{8} \,
 \begin{tikzpicture}[scale=2,baseline={([yshift=-.4ex]current bounding box.center)}]
\node (v1) at (0.0,0.) [Gamma] {}; 
\node (v3) at (0.0,-0.6) [Pi] {}; 
\node (v4) at (0.6,0.) [Pi] {}; 
\node (v5) at (0.,0.6) [Pi] {}; 
\node (l1) at (-0.4,0) {}; 
\node (l2) at (1.,0.) [VEV] {}; 
\node (l3) at (0,-1.0) [VEV] {}; 
\node (l4) at (0,1.) [VEV] {}; 
\draw[double distance=4mm,thick] (l1) -- (v1);
\draw[double distance=4mm,thick] (v1) -- (v3);
\draw[double distance=4mm,thick] (v1) -- (v4);
\draw[double distance=4mm,thick] (v1) -- (v5);
\draw[-] (v4) -- (l2);
\draw[-] (v3) -- (l3);
\draw[-] (v5) -- (l4);
\end{tikzpicture}
+ \frac{1}{16} \,
 \begin{tikzpicture}[scale=2,baseline={([yshift=-.4ex]current bounding box.center)}]
\node (v1) at (0.0,0.) [Gamma] {}; 
\node (v2) at (-0.6,0.) [VR] {\,$\overline{V}^{(4)}$\,}; 
\node (v3) at (0.0,-0.6) [Pi] {}; 
\node (v4) at (0.6,0.) [Pi] {}; 
\node (v5) at (0.,0.6) [Pi] {}; 
\node (l1) at (-1.,0) {}; 
\node (l2) at (1.,0.) [VEV] {}; 
\node (l3) at (0,-1.0) [VEV] {}; 
\node (l4) at (0,1.) [VEV] {}; 
\draw[double distance=4mm,thick] (l1) -- (v2);
\draw[double distance=4mm,thick] (v1) -- (v2);
\draw[double distance=4mm,thick] (v1) -- (v3);
\draw[double distance=4mm,thick] (v1) -- (v4);
\draw[double distance=4mm,thick] (v1) -- (v5);
\draw[-] (v4) -- (l2);
\draw[-] (v3) -- (l3);
\draw[-] (v5) -- (l4);
\end{tikzpicture}
\nonumber \\ & 
+  \,
 \begin{tikzpicture}[scale=2,baseline={([yshift=-.4ex]current bounding box.center)}]
\node (v1) at (0.0,0.) [Gamma] {}; 
\node (v3) at (0.0,-0.6) [Pi] {}; 
\node (v4) at (0.6,0.) [Pi] {}; 
\node (l1) at (-0.4,0) {}; 
\node (l2) at (1.,0.) [VEV] {}; 
\node (l3) at (0,-1.0) [VEV] {}; 
\node (l4) at (0,0.4) [VEV] {}; 
\draw[double distance=4mm,thick] (l1) -- (v1);
\draw (v1) -- (v3);
\draw (v1) -- (v4);
\draw (v3) -- (v4);
\draw[-] (v4) -- (l2);
\draw[-] (v3) -- (l3);
\draw[-] (v1) -- (l4);
\end{tikzpicture}
+ \frac{1}{2} \,
 \begin{tikzpicture}[scale=2,baseline={([yshift=-.4ex]current bounding box.center)}]
\node (v1) at (0.0,0.) [Gamma] {}; 
\node (v2) at (-0.6,0.) [VR] {\,$\overline{V}^{(4)}$\,}; 
\node (v3) at (0.0,-0.6) [Pi] {}; 
\node (v4) at (0.6,0.) [Pi] {}; 
\node (l1) at (-1.,0) {}; 
\node (l2) at (1.,0.) [VEV] {}; 
\node (l3) at (0,-1.0) [VEV] {}; 
\node (l4) at (0,0.4) [VEV] {}; 
\draw[double distance=4mm,thick] (l1) -- (v2);
\draw[double distance=4mm,thick] (v1) -- (v2);
\draw (v1) -- (v3);
\draw (v1) -- (v4);
\draw (v3) -- (v4);
\draw[-] (v4) -- (l2);
\draw[-] (v3) -- (l3);
\draw[-] (v1) -- (l4);
\end{tikzpicture}
\nonumber \\ &
+  \frac{1}{2} \,
 \begin{tikzpicture}[scale=2,baseline={([yshift=-.4ex]current bounding box.center)}]
\node (v1) at (0.0,0.) [Gamma] {}; 
\node (v3) at (0.0,-0.6) [Pi] {}; 
\node (v4) at (0.6,0.) [Pi] {}; 
\node (v5) at (0.,0.6) [Pi] {}; 
\node (l1) at (-0.4,0) {}; 
\node (l2) at (1.,0.) [VEV] {}; 
\node (l3) at (0,-1.0) [VEV] {}; 
\node (l4) at (0,1.) [VEV] {}; 
\draw[double distance=4mm,thick] (l1) -- (v1);
\draw[double distance=4mm,thick] (v1) -- (v5);
\draw (v1) -- (v3);
\draw (v1) -- (v4);
\draw (v3) -- (v4);
\draw[-] (v4) -- (l2);
\draw[-] (v3) -- (l3);
\draw[-] (v5) -- (l4);
\end{tikzpicture}
+ \frac{1}{4} \,
 \begin{tikzpicture}[scale=2,baseline={([yshift=-.4ex]current bounding box.center)}]
\node (v1) at (0.0,0.) [Gamma] {}; 
\node (v2) at (-0.6,0.) [VR] {\,$\overline{V}^{(4)}$\,}; 
\node (v3) at (0.0,-0.6) [Pi] {}; 
\node (v4) at (0.6,0.) [Pi] {}; 
\node (v5) at (0.,0.6) [Pi] {}; 
\node (l1) at (-1.,0) {}; 
\node (l2) at (1.,0.) [VEV] {}; 
\node (l3) at (0,-1.0) [VEV] {}; 
\node (l4) at (0,1.) [VEV] {}; 
\draw[double distance=4mm,thick] (l1) -- (v2);
\draw[double distance=4mm,thick] (v1) -- (v2);
\draw[double distance=4mm,thick] (v1) -- (v5);
\draw (v1) -- (v3);
\draw (v1) -- (v4);
\draw (v3) -- (v4);
\draw[-] (v4) -- (l2);
\draw[-] (v3) -- (l3);
\draw[-] (v5) -- (l4);
\end{tikzpicture}
\nonumber  \\
& +   \,
 \begin{tikzpicture}[scale=2,baseline={([yshift=-.4ex]current bounding box.center)}]
\node (v1) at (0.0,0.) [Gamma] {}; 
\node (v3) at (0.0,-0.6) [Pi] {}; 
\node (v4) at (0.6,0.) [Pi] {}; 
\node (v5) at (0.,0.6) [Pi] {}; 
\node (l1) at (-0.4,0) {}; 
\node (l2) at (1.,0.) [VEV] {}; 
\node (l3) at (0,-1.0) [VEV] {}; 
\node (l4) at (0,1.) [VEV] {}; 
\draw[double distance=4mm,thick] (l1) -- (v1);
\draw (v1) -- (v5);
\draw (v1) -- (v3);
\draw (v5) -- (v4);
\draw (v3) -- (v4);
\draw[-] (v4) -- (l2);
\draw[-] (v3) -- (l3);
\draw[-] (v5) -- (l4);
\end{tikzpicture}
+ \frac{1}{2} \,
 \begin{tikzpicture}[scale=2,baseline={([yshift=-.4ex]current bounding box.center)}]
\node (v1) at (0.0,0.) [Gamma] {}; 
\node (v2) at (-0.6,0.) [VR] {\,$\overline{V}^{(4)}$\,}; 
\node (v3) at (0.0,-0.6) [Pi] {}; 
\node (v4) at (0.6,0.) [Pi] {}; 
\node (v5) at (0.,0.6) [Pi] {}; 
\node (l1) at (-1.,0) {}; 
\node (l2) at (1.,0.) [VEV] {}; 
\node (l3) at (0,-1.0) [VEV] {}; 
\node (l4) at (0,1.) [VEV] {}; 
\draw[double distance=4mm,thick] (l1) -- (v2);
\draw[double distance=4mm,thick] (v1) -- (v2);
\draw (v1) -- (v5);
\draw (v1) -- (v3);
\draw (v5) -- (v4);
\draw (v3) -- (v4);
\draw[-] (v4) -- (l2);
\draw[-] (v3) -- (l3);
\draw[-] (v5) -- (l4);
\end{tikzpicture}
\nonumber \\ & 
+ \frac{1}{2} \,
 \begin{tikzpicture}[scale=2,baseline={([yshift=-.4ex]current bounding box.center)}]
\node (v1) at (0.0,0.) [Gamma] {}; 
\node (v3) at (0.0,-0.6) [Pi] {}; 
\node (l1) at (-0.4,0) {}; 
\node (l2) at (0.4,-0.6) [VEV] {}; 
\node (l3) at (0,-1.0) [VEV] {}; 
\node (l4) at (0,0.4) [VEV] {}; 
\draw[double distance=4mm,thick] (l1) -- (v1);
\draw[double distance=4mm,thick] (v1) -- (v3);
\draw[-] (v3) -- (l2);
\draw[-] (v3) -- (l3);
\draw[-] (v1) -- (l4);
\end{tikzpicture}
+ \frac{1}{4} \,
 \begin{tikzpicture}[scale=2,baseline={([yshift=-.4ex]current bounding box.center)}]
\node (v1) at (0.0,0.) [Gamma] {}; 
\node (v2) at (-0.6,0.) [VR] {\,$\overline{V}^{(4)}$\,}; 
\node (v3) at (0.0,-0.6) [Pi] {}; 
\node (l1) at (-1.,0) {}; 
\node (l2) at (0.4,-0.6) [VEV] {}; 
\node (l3) at (0,-1.0) [VEV] {}; 
\node (l4) at (0,0.4) [VEV] {}; 
\draw[double distance=4mm,thick] (l1) -- (v2);
\draw[double distance=4mm,thick] (v1) -- (v2);
\draw[double distance=4mm,thick] (v1) -- (v3);
\draw[-] (v3) -- (l2);
\draw[-] (v3) -- (l3);
\draw[-] (v1) -- (l4);
\end{tikzpicture}
\nonumber \\ & 
+ \frac{1}{4} \,
 \begin{tikzpicture}[scale=2,baseline={([yshift=-.4ex]current bounding box.center)}]
\node (v1) at (0.0,0.) [Gamma] {}; 
\node (v3) at (0.0,-0.6) [Pi] {}; 
\node (v5) at (0.0,0.6) [Pi] {}; 
\node (l1) at (-0.4,0) {}; 
\node (l2) at (0.4,-0.6) [VEV] {}; 
\node (l3) at (0,-1.0) [VEV] {}; 
\node (l4) at (0,1.) [VEV] {}; 
\draw[double distance=4mm,thick] (l1) -- (v1);
\draw[double distance=4mm,thick] (v1) -- (v3);
\draw[double distance=4mm,thick] (v1) -- (v5);
\draw[-] (v3) -- (l2);
\draw[-] (v3) -- (l3);
\draw[-] (v5) -- (l4);
\end{tikzpicture}
+ \frac{1}{8} \,
 \begin{tikzpicture}[scale=2,baseline={([yshift=-.4ex]current bounding box.center)}]
\node (v1) at (0.0,0.) [Gamma] {}; 
\node (v2) at (-0.6,0.) [VR] {\,$\overline{V}^{(4)}$\,}; 
\node (v3) at (0.0,-0.6) [Pi] {}; 
\node (v5) at (0.0,0.6) [Pi] {}; 
\node (l1) at (-1.,0) {}; 
\node (l2) at (0.4,-0.6) [VEV] {}; 
\node (l3) at (0,-1.0) [VEV] {}; 
\node (l4) at (0,1.) [VEV] {}; 
\draw[double distance=4mm,thick] (l1) -- (v2);
\draw[double distance=4mm,thick] (v1) -- (v2);
\draw[double distance=4mm,thick] (v1) -- (v3);
\draw[double distance=4mm,thick] (v1) -- (v5);
\draw[-] (v3) -- (l2);
\draw[-] (v3) -- (l3);
\draw[-] (v5) -- (l4);
\end{tikzpicture}
\nonumber \\ & 
+  \,
 \begin{tikzpicture}[scale=2,baseline={([yshift=-.4ex]current bounding box.center)}]
\node (v1) at (0.0,0.) [Gamma] {}; 
\node (v3) at (0.0,-0.6) [Pi] {}; 
\node (v4) at (0.6,0.) [Pi] {}; 
\node (l1) at (-0.4,0) {}; 
\node (l2) at (1.,0.) [VEV] {}; 
\node (l3) at (0,-1.0) [VEV] {}; 
\node (l4) at (0.6,0.4) [VEV] {}; 
\draw[double distance=4mm,thick] (l1) -- (v1);
\draw (v1) -- (v3);
\draw (v1) -- (v4);
\draw (v3) -- (v4);
\draw[-] (v4) -- (l2);
\draw[-] (v3) -- (l3);
\draw[-] (v4) -- (l4);
\end{tikzpicture}
+ \frac{1}{2} \,
 \begin{tikzpicture}[scale=2,baseline={([yshift=-.4ex]current bounding box.center)}]
\node (v1) at (0.0,0.) [Gamma] {}; 
\node (v2) at (-0.6,0.) [VR] {\,$\overline{V}^{(4)}$\,}; 
\node (v3) at (0.0,-0.6) [Pi] {}; 
\node (v4) at (0.6,0.) [Pi] {}; 
\node (l1) at (-1.,0) {}; 
\node (l2) at (1.,0.) [VEV] {}; 
\node (l3) at (0,-1.0) [VEV] {}; 
\node (l4) at (0.6,0.4) [VEV] {}; 
\draw[double distance=4mm,thick] (l1) -- (v2);
\draw[double distance=4mm,thick] (v1) -- (v2);
\draw (v1) -- (v3);
\draw (v1) -- (v4);
\draw (v3) -- (v4);
\draw[-] (v4) -- (l2);
\draw[-] (v3) -- (l3);
\draw[-] (v4) -- (l4);
\end{tikzpicture} \,\,.
\label{eq:d3Pi_dphi3_alpha}
\end{align}
While performing the analysis of the diagrams, one may note the following: $\delta \lambda_0$ is still finite and differentiation w.r.t. to three propagators yields $\sim (\alpha_R + \lambda_R \phi_R)^2$ which is also finite. However, all other counterterms, besides $\delta \lambda_0$ are not finite, and hence one needs to be careful when such quantities multiply divergent loop integrals. The explicit form of the counterterms $\delta \alpha_3$ and $\delta \lambda_4$ are obtained by placing the renormalization conditions 
\begin{equation}
	\Gamma^{(3)}(p_{1 \ast},p_{2 \ast},p_{3 \ast}) = -\alpha_R\,,\qquad \Gamma^{(4)}(p_{1 \ast},p_{2 \ast},p_{3 \ast},p_{4 \ast}) = -\lambda_R\,.
\end{equation}
For the numerical evaluation of the three- and
four-point functions, one can then use the same
numerical procedure as outlined above for $G$.


\section{Fermionic Sunset Approximation}
\label{sec:fermions}

Including fermions, we obtain the following effective action
\begin{align}
&\Gamma^{\text{2PI}}_{\text{int}}[\phi_R,G_R, D_R] = 
- \frac{1}{2}\int_{x} (\delta Z_{\phi,0} \square_x + \delta m^2_0) G_R(x,y) \big|_{x=y} 
- \frac{1}{2}\int_x \phi_R(x)  (\delta Z_{\phi,2} \square_x + \delta m^2_2) \phi_R(x)
  \nonumber \\ 
& - \int_x \left[\frac{1}{8}(\lambda_R + \delta \lambda_0) G^2_R(x,x)
   + \frac{1}{4} (\lambda_R + \delta \lambda_2) G_R(x,x) \phi^2_R(x)
   + \frac{1}{4!} (\lambda_R + \delta \lambda_4) \phi^4_R(x)
\right] \nonumber \\
&- \int_x \left[\frac{1}{2}(\alpha_R + \delta \alpha_1) \phi_R(x)G_R(x,x)
   + \frac{1}{3!} (\alpha_R + \delta \alpha_3) \phi^3_R(x)
\right] \nonumber \\
& +\frac{\ii}{12} \int_x \int_y \left(\alpha_R +\lambda_R\phi(x)\right)\left(\alpha_R +\lambda_R\phi(y)\right) \,G^3_R(x,y) \nonumber \\
& +\int_x(i\delta Z_{\psi,0}\myslash{\partial}_x-\delta M_0)D_R(x,y)\big|_{x=y} \nonumber \\
& -\frac{\ii}{2}\int_x \int_y (g_R+\delta g_0)^2 G_R(x,y)\text{tr}[D_R(x,y)D_R(y,x)] - \int_x \delta t_1 \phi_R(x)\nonumber \\
& -\int_x \left[(g_R+\delta g_1)\phi_R(x)\text{tr}[D_R (x,x)]\right] \,.
\label{gammaint_fermions}
\end{align}
where we have already set the corresponding coupling counterterms for the scalar sunset contribution to 0. We retain the counterterm for the Yukawa coupling corresponding to the fermionic sunset diagram, $\delta g_0$, for the moment. Note that as long as the fermionic mass $M_R$ does not vanish, we require an additional trilinear coupling $\alpha_R$ to account for potential divergences in scalar three-point functions with a fermionic loop. We can then obtain the following additional kernels involving fermions and scalars \cite{Reinosa:2005pj}
\begin{equation}
        \Lambda_{\psi \psi}(p_1,p_2,p_3,p_4)_{ab,cd} \equiv -\frac{\delta^2 \Gamma^{\text{2PI}}_{\text{int}}}{\delta D^{ba}_{R}(p)\,\delta D^{cd}_{R}(q)} = i(g_R+\delta g_0)^2 \,\delta_{db}\,G_R(p_3-p_1)\,\delta_{ac}\,,
\end{equation}
    
\begin{align}
        \Lambda^{(4)}_{\psi \phi}(p_1,p_2,p_3,p_4)_{ab} \equiv -2\frac{\delta^2 \Gamma^{\text{2PI}}_{\text{int}}}{\delta D^{ba}_R(p)\,\delta G_R(q)} = 2i(g_R+\delta g_0)^2 \, D_{R}(p_1-p_3)_{ab} \,,
\end{align}

\begin{align}
        \left(\Lambda^{(3)}_{\psi \phi}\right)_{ab} \equiv -\frac{\delta^2 \Gamma^{\text{2PI}}_{\text{int}}}{\delta \phi_R \,\delta D^{ba}_R(p)}
        = -(g_R+\delta g_1) \delta_{ab} \,,
\end{align}
alongside the scalar kernels that we had in the scalar sunset approximation, c.f. \eqref{eq:4ptkalpha1}, \eqref{eq:4ptkalpha2} and \eqref{eq:3ptkalpha}. For now, we have denoted the spinor indices by lowercase Latin characters. The Kronecker deltas refer to the identity matrix in spinor space.

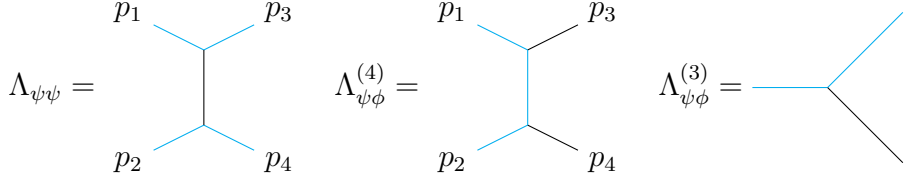
\begin{figure}[t]
\begin{equation*}
\Lambda_{\psi\psi} =
\vcenter{\hbox{\begin{tikzpicture}
	\begin{feynman}
	\vertex (v1) at (-1,1){\(p_1\)};
	\vertex (v2) at (-1,-1){\(p_2\)};
	\vertex (v3) at (1,1){\(p_3\)};
	\vertex (v4) at (1,-1){\(p_4\)};
	\vertex(x1) at (0,0.5);
	\vertex(x2) at (0,-0.5);
	\diagram*{
	(v1) --[cyan](x1),
	(v2) --[cyan](x2),
	(x1) --[cyan](v3),
	(x2) --[cyan](v4),
	(x1) -- (x2),		
	};
	\end{feynman}
\end{tikzpicture}}}
	\quad  \Lambda^{(4)}_{\psi\phi} = 
\vcenter{\hbox{\begin{tikzpicture}
	\begin{feynman}
	\vertex (v1) at (-1,1){\(p_1\)};
	\vertex (v2) at (-1,-1){\(p_2\)};
	\vertex (v3) at (1,1){\(p_3\)};
	\vertex (v4) at (1,-1){\(p_4\)};
	\vertex(x1) at (0,0.5);
	\vertex(x2) at (0,-0.5);
	\diagram*{
	(v1) --[cyan](x1),
	(v2) --[cyan](x2),
	(x1) --(v3),
	(x2) --(v4),
	(x1) -- [cyan](x2),		
	};
	\end{feynman}
\end{tikzpicture}}} 
	\quad  \Lambda^{(3)}_{\psi\phi} = 
\vcenter{\hbox{\begin{tikzpicture}
	\begin{feynman}
	\vertex (v1) at (-1,0);
	\vertex (v2) at (1,1);
	\vertex (v3) at (1,-1);
	\vertex(x) at (0,0);
	\diagram*{
	(v1) --[cyan](x),
	(x) --[cyan](v2),
	(x) --(v3),
	};
	\end{feynman}
\end{tikzpicture}}} 
\end{equation*} 
\caption{Illustration of the various kernels in the fermionic sunset approximation. Cyan lines indicate the fermionic propagator $D_R$.}
\label{fig:kernels_fermions}
\end{figure}

We proceed by defining first the governing BSE for the kernel $\Lambda_{\psi \psi}$ as
\begin{align}
    &V_{\psi\psi} (p_1,p_2,p_3,p_4)_{ab,cd} = \Lambda_{\psi \psi}(p_1,p_2,p_3,p_4)_{ab,cd} \nonumber \\[2mm]
    &\quad +i \int_q \Lambda_{\psi \psi}(p_1,p_2,q+p,-q)_{ab,ef}D_R(q)_{eg} V_{\psi\psi} (q+p,-q,p_3,p_4)_{g h,c d}D_R(p+q)_{hf}\nonumber \\[3mm]
    &= i(g_R+\delta g_0)^2 \delta_{db}\,G_R(p_3-p_1)\,\delta_{ac} \nonumber\\[2mm]
    &\quad - (g_R+\delta g_0)^2 \int_q G_R(q+p_2) D_R(q)_{ae} V_{\psi\psi} (q+p,-q,p_3,p_4)_{ef,cd}D_R(p+q)_{fb} \,,
\end{align}
and $p = p_1 + p_2$. This vertex function is essentially a resummation of ladder diagrams contributing to $t$-channel $\psi \overline{\psi} \rightarrow \psi \overline{\psi}$ scattering via the exchange of a scalar propagator. On inspection, we can easily discern that $V_{\psi \psi}$ is finite by counting the number of propagators. The counterterm $\delta g_0$ hence accounts for a finite renormalization which we determine using the condition
\begin{equation}
    V_{\psi\psi}(p_{1\ast},p_{2\ast},p_{3\ast},p_{4\ast})_{ab,cd} = i g_R^2 \,\delta_{db}\,G_R(p_{3\ast}-p_{1\ast})\,\delta_{ac}\,.
\end{equation}
where we continue to work in the COM frame to give the renormalization conditions. Imposing this and appropriately contracting the spinor indices (with $\sum_{a}\delta_{aa} = 4$) we obtain the following expression for $\delta g_0$
\begin{equation}
    \delta g_0 = -g_R + \frac{g_R}{\sqrt{1 - \frac{1}{4}G^{-1}_R(p_{3\ast}-p_{1\ast})\,F_2}}
\end{equation}
where 
\begin{equation}
    F_2 = i \int_q G_R(q+p_{2 \ast})D_R(q)_{ae}  V_{\psi\psi}(q+ p_{\ast},-q,p_{3\ast},p_{4\ast})_{ef,ab}D_R(p_{\ast}+q)_{fb} \,.
\end{equation}
Note that $\delta g_0$ is a finite renormalization, i.e. $\mathcal{O}(1)$ and therefore, does not modify divergent structures. Thus, for convenience, we choose to set $\delta g_0 = 0$ from this point onward.

Let us consider the BSE for the kernel $\Lambda^{(4)}_{\psi \phi}$, 
\begin{align}
    &V^{(4)}_{\psi\phi} (p_1,p_2,p_3,p_4)_{ab} = \Lambda_{\psi \phi}(p_1,p_2,p_3,p_4)_{ab} \nonumber \\[2mm]
    &\qquad   +i \int_q \Lambda^{(4)}_{\psi \phi}(p_1,p_2,q+p,-q)_{ae}D_R(q)_{ef} V^{(4)}_{\psi\phi}(q+p,-q,p_3,p_4)_{f b}G_R(p+q)
\end{align}
This is again finite by power counting. We will use this now for the three-point kernel $\Lambda^{(3)}_{\psi \phi}$ to define a BSE of the form
\begin{align}
    &V^{(3)}_{\psi\phi} (p_1,p_2,p_3)_{ab} = \left(\Lambda^{(3)}_{\psi \phi}\right)_{ab} + i\, \left(\Lambda^{(3)}_{\psi \phi}\right)_{ac}\int_q D_R(q)_{cd} V^{(4)}_{\psi\phi}(q+p_1,-q,p_2,p_3)_{db}G_R(p_1+q)\,.
\end{align}
We will make use of this now to determine the counterterm $\delta g_1$. With the renormalization condition
\begin{equation}
	V^{(3)}_{\psi\phi}(p_{1\ast})_{(\alpha\beta)} = -g_R\, \delta_{\alpha\beta} \,
\end{equation}
we then obtain
\begin{equation}
	\delta g_1 = - g_R +  \frac{g_R}{1+\frac{1}{4}F_3}\,,
\end{equation}
where 
\begin{equation}
    F_3  =  i \int_q \text{tr}\left[D_R(q)V^{(4)}_{\psi\phi}(q+p_{1\ast},-q,p_{2\ast},p_{3\ast})\right]G_R(q+p_{1\ast})\,,
\end{equation}
and where appropriate contraction of the spinor indices leads to the resultant trace.

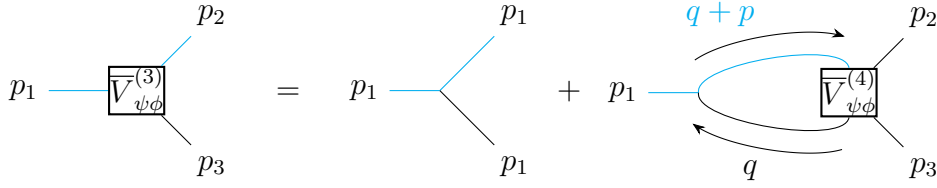
\begin{figure}[t]
\begin{equation*}
\vcenter{\hbox{\begin{tikzpicture}
	\begin{feynman}
	\vertex (v1) at (-1.5,0){\(p_1\)};
	\vertex (v2) at (1,1){\(p_2\)};
	\vertex (v3) at (1,-1){\(p_3\)};;
	\vertex[large, VR] (x) at (0,0) {$\oV^{(3)}_{\psi\phi}$};
	\diagram*{
	(v1) --[cyan](x),
	(x) --[cyan](v2),
	(x) --(v3),
	};
	\end{feynman}
\end{tikzpicture}}} 
	\quad = \quad
\vcenter{\hbox{\begin{tikzpicture}
	\begin{feynman}
	\vertex (v1) at (-1,0){\(p_1\)};
	\vertex (v2) at (1,1){\(p_1\)};
	\vertex (v3) at (1,-1){\(p_1\)};
	\vertex(x) at (0,0);
	\diagram*{
	(v1) --[cyan](x),
	(x) --[cyan](v2),
	(x) --(v3),
	};
	\end{feynman}
\end{tikzpicture}}} 
	\,\, + \,\,
	\vcenter{\hbox{\begin{tikzpicture}
	\begin{feynman}
	\vertex (v1) at (-2,0){\(p_1\)};
	\vertex (v2) at (2,1){\(p_2\)};
	\vertex (v3) at (2,-1){\(p_3\)};
	\vertex (x1) at (-1,0);
	\vertex[large, VR] (x2) at (1,0) {$\oV^{(4)}_{\psi\phi}$};
	\diagram*{
	(v1) -- [cyan](x1),
	(x1) -- [cyan, momentum =\(q+p\),half left, looseness = 0.5] (x2) -- [momentum =\(q\), half left, looseness = 0.5](x1),
	(x2) --  (v2),
	(x2) --  (v3)		
	};
	\end{feynman}
\end{tikzpicture}}}
\end{equation*} 
\caption{Illustration of BSE for the resummation of the three-point vertex $\oV^{(3)}_{\psi\phi}$ using the four-point vertex $\oV^{(4)}_{\psi\phi}$.}
\label{fig:3pt_fermions}
\end{figure}

In dealing with the scalar four-point vertex functions, we have to treat the possibility of divergences introduced by fermionic loops. To this end, with $\oL^{(4)}$ as the base, we build the following ``modified scalar kernel'' using $\Lambda^{(4)}_{\psi\phi}$ and the four-point vertex $V_{\psi\psi}$
\begin{align}
    &\tilde{\Lambda}_{\phi\phi} (p_1,p_2,p_3,p_4) = 
    \oL^{(4)}(p_1,p_2,p_3,p_4) \nonumber \\[2mm]
    &\qquad \qquad  - \ii \int_q D_R(p+q)_{db}\, \Lambda^{(4)}_{\psi\phi}(q+p,-q,p_3,p_4)_{ba}\,D_R(q)_{ac}\,\Lambda^{(4)}_{\psi\phi}(p_1,p_2,p+q-q)_{cd}  \nonumber\\[2mm]
    &\qquad \qquad  + \ii \int_q\int_r \bigg\{ D_R(q+p)_{fb}\,\Lambda^{(4)}_{\psi\phi}(p_1,p_2,p+q,-q)_{ba}\,D_R(q)_{ae}  \nonumber \\[2mm]
    & \qquad\qquad \qquad V^{\psi\psi}(p+q,-q,r+p,-r)_{ef,gh} D_R(r)_{gc}\,\Lambda^{(4)}_{\psi\phi}(r+p,-r,p_3,p_4)_{cd}\,D_R(r+p)_{dh} \bigg\}\nonumber \\[4mm]
    &= \oL^{(4)}(p_1,p_2,p_3,p_4) -4\ii g_R^4 \int_q \text{tr}\left[D_R(q+p)D_R(q+p_1)D_R(q)D_R(q+p_3)\right] \nonumber \\[2mm]
    & \qquad + 4\ii g_R^4 \int_q\int_r \text{tr}\bigg[D_R(q+p)D_R(q+p_1)D_R(q)V^{\psi\psi}(p+q,-q,r+p,-r) \nonumber \\
    & \qquad \qquad \qquad \qquad \qquad \qquad \qquad \qquad  D_R(r)D_R(r+p_3)D_R(r)\bigg]\,,
\end{align}
where in the second equality, we have contracted the spinor indices to obtain the trace. We can now iterate this four-point scalar kernel via its usual BSE
\begin{align}
    \overline{V}^{(4)}(p_1,p_2,p_3,p_4) &= \tilde{\Lambda}_{\phi\phi}(p_1,p_2,p_3,p_4) \nonumber \\[2mm]
    &\quad + \frac{\ii}{2}\int_q \tilde{\Lambda}_{\phi\phi}(p_1,p_2,q+p,-q)\, G_R(q) G_R(p+q)\,\overline{V}^{(4)}(q+p,-q,p_3,p_4) \,.
    \label{eqn:modvertex_fermions}
\end{align}
The quartic scalar coupling divergences generated will then be absorbed into the counterterm $\delta\lambda_0$. Imposing the same renormalization condition as in the scalar sunset approximation, c.f. \eqref{eq:v4alpha_rc}, we solve for the counterterm to obtain
\begin{align}
    \delta \lambda_0 &= -\lambda_R + \frac{\lambda_R -  (\alpha_R+\lambda_R\phi_R)^2\, I_2}{1+\frac{1}{2}I_1}  \nonumber \\[2mm]
    &\qquad \qquad +4 g_R^4 \left[\frac{F_4-\frac{1}{2}F_{4V}}{1+\frac{1}{2}I_1}\right] -4 g_R^4 \left[\frac{F_V-\frac{1}{2}F_{VV}}{1+\frac{1}{2}I_1}\right]\,,
\end{align}
where $I_{1,2}$ are the same integrals in \eqref{eqn:vertexintegral1} and \eqref{eqn:vertexintegral2}, defined at the renormalization point, and the following notation has been introduced for the new loop integrals pertaining to the fermions
\begin{equation}
    F_4 = \ii\int_q \text{tr}\left[D_R(q+p_{\ast})D_R(q+p_{2 \ast})D_R(q)D_R(q+p_{4 \ast})\right]\,,
\end{equation}
    
\begin{equation}
    F_{4V} = \int_q \int_r \text{tr}\left[D_R(q+p_{\ast})D_R(q+p_{2 \ast})D_R(q)D_R(q-r)\right]G_R(r)G_R(p_{\ast}+r)\overline{V}^{(4)}(r+p_{\ast},p_{3\ast})\,,
\end{equation}
    
\begin{align}
    F_{V} = \ii\int_q \int_r \text{tr}\bigg[D_R(q+p_{\ast})D_R(q+p_{2 \ast})D_R(q)&V_{\psi\psi}(p_{\ast}+q,r+p_{\ast}) \nonumber \\
    &D_R(r)D_R(r+p_{4 \ast})D_R(r+p_{\ast})\bigg]\,,
\end{align}
    
\begin{align}
    F_{VV} &= \int_k \int_q \int_r \bigg\{\text{tr}\bigg[D_R(q+p_{\ast})D_R(q+p_{2 \ast})D_R(q)V_{\psi\psi}(p_{\ast}+q,r+p_{\ast})\nonumber \\
& \quad \quad \quad D_R(r)D_R(r-k)D_R(r+p_{\ast})\bigg] 
 G_R(k)G_R(k+p_{\ast})\overline{V}^{(4)}(k+p_{\ast},p_{3 \ast})\bigg\}\,.
\end{align}  
Note that $F_V$ and $F_{VV}$ are both finite, and $F_4$ and $F_{4V}$ are logarithmically divergent. The remaining counterterms are $\delta \lambda_2$ and $\delta \alpha_1$ are obtained in the exact same manner as in the scalar sunset case, and the expressions are the same \eqref{eqn:dellambda2_tri} and \eqref{eqn:delalpha1}, with the vertex function now being defined with the modified four-point scalar kernel to include the contributions from fermions.

We can now turn to the formulation of the gap equations to determine the form of the fermionic and scalar propagators. These are given by
\begin{align}
    iG^{-1}_R(p) &= p^2 -m^2_R- \ii\overline{\Pi}(p) \nonumber \\[2mm]
    &= (p^2 - m^2_R) + (\delta Z_{\phi,0} \, p^2 - \delta m^2_0) - (\alpha_R+\delta\alpha_1)\phi_R  - \frac{(\lambda_R + \delta \lambda_2)}{2}  \phi_R^2 \nonumber \\[2mm]
    &\qquad  - \frac{(\lambda_R + \delta \lambda_0)}{2}\, \mathcal{T} + \frac{(\alpha_R + \lambda_R \phi_R)^2}{2}\, \mathcal{I}(p) - \ii g^2_R\,\int_q \text{tr}[D_R(q)D_R(p+q)] \nonumber \\[2mm]
    &\equiv (p^2 - m^2_R) + (\delta Z_{\phi,0} \, p^2 - \delta m^2_0) - (\alpha_R+\delta\alpha_1)\phi_R  - \frac{(\lambda_R + \delta \lambda_2)}{2}  \phi_R^2 \nonumber \\[2mm]
    &\qquad  - \frac{(\lambda_R + \delta \lambda_0)}{2}\, \mathcal{T} + \frac{(\alpha_R + \lambda_R \phi_R)^2}{2}\, \mathcal{I}(p) - g^2_R\left[p^2 \mathcal{F}_1(p) + \mathcal{F}_2(p)\right]
    \label{eqn:scalargap_os}
\end{align}

\begin{align}
    iD^{-1}_R(p) &= \myslash{p} - M_R-\ii\overline{\Sigma}(\myslash{p}) \nonumber \\[2mm]
    &= \myslash{p} - M_R + (\delta Z_{\psi,0}\,\myslash{p} -\delta M_0) -  (g_R+\delta g_1)\phi_R - \ii g_R^2\int_q D_R(p+q)G_R(q) \nonumber \\[2mm]
    &\equiv \myslash{p} - M_R + (\delta Z_{\psi,0}\,\myslash{p} -\delta M_0)  -  (g_R+\delta g_1)\phi_R - g_R^2 \left[X(p) \myslash{p} + Y(p)\right] \,,
    \label{eqn:fermiongap_os}
\end{align}
and are evidently coupled. For the loop integrals related to fermions, we have decomposed them into forms based on possible Lorentz structures. We now impose the appropriate on-shell renormalization conditions to obtain the counterterms related to the propagators. For the scalar propagator, we have
\begin{equation}
    \delta Z_{\phi,0} =  -\frac{(\alpha_R+\lambda_R \phi_R)^2}{2}\,\frac{\partial \mathcal{I}(p)}{\partial p^2}\bigg|_{p^2=m^2_R} + g_R^2 \left[\mathcal{F}_1(p) + m^2_R\frac{\partial \mathcal{F}_1(p)}{\partial p^2}  + \frac{\partial \mathcal{F}_2(p)}{\partial p^2}\right]\bigg|_{p^2=m^2_R} \,,
    \label{eqn:delZ_scalar}
\end{equation}
\begin{align}
    \delta m^2_0 &=  \delta Z_{\phi,0} m^2_R - (\alpha_R+\delta\alpha_1)\phi_R - \frac{(\lambda_R + \delta \lambda_2)}{2} \phi_R^2  -\frac{(\lambda_R+\delta \lambda_0)}{2}\, \mathcal{T}  \nonumber \\[2mm]
    &\qquad + \frac{(\alpha_R +\lambda_R \phi_R)^2}{2}\,\mathcal{I}(p)\big|_{p^2=m^2_R} - g_R^2 \left[m^2_R \mathcal{F}_1(p)  + \mathcal{F}_2(p)\right]\big|_{p^2=m^2_R}\,,
\end{align}
and for the fermionic propagator
\begin{equation}
    \delta Z_{\psi,0} = g_R^2 X(p)\big|_{p^2 = M^2_R}  + 2g_R^2\, M_R\,\left[M_R \frac{\partial X(p)}{\partial p^2} + \frac{\partial Y(p)}{\partial p^2}\right] \bigg|_{p^2=M^2_R}\,,
    \label{eqn:delZ_ferm}
\end{equation}

\begin{align}
    \delta M_0 &=  \delta Z_{\psi,0} M_R - (g_R+\delta g_1)\phi_R - g_R^2 [M_R \,X(p) + Y(p)]\big|_{p^2 = M^2_R}   \,\,.
\end{align}
Note that the counterterms for the scalar and fermionic wave function renormalizations are not finite, due to the loop integrals not always being subtracted. 

\begin{figure}[h!]
\centering
\includegraphics[width=1.0\textwidth, keepaspectratio]{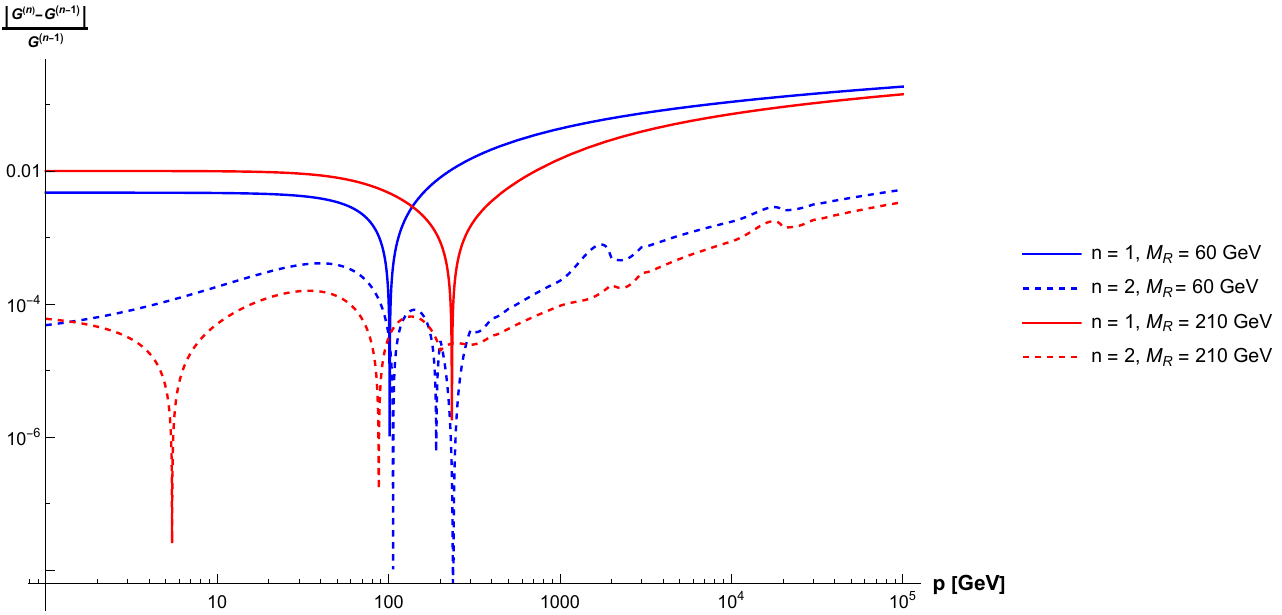}
\caption{The relative difference between the two 
successive iterations of the scalar propagator 
for two choices of the renormalised fermionic mass, $M_R$. 
We have set $g_R = 2$, $\lambda_R = 0.5$,  \mbox{$\alpha_R = 50$~GeV}, $\phi_R \approx 227.5$~GeV and $m_R = 100 \,\text{GeV}$ in all cases. The UV cutoff was taken to be \mbox{$\Lambda = 10^5\,\text{GeV}$}.}
\label{fig:scalardiffyukawa}
\end{figure}

\begin{figure}[h!]
\centering
\includegraphics[width=1.0\textwidth, keepaspectratio]{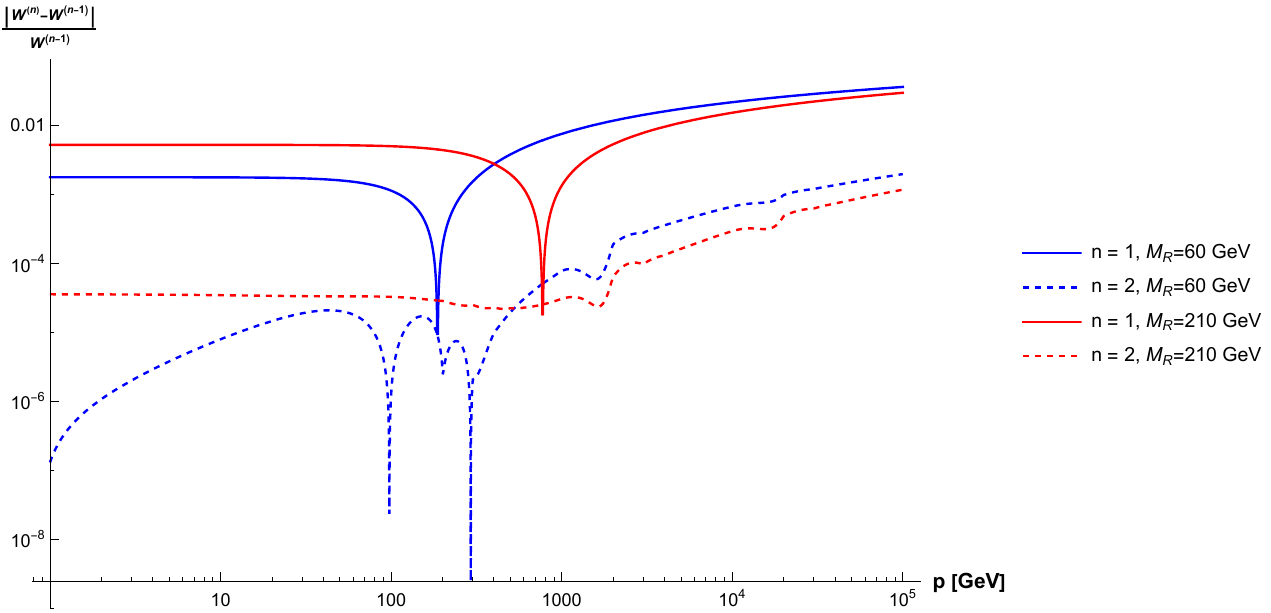}
\caption{The relative difference between the two successive iterations of the function $W(p)$ for two choices of the renormalised fermionic mass, $M_R$. All parameters are set as in Fig. \ref{fig:scalardiffyukawa}.}
\label{fig:fermionwdiff}
\end{figure}

\begin{figure}[h!]
\centering
\includegraphics[width=1.0\textwidth, keepaspectratio]{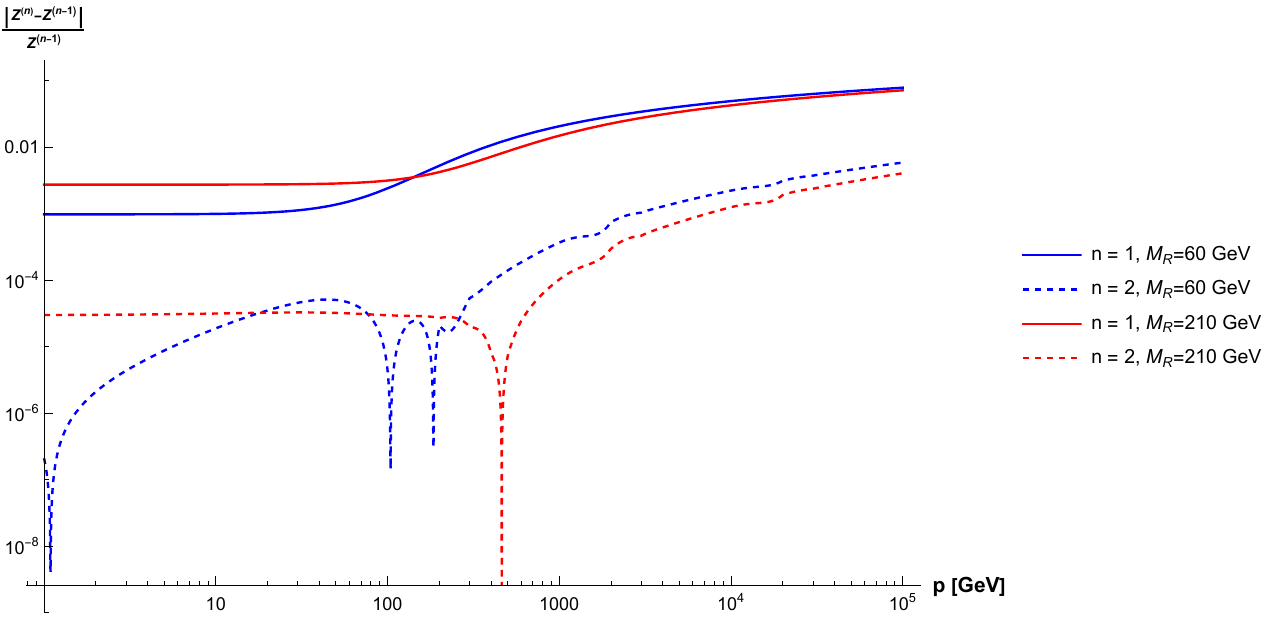}
\caption{The relative difference between the two successive iterations of the function $Z(p)$ for two choices of the renormalised fermionic mass, $M_R$. All parameters are set as in Fig. \ref{fig:scalardiffyukawa}.}
\label{fig:fermionzdiff}
\end{figure}

We proceed now in the same manner as in the scalar sunset approximation: we substitute these counterterms into the gap equations to obtain the following coupled integral equations to solve for the propagators
\begin{align}
    &iG^{-1}_R(p) = (p^2-m^2_R)\left[1 -\frac{(\alpha_R +\lambda_R \phi_R)^2}{2}\, \frac{\partial \mathcal{I}(q)}{\partial q^2} +g^2_R\left(m^2_R \frac{\partial \mathcal{F}_1(q)}{\partial q^2}  +\frac{\partial \mathcal{F}_1(q)}{\partial q^2}\right)\right]\bigg|_{q^2=m^2_R} \nonumber \\[2mm] 
    & \qquad +\frac{(\alpha_R +\lambda_R \phi_R)^2}{2}\left[\mathcal{I}(p) - \mathcal{I}(q)\big|_{q^2=m^2_R}\right] \nonumber \\[2mm] 
    & \qquad - g^2_R \left[p^2\left(\mathcal{F}_1(p) - \mathcal{F}_1(q)\big|_{q^2=m^2_R}\right) + \left(\mathcal{F}_2(p) - \mathcal{F}_2(q)\big|_{q^2=m^2_R}\right)\right]\,,
    \label{scalargapeqn}
\end{align}

\begin{align}
    iD^{-1}_R(p) &= (\myslash{p}-M_R)\left[1+2g_R^2 M_R\,\left(M_R \frac{\partial X(q)}{\partial q^2} + \frac{\partial Y(q)}{\partial q^2}\right)\right]\bigg|_{q^2=M^2_R} \nonumber \\[2mm]
    &\qquad  - g^2_R \left[\myslash{p} \left(X(p)-X(q)\big|_{q^2=M^2_R}\right) + \left(Y(p^2)-Y(q)\big|_{q^2=M^2_R}\right)\right] \nonumber \\[2mm]
    &\equiv W(p)\myslash{p} - Z(p) \,\,.
    \label{fermiongapeqn}
\end{align}
These are manifestly finite as divergences drop out due to the subtraction from a fixed point or differentiation of the divergent functions. In the last step for the fermionic propagator, we have defined the following quantities 
\begin{align}
    &W(p) = 1 - g^2_R \left(X(p)-X(q)\big|_{q^2=M^2_R}\right) + 2 g^2_R M_R \left(M_R \frac{\partial X(q)}{\partial q^2} + \frac{\partial Y(q)}{\partial q^2}\right)\bigg|_{q^2=M^2_R} \,, \\[3mm]
    &Z(p) = M_R -   g^2_R \left(Y(p)-Y(q)\big|_{q^2=M^2_R}\right) -2 g^2_R M^2_R \left(M_R \frac{\partial X(q)}{\partial q^2} + \frac{\partial Y(q)}{\partial q^2}\right)\bigg|_{q^2=M^2_R} \,,
\end{align}
according to which we can explicitly write down the expressions for $X(p)$ and $Y(p)$,
\begin{equation}
	X(p) = i\int_q W(p+q) G_R(q) \,,\qquad Y(p) = i \int_q Z(p+q) G_R(q)\,.
\end{equation}
For the trace that appears in the scalar propagator, we resolve this as
\begin{align}
    &\text{tr}[D_R(q)D_R(p+q)] = \text{tr}\left\{\frac{i}{W(q)\myslash{q} - Z(q)}\, \frac{i}{W(p+q)(\myslash{p}+\myslash{q}) - Z(p+q)} \right\} \nonumber \\[2mm]
    &= \frac{\text{tr}\left\{[W(q)\myslash{q} + Z(q)][W(p+q)(\myslash{p}+\myslash{q}) + Z(p+q)] \right\}}{[W^2(q)q^2 - Z^2(q)]\,[W^2(p+q)(p+q)^2 - Z^2(p+q)]} \nonumber \\[2mm]
    &= 4 \,\frac{(p\cdot q + q^2)W(q)W(p+q) + Z(q)Z(p+q)}{[W^2(q)q^2 - Z^2(q)]\,[W^2(p+q)(p+q)^2 - Z^2(p+q)]} \nonumber \\[2mm]
    &= p^2\left\{\frac{-2\,W(q) W(p+q)}{[W^2(q)q^2 -Z^2(q)][W^2(p+q)(p+q)^2 -Z^2(p+q)]}\right\}\nonumber\\[2mm] 
    &\qquad + 2\left\{\frac{((p+q)^2 -q^2)W(q) W(p+q)+ Z(q) Z(p+q)}{[W^2(q)q^2 -Z^2(q)][W^2(p+q)(p+q)^2 -Z^2(p+q)]}\right\} 
\end{align}
where in the second last step we have used $ 2 p\cdot q = (p+q)^2 - p^2 - q^2$. This finally gives the expressions for the functions $\mathcal{F}_1$ and $\mathcal{F}_2$
\begin{align}
	&\mathcal{F}_1(p) = - 2\ii \int_q \frac{W(q) W(p+q)}{[W^2(q)q^2 -Z^2(q)][W^2(p+q)(p+q)^2 -Z^2(p+q)]}\,, \\[2mm]
	&\mathcal{F}_2(p) =  2\ii \int_q \frac{((p+q)^2 -q^2)W(q) W(p+q)+ Z(q) Z(p+q)}{[W^2(q)q^2 -Z^2(q)][W^2(p+q)(p+q)^2 -Z^2(p+q)]}\,.
\end{align}
We now have the entire setup to solve the gap equations, which we approach in the same iterative manner outlined for the scalar sunset approximation, beginning with the free propagators. This yields
\begin{align}
    &X^{(1)}(p^2) =  \frac{1}{16\pi^2}\left[B_1(p^2,m^2_R,M^2_R)+B_0(p^2,m^2_R,M^2_R)\right] \,,\\[4mm]
    &Y^{(1)}(p^2) = \frac{1}{16\pi^2}\left[M_R \, B_0(p^2,m^2_R,M^2_R)\right] \,, \\[4mm]
    &i(G^{-1}_R(p))^{(1)} = p^2 \bigg\{1 -\frac{g_R^2}{8\pi^2}\left[B_0(p^2,M^2_R,M^2_R)-B_0(m^2_R,M^2_R,M^2_R)\right] \nonumber \\[2mm]
    & \qquad \qquad - \frac{g_R^2}{4\pi^2}\left(M^2_R-\frac{m^2_R}{2}\right)\dot{B}_0(m^2_R,M^2_R,M^2_R) -\frac{(\alpha_R +\lambda_R \phi_R)^2}{32\pi^2}\, \dot{B}_0(m^2_R,m^2_R,m^2_R)\bigg\} \nonumber \\[2mm]
    & -m^2_R \bigg\{1- \frac{g_R^2}{8\pi^2}\frac{M^2_R}{m^2_R}\left[B_0(p^2,M^2_R,M^2_R)-B_0(m^2_R,M^2_R,M^2_R)\right]
    \nonumber \\[2mm]
    & \qquad \qquad -\frac{g_R^2}{4\pi^2}\left(M^2_R-\frac{m^2_R}{2}\right)\dot{B}_0(m^2_R,M^2_R,M^2_R) +\frac{(\alpha_R +\lambda_R \phi_R)^2}{32\pi^2}\, \dot{B}_0(m^2_R,m^2_R,m^2_R)\bigg\} \nonumber \\[2mm]
    &\qquad \qquad + \frac{(\alpha_R +\lambda_R \phi_R)^2}{32\pi^2}\left[B_0(p^2,m^2_R,m^2_R)-B_0(m^2_R,m^2_R,m^2_R)\right]\,,
\end{align}
which are converted to Euclidean space for the next iteration. We focus on the case of a large Yukawa coupling, $g_R = 2$, and set smaller couplings pertaining to scalars. We again see that the relative difference between successive iterations drops by about two orders of magnitude for the scalar propagator (Fig. \ref{fig:scalardiffyukawa}) and the functions $W(p)$ (Fig. \ref{fig:fermionwdiff}) and $Z(p)$ (Fig. \ref{fig:fermionzdiff}), demonstrating the convergence of this approach. The spikes correspond to cases where the corresponding difference vanishes.

We now outline the procedure to obtain the remaining counterterms, which pertain to the scalar field. We must start with the generic expression involving field derivatives of the effective action and include the relevant modifications due to the presence of fermions. 

Starting with the scalar one-point function, we have 
\begin{align}
    \frac{\delta\Gamma^{\text{2PI}}}{\delta\phi_R(x_1)} = &\frac{\delta \Gamma^{\text{2PI}}}{\delta \phi_R(x_1)}\bigg|_{\phi_R,\,G_R,\,D_R}  + \int_{y_1,y_2}\frac{\delta\Gamma^{\text{2PI}}}{\delta G_R(y_1,y_2)}\bigg|_{\phi_R,\,G_R,\,D_R} \frac{\delta G_R(y_1,y_2)}{\delta\phi_R(x_1)}  \nonumber \\
    &+   \int_{y_1,y_2}\text{tr}\bigg\{\frac{\delta\Gamma^{\text{2PI}}}{\delta D_R(y_1,y_2)}\bigg|_{\phi_R,\,G_R,\,D_R} \frac{\delta D_R(y_1,y_2)}{\delta\phi_R(x_1)}\bigg\} \nonumber 
\end{align}
where the second and third terms are obtained from the chain rule. The stationarity conditions of $\Gamma^{\text{2PI}}$,
\begin{equation}
	\frac{\delta\Gamma^{\text{2PI}}}{\delta G_R}\bigg|_{\phi_R,\,G_R,\,D_R} = \frac{\delta\Gamma_{\text{2PI}}}{\delta D_R}\bigg|_{\phi_R,\,G_R,\,D_R} \stackrel{!}{=} 0 \,,
\end{equation}
cause these to drop out. We thus obtain, by converting to momentum space, 
\begin{align}
&    \Gamma^{(1)} = \frac{\delta \Gamma^{\text{2PI}}}{\delta \phi_R}\bigg|_{p^2 =0} = -\delta t_1 - (m^2_R +\delta m^2_2) \phi_R - \frac{(\alpha_R + \delta \alpha_3)}{2}\phi^2_R - \frac{(\lambda_R + \delta \lambda_4)}{6}\phi^3_R \nonumber \\[2mm]
    &\quad - \frac{1}{2}\left[(\alpha_R+\delta \alpha_1) + (\lambda_R + \delta \lambda_2)\phi_R\right]\,\mathcal{T} + \frac{\lambda_R \left(\alpha_R + \lambda_R \phi_R\right)}{6} \,\mathcal{S} -(g_R+\delta g_1)\int_q \text{tr}\left[D_R(q)\right] \stackrel{!}{=} 0   \,\,,
    \label{eqn:gamma1_mincond}
\end{align}
which is the same minimisation condition as in the scalar sunset, but with a new contribution from the fermionic tadpole, which is the last term in the second line. 

Consider the physical two-point function of the scalar field,
\begin{align}
    &\Gamma^{(2)}(x_1,x_2) \equiv \frac{\delta^2\Gamma^{\text{2PI}}}{\delta\phi(x_1)\delta\phi(x_2)} \nonumber\\[2mm]
    &= \frac{\delta^2 \Gamma^{\text{2PI}}}{\delta \phi_R(x_1)\delta \phi_R(x_2)}\bigg|_{\phi_R,\,G_R,\,D_R} + \int_{y_1,...,y_4}\frac{\delta^2\Gamma^{\text{2PI}}}{\delta\phi_R(x_1)\delta G_R(y_1,y_2)}\bigg|_{\phi_R,\,G_R,\,D_R}\frac{\delta G_R(y_1,y_2)}{\delta\phi_R(x_2)} \nonumber \\[3mm]
    &\qquad \qquad \qquad \qquad \qquad \qquad  +   \int_{y_1,...,y_4}\text{tr}\bigg\{\frac{\delta^2\Gamma^{\text{2PI}}}{\delta \phi(x_1)\delta D_R(y_1,y_2)}\bigg|_{\phi_R,\,G_R,\,D_R}\frac{\delta D_R(y_1,y_2)}{\delta\phi_R(x_2)}\bigg\} \nonumber  \\[4mm]
    &= iG^{-1}_{0,R} + \frac{\delta^2\Gamma^{\text{2PI}}_{\text{int}}}{\delta\phi_R(x_1)\delta\phi_R(x_2)}\bigg|_{\phi_R,\,G_R,\,D_R} \nonumber \\[3mm]
    &\qquad \qquad + \int_{y_1,...,y_4}\frac{\delta^2\Gamma^{\text{2PI}}_{\text{int}}}{\delta\phi_R(x_1)\delta G_R(y_1,y_2)}\bigg|_{\phi_R,\,G_R,\,D_R}G_R(y_1,y_3)\frac{\delta\overline{\Pi}(y_3,y_4)}{\delta\phi_R(x_2)}G_R(y_4,y_2) \nonumber \\[3mm]
    &\qquad \qquad +\int_{y_1,...,y_4}\text{tr}\bigg\{\frac{\delta^2\Gamma^{\text{2PI}}_{\text{int}}}{\delta \phi_R(x_1)\delta D_R(y_1,y_2)}\bigg|_{\phi_R,\,G_R,\,D_R}D_R(y_1,y_3)\frac{\delta \overline{\Sigma}(y_3,y_4)}{\delta\phi_R(x_2)}D_R(y_4,y_2)\bigg\}  \,.
    \label{eqn:scal2pt_ferm}
\end{align}
The last line gives explicit fermionic contributions by virtue of $\delta \overline{\Sigma}/\delta \phi_R$. In a similar manner to \eqref{eq:dPidphi}, it can be expanded out using the following diagrammatic equation 
\begin{equation}
\frac{\delta \overline{\Sigma}(x_1,x_2)}{\delta \phi_R(x_3)} \equiv
\begin{tikzpicture}[scale=2,baseline={([yshift=-.4ex]current bounding box.center)}]
\node (threepoint) at (3,0) [Sigma] {}; 
\coordinate (tra) at (2.6,0);
\draw[double distance=4mm,thick,cyan] (threepoint) -- (tra) node[right,above]{\hspace*{-2mm}1\,}
node[right,below]{\hspace*{-2mm}2\,};
\node (l3) at (3.4,0) [VEV] {}; 
\draw[-] (threepoint) -- (l3) node[right] {3} ;
 \end{tikzpicture}
=
 \begin{tikzpicture}[scale=2,baseline={([yshift=-.4ex]current bounding box.center)}]
\node (threepoint) at (3,0) [Gamma] {}; 
\coordinate (tra) at (2.6,0);
\draw[double distance=4mm,thick,cyan] (threepoint) -- (tra) node[right,above]{\hspace*{-2mm}1\,}
node[right,below]{\hspace*{-2mm}2\,};
\node (l3) at (3.4,0) [VEV] {}; 
\draw[-] (threepoint) -- (l3) node[right] {3} ;
 \end{tikzpicture}
+ \quad
\begin{tikzpicture}[scale=2,baseline={([yshift=-.4ex]current bounding box.center)}]
\node (fourpoint) at (3,0) [VR] {\,$V_{\psi\psi}$\,}; 
\node (threepoint) at (3.5,0) [Gamma] {}; 
\coordinate (tra) at (2.6,0);
\draw[double distance=4mm,thick,cyan] (fourpoint) -- (tra) node[right,above]{\hspace*{-2mm}1\,}
node[right,below]{\hspace*{-2mm}2\,};
\node (l3) at (4,0) [VEV] {}; 
\draw[-] (threepoint) -- (l3) node[right] {3} ;
\draw[double distance=4mm,thick,cyan] (fourpoint) -- (threepoint);\,,
\end{tikzpicture}
+ \quad
\begin{tikzpicture}[scale=2,baseline={([yshift=-.4ex]current bounding box.center)}]
\node (fourpoint) at (3,0) [VR] {\,$V^{(4)}_{\psi\phi}$\,}; 
\node (threepoint) at (3.5,0) [Gamma] {}; 
\coordinate (tra) at (2.6,0);
\draw[double distance=4mm,thick,cyan] (fourpoint) -- (tra) node[right,above]{\hspace*{-2mm}1\,}
node[right,below]{\hspace*{-2mm}2\,};
\node (l3) at (4,0) [VEV] {}; 
\draw[-] (threepoint) -- (l3) node[right] {3} ;
\draw[double distance=4mm,thick] (fourpoint) -- (threepoint);\,,
\end{tikzpicture}
\end{equation}
where we have the new building blocks involving fermions. cyan lines indicate fermionic propagators. Note the appearance of the vertex functions $V_{\psi\psi}$ and $V_{\psi\phi}$ now, which appear as a solution to the self-energy equation of $\delta \overline{\Sigma}/\delta \phi_R$, in the similar manner how $\overline{V}^{(4)}$ appeared as solution to $\delta \overline{\Pi}/\delta \phi_R$. We stress that the $\oV^{(4)}$ used is \eqref{eqn:modvertex_fermions} with the modified four-point scalar kernel. Furthermore, note that correspondingly $\delta \overline{\Pi}/\delta \phi_R$ is modified from \eqref{eq:dPidphi} and gains the additional part
\begin{equation}
\begin{tikzpicture}[scale=2,baseline={([yshift=-.4ex]current bounding box.center)}]
\node (threepoint) at (3,0) [Pi] {}; 
\coordinate (tra) at (2.6,0);
\draw[double distance=4mm,thick] (threepoint) -- (tra) node[right,above]{\hspace*{-2mm}1\,}
node[right,below]{\hspace*{-2mm}2\,};
\node (l3) at (3.4,0) [VEV] {}; 
\draw[-] (threepoint) -- (l3) node[right] {3} ;
 \end{tikzpicture}
=
 \begin{tikzpicture}[scale=2,baseline={([yshift=-.4ex]current bounding box.center)}]
\node (threepoint) at (3,0) [Gamma] {}; 
\coordinate (tra) at (2.6,0);
\draw[double distance=4mm,thick] (threepoint) -- (tra) node[right,above]{\hspace*{-2mm}1\,}
node[right,below]{\hspace*{-2mm}2\,};
\node (l3) at (3.4,0) [VEV] {}; 
\draw[-] (threepoint) -- (l3) node[right] {3} ;
 \end{tikzpicture}
+ \quad
\begin{tikzpicture}[scale=2,baseline={([yshift=-.4ex]current bounding box.center)}]
\node (fourpoint) at (3,0) [VR] {\,$\oV^{(4)}$\,}; 
\node (threepoint) at (3.5,0) [Gamma] {}; 
\coordinate (tra) at (2.6,0);
\draw[double distance=4mm,thick] (fourpoint) -- (tra) node[right,above]{\hspace*{-2mm}1\,}
node[right,below]{\hspace*{-2mm}2\,};
\node (l3) at (4,0) [VEV] {}; 
\draw[-] (threepoint) -- (l3) node[right] {3} ;
\draw[double distance=4mm,thick] (fourpoint) -- (threepoint);\,,
\end{tikzpicture}
+ \quad
\begin{tikzpicture}[scale=2,baseline={([yshift=-.4ex]current bounding box.center)}]
\node (fourpoint) at (3,0) [VR] {\,$V^{(4)}_{\phi\psi}$\,}; 
\node (threepoint) at (3.5,0) [Gamma] {}; 
\coordinate (tra) at (2.6,0);
\draw[double distance=4mm,thick] (fourpoint) -- (tra) node[right,above]{\hspace*{-2mm}1\,}
node[right,below]{\hspace*{-2mm}2\,};
\node (l3) at (4,0) [VEV] {}; 
\draw[-] (threepoint) -- (l3) node[right] {3} ;
\draw[double distance=4mm,thick,cyan] (fourpoint) -- (threepoint);\,,
\end{tikzpicture}
\end{equation}
where $V^{(4)}_{\phi\psi}$ is the transpose of the vertex function $V^{(4)}_{\psi\phi}$. We continue now with the diagrammatic analysis and look at the 3-point and 4-point functions, required to obtain the counterterms $\delta \alpha_3$ and $\delta \lambda_4$. Besides the scalar contributions in \eqref{eq:Gamma3_alpha} and \eqref{eq:Gamma4_alpha}, we have in addition the following ones from the fermions
\begin{align}
\Gamma^{(3)}\bigg|_{\text{fermions}} ~=~ 
\begin{tikzpicture}[scale=2,baseline={([yshift=-.4ex]current bounding box.center)}]
\node (v1) at (0,0.3) [Gamma] {}; 
\node (v2) at (0,-0.3) [Sigma] {}; 
\node (v3) at (0.6,0.3) [Sigma] {}; 
\node (l1) at (-0.4,0.3) [VEV] {}; 
\node (l2) at (0.9,0.3) [VEV] {}; 
\node (l3) at (0.,-0.7) [VEV] {}; 
\draw[-,cyan] (v1) -- (v2);
\draw[-,cyan] (v1) -- (v3);
\draw[-,cyan] (v2) -- (v3);
\draw[-] (v1) -- (l1);
\draw[-] (v3) -- (l2);
\draw[-] (v2) -- (l3);
\end{tikzpicture}
+ \,\,
 \begin{tikzpicture}[scale=2,baseline={([yshift=-.4ex]current bounding box.center)}]
\node (v1) at (0,0.3) [Gamma] {}; 
\node (v2) at (0.6,0.3) [Sigma] {}; 
\node (l1) at (-0.4,0.3) [VEV] {}; 
\node (l2) at (0.9,0.3) [VEV] {}; 
\node (l3) at (0.6,-0.1) [VEV] {}; 
\draw[double distance=4mm,thick,cyan] (v1) -- (v2);
\draw[-] (v1) -- (l1);
\draw[-] (v2) -- (l2);
\draw[-] (v2) -- (l3);
\end{tikzpicture} \,\,\,\,,
\label{eq:Gamma3_fermions}
\end{align}

\begin{align}
\Gamma^{(4)}\bigg|_{\text{fermions}} ~=~ 
 \begin{tikzpicture}[scale=2,baseline={([yshift=-.4ex]current bounding box.center)}]
\node (v1) at (0,0.3) [Gamma] {}; 
\node (v2) at (0,-0.3) [Sigma] {}; 
\node (v3) at (0.6,0.3) [Sigma] {}; 
\node (v4) at (0.0,0.9) [Sigma] {}; 
\node (l1) at (-0.4,0.3) [VEV] {}; 
\node (l2) at (0.,1.3) [VEV] {}; 
\node (l3) at (0.9,0.3) [VEV] {}; 
\node (l4) at (0.,-0.7) [VEV] {}; 
\draw[-,cyan] (v1) -- (v2);
\draw[-,cyan] (v4) -- (v3);
\draw[-,cyan] (v2) -- (v3);
\draw[-,cyan] (v1) -- (v4);
\draw[-] (v1) -- (l1);
\draw[-] (v4) -- (l2);
\draw[-] (v3) -- (l3);
\draw[-] (v2) -- (l4);
\end{tikzpicture}
+
 \begin{tikzpicture}[scale=2,baseline={([yshift=-.4ex]current bounding box.center)}]
\node (v1) at (0,0.3) [Gamma] {}; 
\node (v3) at (0.6,0.3) [Sigma] {}; 
\node (v4) at (0.0,0.9) [Sigma] {}; 
\node (l1) at (-0.4,0.3) [VEV] {}; 
\node (l2) at (0.,1.3) [VEV] {}; 
\node (l3) at (0.9,0.3) [VEV] {}; 
\node (l4) at (0.6,-0.1) [VEV] {}; 
\draw[-,cyan] (v1) -- (v3);
\draw[-,cyan] (v4) -- (v3);
\draw[-,cyan] (v1) -- (v4);
\draw[-] (v1) -- (l1);
\draw[-] (v4) -- (l2);
\draw[-] (v3) -- (l3);
\draw[-] (v3) -- (l4);
\end{tikzpicture}
+
 \begin{tikzpicture}[scale=2,baseline={([yshift=-.4ex]current bounding box.center)}]
\node (v1) at (0,0.3) [Gamma] {}; 
\node (v2) at (0.6,0.3) [Sigma] {}; 
\node (l1) at (-0.4,0.3) [VEV] {}; 
\node (l2) at (0.6,0.7) [VEV] {}; 
\node (l3) at (0.9,0.3) [VEV] {}; 
\node (l4) at (0.6,-0.1) [VEV] {}; 
\draw[double distance=4mm,thick,cyan] (v1) -- (v2);
\draw[-] (v1) -- (l1);
\draw[-] (v2) -- (l2);
\draw[-] (v2) -- (l3);
\draw[-] (v2) -- (l4);
\end{tikzpicture} \,\,\,.
\label{eq:Gamma4_fermions}
\end{align}
A factor of $\frac{1}{2}$ does not appear for the fermionic self-energy insertions as this quantity is not defined with a factor of 2 (compare \eqref{eq:def_Pi} and \eqref{eq:def_Sigma}). Now, we list the non-vanishing fermionic contributions to the derivatives of the scalar self-energy which would be inserted in to the above equations
\begin{align}
 \begin{tikzpicture}[scale=2,baseline={([yshift=-.4ex]current bounding box.center)}]
\node (v1) at (0.0,0.) [Pi] {}; 
\node (l1) at (-0.4,0) {}; 
\node (l2) at (0.4,0.) [VEV] {}; 
\node (l3) at (0,-0.4) [VEV] {}; 
\draw[double distance=4mm,thick] (l1) -- (v1);
\draw[-] (v1) -- (l2);
\draw[-] (v1) -- (l3);
\end{tikzpicture}
\Bigg|_{\text{fermions}} ~=~ &  
+  \,
 \begin{tikzpicture}[scale=2,baseline={([yshift=-.4ex]current bounding box.center)}]
\node (v1) at (0.0,0.) [Gamma] {}; 
\node (v3) at (0.0,-0.6) [Sigma] {}; 
\node (v4) at (0.6,0.) [Sigma] {}; 
\node (l1) at (-0.4,0) {}; 
\node (l2) at (1.,0.) [VEV] {}; 
\node (l3) at (0,-1.0) [VEV] {}; 
\draw[double distance=4mm,thick] (l1) -- (v1);
\draw[-,cyan] (v1) -- (v3);
\draw[-,cyan] (v1) -- (v4);
\draw[-,cyan] (v3) -- (v4);
\draw[-] (v4) -- (l2);
\draw[-] (v3) -- (l3);
\end{tikzpicture}
+ \,
 \begin{tikzpicture}[scale=2,baseline={([yshift=-.4ex]current bounding box.center)}]
\node (v1) at (0.0,0.) [Gamma] {}; 
\node (v2) at (-0.6,0.) [VR] {\,$V^{(4)}_{\phi\psi}$\,}; 
\node (v3) at (0.0,-0.6) [Sigma] {}; 
\node (v4) at (0.6,0.) [Sigma] {}; 
\node (l1) at (-1.,0) {}; 
\node (l2) at (1.,0.) [VEV] {}; 
\node (l3) at (0,-1.0) [VEV] {}; 
\draw[double distance=4mm,thick] (l1) -- (v2);
\draw[double distance=4mm,thick,cyan] (v1) -- (v2);
\draw[-,cyan] (v1) -- (v3);
\draw[-,cyan] (v1) -- (v4);
\draw[-,cyan] (v3) -- (v4);
\draw[-] (v4) -- (l2);
\draw[-] (v3) -- (l3);
\end{tikzpicture}
\label{eq:d2Pi_dphi2_fermions}
\end{align}

\begin{align}
 \begin{tikzpicture}[scale=2,baseline={([yshift=-.4ex]current bounding box.center)}]
\node (v1) at (0.0,0.) [Pi] {}; 
\node (l1) at (-0.4,0) {}; 
\node (l2) at (0.4,0.) [VEV] {}; 
\node (l3) at (0,-0.4) [VEV] {}; 
\node (l4) at (0,0.4) [VEV] {}; 
\draw[double distance=4mm,thick] (l1) -- (v1);
\draw[-] (v1) -- (l2);
\draw[-] (v1) -- (l3);
\draw[-] (v1) -- (l4);
\end{tikzpicture}\Bigg|_{\text{fermions}}
~=~ & 
 \begin{tikzpicture}[scale=2,baseline={([yshift=-.4ex]current bounding box.center)}]
\node (v1) at (0.0,0.) [Gamma] {}; 
\node (v3) at (0.0,-0.6) [Sigma] {}; 
\node (v4) at (0.6,0.) [Sigma] {}; 
\node (v5) at (0.,0.6) [Sigma] {}; 
\node (l1) at (-0.4,0) {}; 
\node (l2) at (1.,0.) [VEV] {}; 
\node (l3) at (0,-1.0) [VEV] {}; 
\node (l4) at (0,1.) [VEV] {}; 
\draw[double distance=4mm,thick] (l1) -- (v1);
\draw[-,cyan] (v1) -- (v5);
\draw[-,cyan] (v1) -- (v3);
\draw[-,cyan] (v5) -- (v4);
\draw[-,cyan] (v3) -- (v4);
\draw[-] (v4) -- (l2);
\draw[-] (v3) -- (l3);
\draw[-] (v5) -- (l4);
\end{tikzpicture}
+
 \begin{tikzpicture}[scale=2,baseline={([yshift=-.4ex]current bounding box.center)}]
\node (v1) at (0.0,0.) [Gamma] {}; 
\node (v2) at (-0.6,0.) [VR] {\,$V^{(4)}_{\phi\psi}$\,}; 
\node (v3) at (0.0,-0.6) [Sigma] {}; 
\node (v4) at (0.6,0.) [Sigma] {}; 
\node (v5) at (0.,0.6) [Sigma] {}; 
\node (l1) at (-1.,0) {}; 
\node (l2) at (1.,0.) [VEV] {}; 
\node (l3) at (0,-1.0) [VEV] {}; 
\node (l4) at (0,1.) [VEV] {}; 
\draw[double distance=4mm,thick] (l1) -- (v2);
\draw[double distance=4mm,thick,cyan] (v1) -- (v2);
\draw[-,cyan] (v1) -- (v5);
\draw[-,cyan] (v1) -- (v3);
\draw[-,cyan] (v5) -- (v4);
\draw[-,cyan] (v3) -- (v4);
\draw[-] (v4) -- (l2);
\draw[-] (v3) -- (l3);
\draw[-] (v5) -- (l4);
\end{tikzpicture}
\nonumber \\ & 
+  \,
 \begin{tikzpicture}[scale=2,baseline={([yshift=-.4ex]current bounding box.center)}]
\node (v1) at (0.0,0.) [Gamma] {}; 
\node (v3) at (0.0,-0.6) [Sigma] {}; 
\node (v4) at (0.6,0.) [Sigma] {}; 
\node (l1) at (-0.4,0) {}; 
\node (l2) at (1.,0.) [VEV] {}; 
\node (l3) at (0,-1.0) [VEV] {}; 
\node (l4) at (0.6,0.4) [VEV] {}; 
\draw[double distance=4mm,thick] (l1) -- (v1);
\draw[-,cyan] (v1) -- (v3);
\draw[-,cyan] (v1) -- (v4);
\draw[-,cyan] (v3) -- (v4);
\draw[-] (v4) -- (l2);
\draw[-] (v3) -- (l3);
\draw[-] (v4) -- (l4);
\end{tikzpicture}
+
 \begin{tikzpicture}[scale=2,baseline={([yshift=-.4ex]current bounding box.center)}]
\node (v1) at (0.0,0.) [Gamma] {}; 
\node (v2) at (-0.6,0.) [VR] {\,$V^{(4)}_{\phi\psi}$\,}; 
\node (v3) at (0.0,-0.6) [Sigma] {}; 
\node (v4) at (0.6,0.) [Sigma] {}; 
\node (l1) at (-1.,0) {}; 
\node (l2) at (1.,0.) [VEV] {}; 
\node (l3) at (0,-1.0) [VEV] {}; 
\node (l4) at (0.6,0.4) [VEV] {}; 
\draw[double distance=4mm,thick] (l1) -- (v2);
\draw[double distance=4mm,thick,cyan] (v1) -- (v2);
\draw[-,cyan] (v1) -- (v3);
\draw[-,cyan] (v1) -- (v4);
\draw [-,cyan](v3) -- (v4);
\draw[-] (v4) -- (l2);
\draw[-] (v3) -- (l3);
\draw[-] (v4) -- (l4);
\end{tikzpicture} \,\,,
\label{eq:d3Pi_dphi3_fermions}
\end{align}
 
and finally, those of the fermionic self-energy
 \begin{align}
 \begin{tikzpicture}[scale=2,baseline={([yshift=-.4ex]current bounding box.center)}]
\node (v1) at (0.0,0.) [Sigma] {}; 
\node (l1) at (-0.4,0) {}; 
\node (l2) at (0.4,0.) [VEV] {}; 
\node (l3) at (0,-0.4) [VEV] {}; 
\draw[double distance=4mm,thick,cyan] (l1) -- (v1);
\draw[-] (v1) -- (l2);
\draw[-] (v1) -- (l3);
\end{tikzpicture}
~=~ & \,\,
 \begin{tikzpicture}[scale=2,baseline={([yshift=-.4ex]current bounding box.center)}]
\node (v1) at (0.0,0.) [Gamma] {}; 
\node (v3) at (0.0,-0.6) [Sigma] {}; 
\node (v4) at (0.6,0.) [Pi] {}; 
\node (l1) at (-0.4,0) {}; 
\node (l2) at (1.,0.) [VEV] {}; 
\node (l3) at (0,-1.0) [VEV] {}; 
\draw[double distance=4mm,thick,cyan] (l1) -- (v1);
\draw[double distance=4mm,thick,cyan] (v1) -- (v3);
\draw[double distance=4mm,thick] (v1) -- (v4);
\draw[-] (v4) -- (l2);
\draw[-] (v3) -- (l3);
\end{tikzpicture}
+ \,
 \begin{tikzpicture}[scale=2,baseline={([yshift=-.4ex]current bounding box.center)}]
\node (v1) at (0.0,0.) [Gamma] {}; 
\node (v2) at (-0.6,0.) [VR] {\,$V_{\psi\psi}$\,}; 
\node (v3) at (0.0,-0.6) [Sigma] {}; 
\node (v4) at (0.6,0.) [Pi] {}; 
\node (l1) at (-1.,0) {}; 
\node (l2) at (1.,0.) [VEV] {}; 
\node (l3) at (0,-1.0) [VEV] {}; 
\draw[double distance=4mm,thick,cyan] (l1) -- (v2);
\draw[double distance=4mm,thick,cyan] (v1) -- (v2);
\draw[double distance=4mm,thick,cyan] (v1) -- (v3);
\draw[double distance=4mm,thick] (v1) -- (v4);
\draw[-] (v4) -- (l2);
\draw[-] (v3) -- (l3);
\end{tikzpicture}
+ \,
 \begin{tikzpicture}[scale=2,baseline={([yshift=-.4ex]current bounding box.center)}]
\node (v1) at (0.0,0.) [Gamma] {}; 
\node (v2) at (-0.6,0.) [VR] {\,$V^{(4)}_{\psi\phi}$\,}; 
\node (v3) at (0.0,-0.6) [Sigma] {}; 
\node (v4) at (0.6,0.) [Sigma] {}; 
\node (l1) at (-1.,0) {}; 
\node (l2) at (1.,0.) [VEV] {}; 
\node (l3) at (0,-1.0) [VEV] {}; 
\draw[double distance=4mm,thick,cyan] (l1) -- (v2);
\draw[double distance=4mm,thick] (v1) -- (v2);
\draw[double distance=4mm,thick,cyan] (v1) -- (v3);
\draw[double distance=4mm,thick,cyan] (v1) -- (v4);
\draw[-] (v4) -- (l2);
\draw[-] (v3) -- (l3);
\end{tikzpicture}
\label{eq:d2sig_dphi2}
\end{align}

\begin{align}
 \begin{tikzpicture}[scale=2,baseline={([yshift=-.4ex]current bounding box.center)}]
\node (v1) at (0.0,0.) [Sigma] {}; 
\node (l1) at (-0.4,0) {}; 
\node (l2) at (0.4,0.) [VEV] {}; 
\node (l3) at (0,-0.4) [VEV] {}; 
\node (l4) at (0,0.4) [VEV] {}; 
\draw[double distance=4mm,thick,cyan] (l1) -- (v1);
\draw[-] (v1) -- (l2);
\draw[-] (v1) -- (l3);
\draw[-] (v1) -- (l4);
\end{tikzpicture}
= & \,\,
 \begin{tikzpicture}[scale=2,baseline={([yshift=-.4ex]current bounding box.center)}]
\node (v1) at (0.0,0.) [Gamma] {}; 
\node (v3) at (0.0,-0.6) [Sigma] {}; 
\node (v5) at (0.0,0.6) [Pi] {}; 
\node (l1) at (-0.4,0) {}; 
\node (l2) at (0.4,-0.6) [VEV] {}; 
\node (l3) at (0,-1.0) [VEV] {}; 
\node (l4) at (0,1.) [VEV] {}; 
\draw[double distance=4mm,thick,cyan] (l1) -- (v1);
\draw[double distance=4mm,thick,cyan] (v1) -- (v3);
\draw[double distance=4mm,thick] (v1) -- (v5);
\draw[-] (v3) -- (l2);
\draw[-] (v3) -- (l3);
\draw[-] (v5) -- (l4);
\end{tikzpicture}
+ \,
 \begin{tikzpicture}[scale=2,baseline={([yshift=-.4ex]current bounding box.center)}]
\node (v1) at (0.0,0.) [Gamma] {}; 
\node (v2) at (-0.6,0.) [VR] {\,$V_{\psi\psi}$\,}; 
\node (v3) at (0.0,-0.6) [Sigma] {}; 
\node (v5) at (0.0,0.6) [Pi] {}; 
\node (l1) at (-1.,0) {}; 
\node (l2) at (0.4,-0.6) [VEV] {}; 
\node (l3) at (0,-1.0) [VEV] {}; 
\node (l4) at (0,1.) [VEV] {}; 
\draw[double distance=4mm,thick,cyan] (l1) -- (v2);
\draw[double distance=4mm,thick,cyan] (v1) -- (v2);
\draw[double distance=4mm,thick,cyan] (v1) -- (v3);
\draw[double distance=4mm,thick] (v1) -- (v5);
\draw[-] (v3) -- (l2);
\draw[-] (v3) -- (l3);
\draw[-] (v5) -- (l4);
\end{tikzpicture}
+ \,
 \begin{tikzpicture}[scale=2,baseline={([yshift=-.4ex]current bounding box.center)}]
\node (v1) at (0.0,0.) [Gamma] {}; 
\node (v2) at (-0.6,0.) [VR] {\,$V^{(4)}_{\psi\phi}$\,}; 
\node (v3) at (0.0,-0.6) [Sigma] {}; 
\node (v5) at (0.0,0.6) [Sigma] {}; 
\node (l1) at (-1.,0) {}; 
\node (l2) at (0.4,-0.6) [VEV] {}; 
\node (l3) at (0,-1.0) [VEV] {}; 
\node (l4) at (0,1.) [VEV] {}; 
\draw[double distance=4mm,thick,cyan] (l1) -- (v2);
\draw[double distance=4mm,thick] (v1) -- (v2);
\draw[double distance=4mm,thick,cyan] (v1) -- (v3);
\draw[double distance=4mm,thick,cyan] (v1) -- (v5);
\draw[-] (v3) -- (l2);
\draw[-] (v3) -- (l3);
\draw[-] (v5) -- (l4);
\end{tikzpicture}\,.
\label{eq:d3sig_dphi3}
\end{align}

\noindent With these, one can analyse the scalar $n$-point functions with fermionic contributions and obtain the field counterterms. Moreover, we note that this diagrammatic analysis extends to higher truncations of the 2PI effective action involving fermions.

\section{Conclusions and Outlook}
\label{sec:outlook}

We have investigated an on-shell scheme for the
2PI formalism with a particular focus on the
equations of motion, the motivation being that
one can obtain from these in principle the
transport equations which are relevant, for example, while studying
cosmological phase transitions. After an outline of 
the generic procedure, we have
revisited in a first step the so-called Hartree approximation
as one can obtain in this case analytic formulas for
all interesting quantities. 
We have given the
relation between the counterterms in the broken and
unbroken phases. Moreover, we have given 
the formulas for three- and four-point functions
in the broken phase. A particular feature of this approximation 
is, that all counterterms are finite and the resumed two-point function has the same form as the tree-level one. Neither of these hold when going beyond
this approximation.

In a second step, we have first allowed for an
additional independent 
trilinear scalar coupling in the classical 
action which induces the so-called scalar sunset 
diagram at two-loop level. We have used this toy model 
to give the explicit procedure on how to obtain the 
on-shell counterterms in a more complicated system. Moreover, 
we have shown that the two-point function $G$ can
be evaluated numerically in a fast converging iteration. This in turn serves as input for the
calculation of the renormalized three- and four-point functions. 

For a Yukawa theory, the equations of motion
for the scalar and fermionic two-point functions
are coupled. We have demonstrated that the numerical
procedure used in the pure scalar case works also
for this coupled system even for $\mathcal{O}(1)$ couplings.
We have given explicitly the counterterms for
the wave function and mass renormalization, whereas
in the case of the coupling counterterms, we give a 
diagrammatic form which can easily be translated into 
formulae, which are however very lengthy. 

The procedure
outlined in this paper can easily be extended to the
case of multiple scalars and fermions. It
can also be extended to include gauge fields
which we will discuss in a forthcoming paper.
From our results, one can easily get the counterterms
for other renormalization schemes such as
$\overline{\text{MS}}$. This scheme is widely
used for the calculation of the effective potential,
which is an important tool for the study of phase transitions. 

\section*{Acknowledgements}
We thank K.~Kainulainen and O.~Koskivaara for discussions and
Ch.~Gross for collaboration in the early stage of 
this project. This work has been supported by the 
DFG, project nrs.\ PO-1337/8-1 and HI 744/10-1.


\bibliographystyle{h-physrev5}
\bibliography{lit}

\end{document}